\documentclass[useAMS,usenatbib]{mn2e}

\usepackage{graphicx}
\usepackage{multirow}

\newcommand{\brg}{Br~$\gamma$} %2.1661 microns
\newcommand{\water}{H$_{2}$O}
\newcommand{\h}{H$_{2}$}
\newcommand{\lsol}{L$_{\odot}$}

\newcommand{\msol}{M$_{\odot}$}

\newcommand{\flux}{Wm$^{-2}\umu$m$^{-1}$}

\newcommand{\aj}[1]{AJ}
\newcommand{\mnras}[1]{MNRAS}
\newcommand{\apj}[1]{ApJ}
\newcommand{\apjs}[1]{ApJS}
\newcommand{\apjl}[1]{ApJ}
\newcommand{\aap}[1]{A\&A}
\newcommand{\aaps}[1]{A\&AS}
\newcommand{\nat}[1]{Nature}
\newcommand{\araa}[1]{ARA\&A}
\newcommand{\baas}[1]{BAAS}

\title{The RMS Survey: Near-IR Spectroscopy of Massive Young Stellar Objects}

\author[H.D.B. Cooper et. al.]{H.D.B. Cooper,$^1$ S.L. Lumsden,$^1$
R.D. Oudmaijer,$^1$ M.G. Hoare,$^1$ A.J. Clarke,$^1$ \newauthor
J.S. Urquhart,$^2$ J.C. Mottram,$^3$ T.J.T. Moore,$^4$ B. Davies$^4$ \\
$^1$School of Physics \& Astronomy, University of Leeds, Leeds, LS2
9JT, UK \\
$^2$Max-Planck-Institut für Radioastronomie, 53121 Bonn, Germany \\
$^3$Leiden Observatory, Leiden University, 2300 RA, Leiden,
Netherlands \\
$^4$Astrophysics Research Institute, Liverpool John Moores University, Twelve Quays House, Egerton Wharf, Birkenhead CH41 1LD, UK
}

\begin{document}

\date{\today}

\maketitle
\label{firstpage}

\begin{abstract}
Near-infrared H- and K-band spectra are presented for 247 objects,
selected from the Red MSX Source (RMS) survey as potential young
stellar objects (YSOs). 195 ($\sim 80\%$) of the targets are YSOs,
of which 131 are massive YSOs (${\rm L}_{{\rm BOL}} >
5~\times~10^{3}$~\lsol, ${\rm M} > 8$~\msol). This is the largest
spectroscopic study of massive YSOs to date, providing a valuable
resource for the study of massive star formation. In this paper we
present our exploratory analysis of the data. The YSOs observed have
a wide range of embeddedness ($2.7 < A_{V} < 114$), demonstrating
that this study covers minimally obscured objects right through to
very red, dusty sources. Almost all YSOs show some evidence for
emission lines, though there is a wide variety of observed
properties. The most commonly detected lines are \brg, \h,
fluorescent Fe~II, CO bandhead, [Fe~II] and He~I 2--1 $^1$S--$^1$P,
in order of frequency of occurrence. In total, $\sim 40\%$ of the
YSOs display either fluorescent Fe~II 1.6878~$\umu$m or CO bandhead
emission (or both), indicative of a circumstellar disc; however, no
correlation of the strength of these lines with bolometric
luminosity was found. We also find that $\sim 60\%$ of the sources
exhibit [Fe~II] or \h\ emission, indicating the presence of an
outflow. Three quarters of all sources have \brg\ in emission. A
good correlation with bolometric luminosity was observed for both
the \brg\ and \h\ emission line strengths, covering 1~\lsol~$< {\rm
L}_{{\rm BOL}} < 3.5~\times~10^{5}$~\lsol. This suggests that the
emission mechanism for these lines is the same for low-,
intermediate-, and high-mass YSOs, i.e. high-mass YSOs appear to
resemble scaled-up versions of low-mass YSOs.

\end{abstract}


\section{Introduction}\label{sec:intro}
Massive stars ($>8$ \msol, i.e. $>5~\times~10^{3}$~\lsol) have a profound
influence on their surroundings due to their intense ionizing
radiation, strong stellar winds and powerful outflows. However,
despite the fact that low-mass star formation is broadly
understood \citep{1987ShuAdamsLizano}, relatively little is known
about the conditions and processes that influence high-mass star
formation. The study of massive star formation is hindered by the
rarity of massive stars, their short evolutionary time-scales, and
the dust extinction which renders the stars invisible at short
wavelengths for most of their formation. 

The massive young stellar object (MYSO) phase is of particular
interest for the study of massive star formation. In this relatively
brief phase (10$^{4-5}$ years: \citealt{2011Davies-accretion,
2011Mottram-luminosity}), fusion has begun in the central star, but
it has not yet ionized its surroundings to form an H~II region.
Accretion may be disrupted or completely halted, in which case the
final mass of the star would be set in this phase. MYSOs typically
have strong, ionized stellar winds \citep{1995MNRAS.272..346B} and
bipolar molecular outflows \citep{1996ApJ...472..225S}, both of
which have an important feedback effect on the surrounding molecular
cloud. 

Whilst these properties mean that the MYSO stage is crucial to the
understanding of massive star formation, their observation is not
without difficulties. Dust extinction hinders optical and IR
studies, their lack of H~II regions means that radio emission is
either weak or non-existent, and although MYSOs are often associated
with masers, no single maser transition can be guaranteed to be
present \citep{2002A&A...390..289B}.

Most of the bolometric luminosity of MYSOs is found in the thermal
IR, so that is where they are most easily observed. Previous attempts
to produce a large sample of MYSOs have been made using IRAS (e.g.
\citealt{1996A&A...308..573M, 2002ApJ...566..931S}). The low spatial
resolution of IRAS (3\arcmin\ $\times$ 5\arcmin\ at 100~$\umu$m)
meant that in star clusters and in the dense environment of the
Galactic Plane sources were often confused. Since most massive stars
form in clusters and exist within the Galactic Plane, the MYSOs in
these studies may not be representative of all massive stars. The
small number of MYSOs that have been identified so far means that
there is a lack of statistical evidence with which to improve our
understanding of high-mass star formation. 

The Red MSX-Source Survey (RMS; \citealt{2005Hoare-RMS-conference,
2008Urquhart-rms-general}) is a valuable tool with which to address
these issues, having taken both imaging and spectroscopic data from
near-IR to radio wavelengths on a comprehensive, uniformly selected
sample of MYSOs. The MSX point source catalogue
\citep{2003AAS...203.5708E} and near-IR 2 Micron All Sky Survey
(2MASS; \citealt{2006AJ....131.1163S}) data were used to find sources
with IR excess. The MSX survey has a much better spatial resolution
than IRAS, thus eliminating the bias against dense environments
\citep{2004Hoare}. The reddest, most embedded objects were selected
as targets using colour cuts defined by \citet{2002Lumsden}. However,
ultra-compact H~II (UCHII) regions, planetary nebulae (PN) and dusty
evolved stars also have similar colours to MYSOs, so classification
of the sources has been a major part of the project. A campaign of
follow-up observations has been undertaken to achieve this
(see \citealt{2008Urquhart-rms-general}). We combined radio, mid-IR, and
near-IR images with kinematic distances and far-IR data to
characterise our sources \citep{2007Mottram-mid-ir,
2007Urquhart-radio-S, 2007Urquhart-13CO-S, 2008Urquhart-13CO-N,
2009Urquhart-radio-N, 2009Urquhart-H2O, 2010Mottram-far-ir-photometry,
2011Mottram-luminosity, 2011Urquhart-ammonia,
2011Urquhart-dist-prop, 2012Urquhart-KDA}. The full classification
criteria are discussed in Lumsden et al. (in preparation).

Near-IR spectroscopy is the final classification stage, separating
the MYSOs from the remaining sources. To date, $\sim 90 \%$ of the
$\sim 2000$ objects from the original survey have been classified,
identifying $\sim 600$ MYSOs and $\sim 500$ UCHII
regions.\footnote{See RMS data base http://www.ast.leeds.ac.uk/RMS/,
Lumsden et al. (in preparation)} In total, we have spectroscopic observations
of $\sim 450$ objects. The data for the $\sim 200$ objects observed
in the Southern Hemisphere are still in the analysis process, and
will be presented in a future paper. In this paper we present near-IR spectra and
classifications of 247 objects observed in the Northern Hemisphere,
i.e. $\sim 12\%$ of the objects in the survey. 

%%%%%%%%%%%%%%%%%%%%%%%%%%%%%%%%%%%%%%%%
\section{Observations}
Spectroscopic observations of the young stellar object candidates
were made using the UIST instrument at the United Kingdom Infra-Red
Telescope (UKIRT) observatory from 2002 to 2008. 247 objects were
successfully observed over 84 nights. Sources were selected
from the $\sim 2000$ candidate MYSOs found using the MSX catalogue
in the preceding stages of the RMS survey. Initially, no radio or
luminosity data were available, so the criteria for the
spectroscopic observations were that the objects satisfy the colour
cuts laid out in \citet{2002Lumsden}, and that they were bright
enough to be seen at the K-band in 2MASS to enable correct
identification of the counterpart. Once luminosity estimates were
available, a loose luminosity cut of L$_{21\umu{\rm m}}\ge500$
~\lsol\ (which corresponds to L$_{{\rm BOL}}\ge10000$~\lsol, see
\citealt{2011Mottram-luminosity}) was imposed. Therefore fewer
low-luminosity sources were observed in the later runs. We targeted
our source selection to observe all objects we believed to be YSOs
with the appropriate luminosity, plus the 10\% of the $\sim 2000$
candidates whose classifications were ambiguous. A full list of
observational parameters for each individual object is shown in
Tables A1 and A2 (see Appendix A). 

UIST is a 1--5~$\umu$m imager-spectrometer with a 1024$\times$1024
InSb array and a camera with 0.12\arcsec/pixel resolution in
spectroscopy mode. The HK grism (1.395--2.506~$\umu$m) was used,
with a 0.48\arcsec$\times$120\arcsec\ long slit. This produces a
fairly low spectral resolution of $\lambda$/$\Delta\lambda\sim$500,
but meant that in ideal conditions a high signal-noise ratio could be
achieved with relatively short exposure time. This allowed
identification of faint lines whilst maximising the number of objects
that could be observed in the time awarded. In practise a large
number of the observations were done in poor weather conditions and with
typical seeing of 1\arcsec, so the signal-noise ratio was not always
very high, and flux calibration was unreliable on some nights.
Exposure times for each object are shown in
Table A1. 

Many of the observations were taken before the near-IR counterparts
had been accurately identified. As a result, finding charts were
frequently done at the telescope using UIST in imaging mode so that
the counterpart could be identified prior to taking the spectrum. In
a few instances, such spectra were repeated later if subsequent high
spatial resolution thermal IR images suggested a different source
was the correct counterpart. The position angle for the slit (also
shown in Table A1) was derived from such images, as well as from
2MASS images for brighter, isolated sources, or
GLIMPSE\footnote{GLIMPSE: the Galactic Legacy Infrared Mid-Plane
Survey Extraordinaire -- a Spitzer legacy science programme which
imaged the Galactic Mid-Plane at 3.6, 4.5, 5.8, and 8~$\umu$m.} for
complex embedded sources. The aim was to avoid very bright,
unrelated stars nearby, and if possible to align the slit on other
interesting objects and extended emission nearby. This proved a
useful technique as several fields of view had more than one MYSO
within the field, and many have extended emission. In the tables,
multiple objects within the same field of view are denoted by
G*** -- 1 and G*** -- 2 etc.

Standard stars with spectral types ranging from A0 -- B9.5 were
observed before each object for flux calibration, and the objects
were nodded along the slit to minimise the effect of bad pixels and
aid sky subtraction. The spectrum of an argon lamp was taken each
night for wavelength calibration.
 
\begin{figure*}
\begin{minipage}{\textwidth}
\centering
\includegraphics[width=84mm]{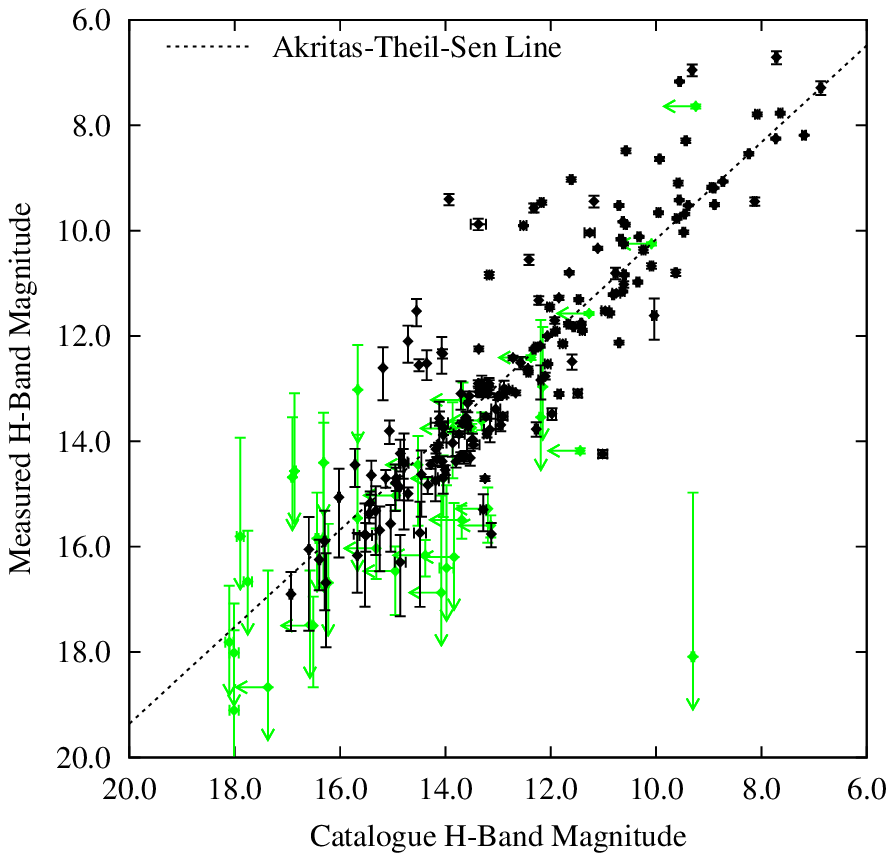}
\hspace{0.7cm}
\includegraphics[width=84mm]{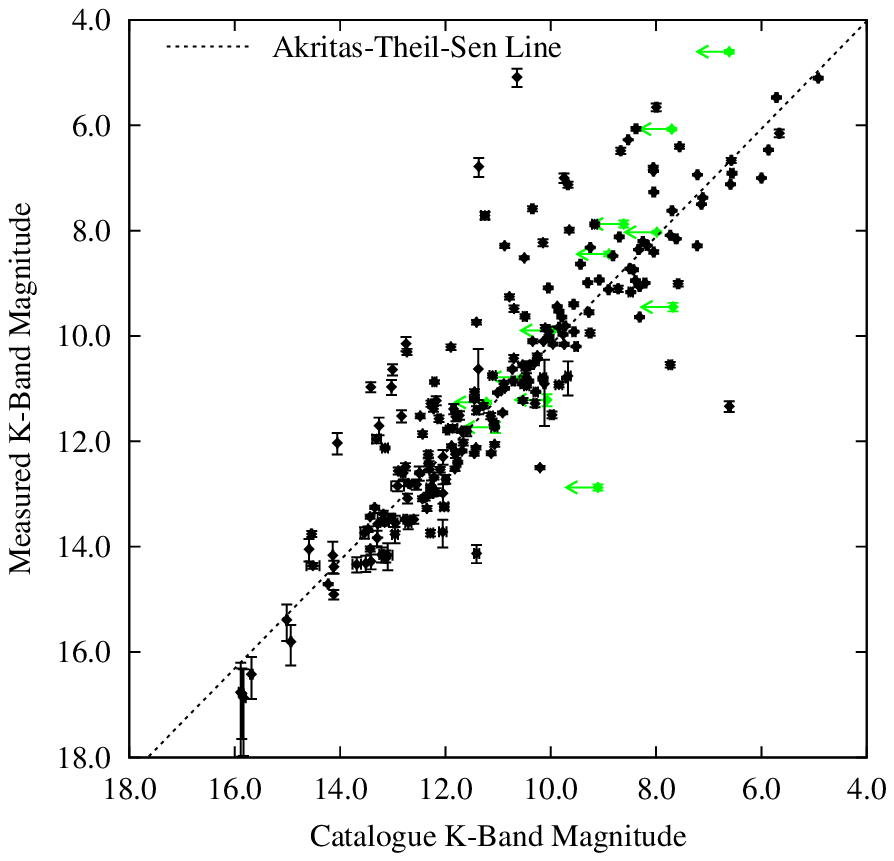}
\caption{Measured vs. catalogue magnitudes for both H- and K-bands.
The catalogue magnitudes were taken from either UKIDSS or 2MASS
depending on the object and the quality of the catalogue data. 
Detections are denoted by black diamonds and limits by green arrows.
The dotted lines represent the ATS regressions. As expected, a
strong correlation was found between the measured and catalogue
magnitudes (see text for more details).
}\label{fig1}
\end{minipage}
\end{figure*}
 
The standard star and object spectra were all reduced using the UKIRT
pipeline {\sc Orac-Dr} and the Starlink package {\sc Figaro}. To
prevent confusion between emission lines in the object spectra and
the H~I absorption lines in the standard star spectra, the lines in
the standards were fitted with Gaussian functions and removed before
flux calibrating the object spectra. Flux calibration was done using the {\sc Figaro} routine
{\sc irflux}. This routine uses a black-body model for the standard
star, normalised to the star's flux at K. The K-band magnitudes and
spectral types of the standard stars were taken from the SIMBAD
data base. The temperatures corresponding to the spectral types were
taken from \citet{schmidt-kaler}.

\section{Photometry} \label{sec:magnitudes}
We compared our magnitude data with the 2MASS Point Source Catalogue
(2MASS PSC; \citealt{2006AJ....131.1163S}) and UKIDSS GPS data base
release 8.\footnote{The UKIDSS project is defined in
\citet{2007MNRAS.379.1599L}. UKIDSS uses the UKIRT Wide Field Camera
(WFCAM; \citealt{2007A&A...467..777C}) and a photometric system
described in \citet{2006MNRAS.367..454H}. The pipeline processing and
science archive are described in \citet{2008eic..work..541I} and
\citet{2008MNRAS.384..637H}.} The choice of catalogue used for
analysis was decided on an object-by-object basis. Some objects are
listed in both catalogues, others in only one, and some do not have
any near-IR magnitude data available. Where available, UKIDSS GPS
magnitudes were preferred due to their improved sensitivity and
resolution. However, due to its incomplete status, not all the
objects we observed were covered by UKIDSS, and those that were do
not always have data for all of J-, H- and K-bands. In addition, the
UKIDSS camera, WFCAM \citep{2007A&A...467..777C} tends to saturate
for the brighter objects. For objects with incomplete, low quality,
or saturated WFCAM data 2MASS was used. Overall, for our 247 sources,
125 were compared to 2MASS photometry, 108 to UKIDSS, and 14 had no
magnitude data available in either catalogue. The data for each
object are found in Table A2. 
 
The magnitudes were also measured from the spectra using the {\sc Figaro}
function {\sc istat}, which gives the mean and total flux and the
r.m.s. ($\sigma$) across a user-defined wavelength range. The ranges
used to measure the total flux were 1.5365--1.7875~$\umu$m and
2.028--2.290~$\umu$m for H and K respectively. These correspond to
the wavelength ranges for the 2MASS magnitude measurements, as
described in \citet{2003AJ....126.1090C}. UKIDSS magnitudes are
calibrated to 2MASS \citep{2009MNRAS.394..675H}, so the same ranges
were used to measure the spectral magnitude regardless of which
catalogue magnitudes were used for comparison. The continuum noise
($\sigma$) was measured using 0.01~$\umu$m ranges around 1.6~$\umu$m 
and 2.2~$\umu$m for H and K respectively. These ranges are
clear of emission lines and of distortion from atmospheric features,
both of which can cause an over-estimation of $\sigma$. The fluxes
were then converted into magnitudes, using the H and K zero-points
$1.133~\times~10^{-9}$ and $4.283~\times~10^{-10}$ \flux\ for 2MASS
sources \citep{2003AJ....126.1090C}, and $1.131~\times~10^{-9}$ and
$4.271~\times~10^{-10}$ \flux\ for UKIDSS sources
\citep{2009MNRAS.394..675H}. The errors on the magnitudes given in
Table A2 are $3\sigma$ errors. Some of the sources were not detected at
H, in which case their H-band magnitudes could not be calculated.
However, most spectra were bright enough at H that their magnitudes
could be measured, and all were bright enough at K.
 
\begin{figure*}
\begin{minipage}{84mm}
\includegraphics[width=84mm]{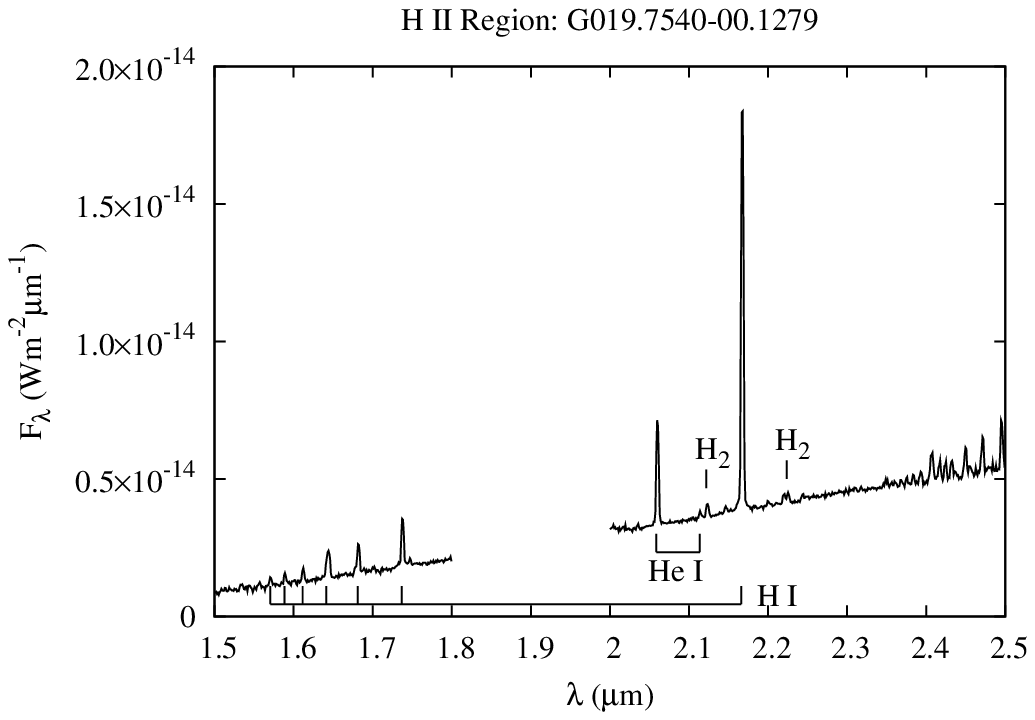}
\end{minipage}
\hspace{0.7cm}
\begin{minipage}{84mm}
\centering
\includegraphics[width=84mm]{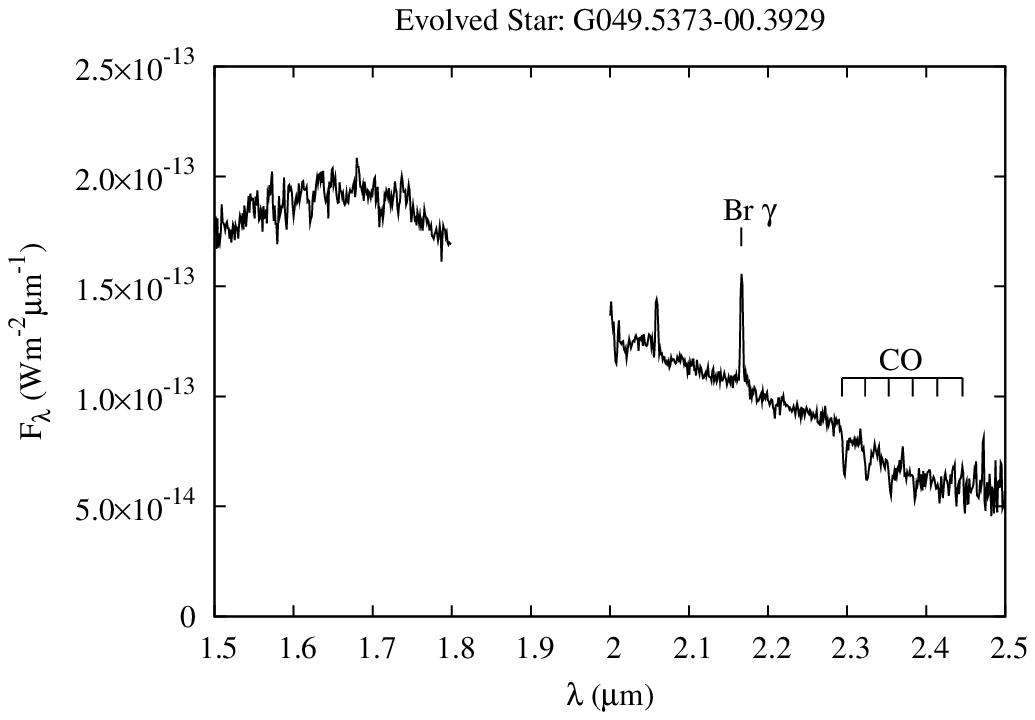}
\end{minipage}
\hspace{0.7cm}
\begin{minipage}{84mm}
\includegraphics[width=84mm]{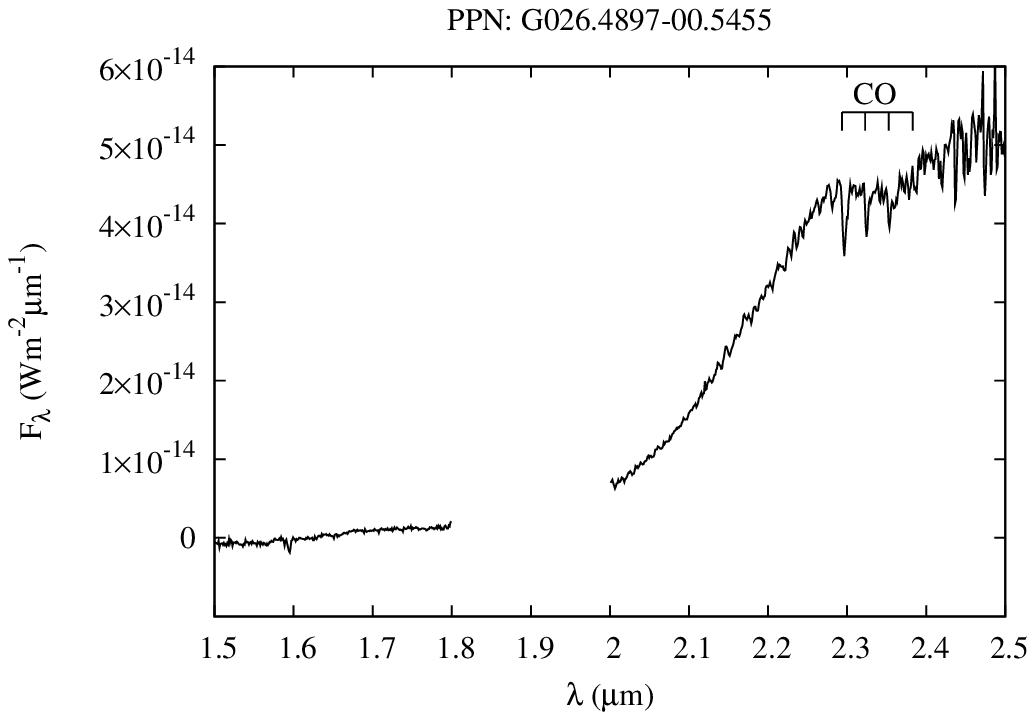}
\end{minipage}
\hspace{0.7cm}
\begin{minipage}{84mm}
\includegraphics[width=84mm]{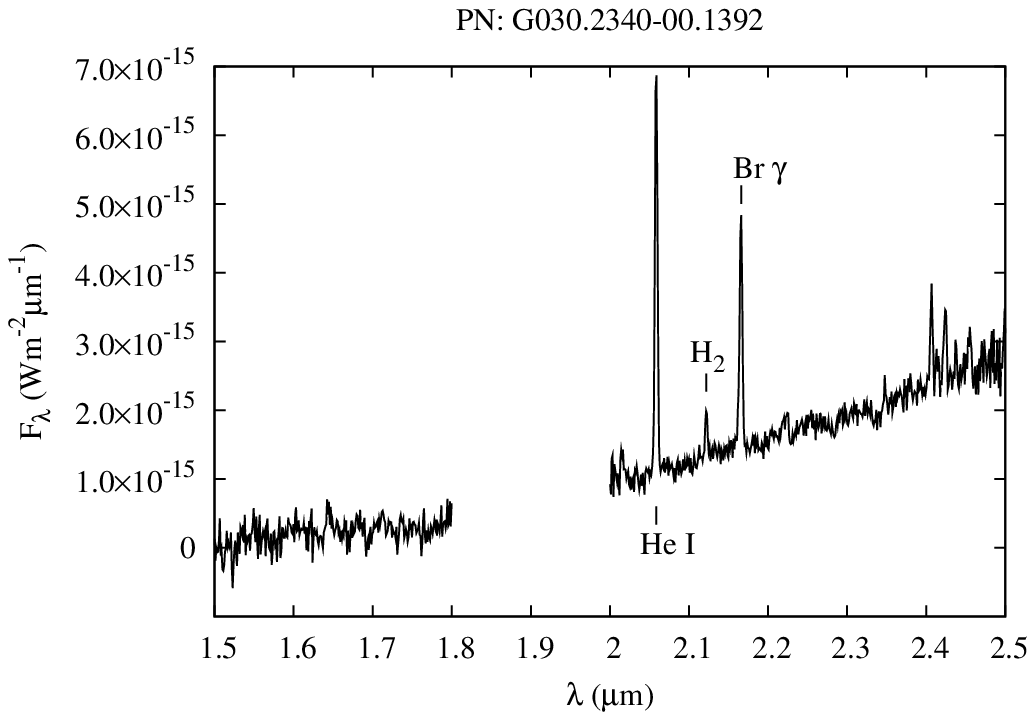}
\end{minipage}
\hspace{0.7cm}
\begin{minipage}{84mm}
\includegraphics[width=84mm]{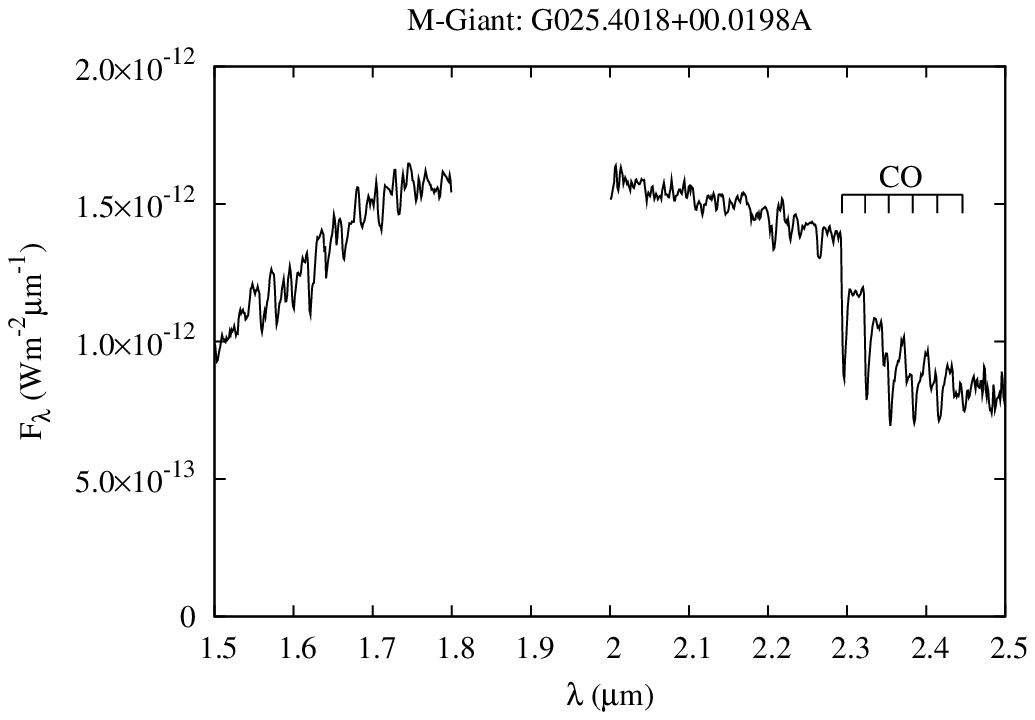}
\end{minipage}
\hspace{0.7cm}
\begin{minipage}{84mm}
\includegraphics[width=84mm]{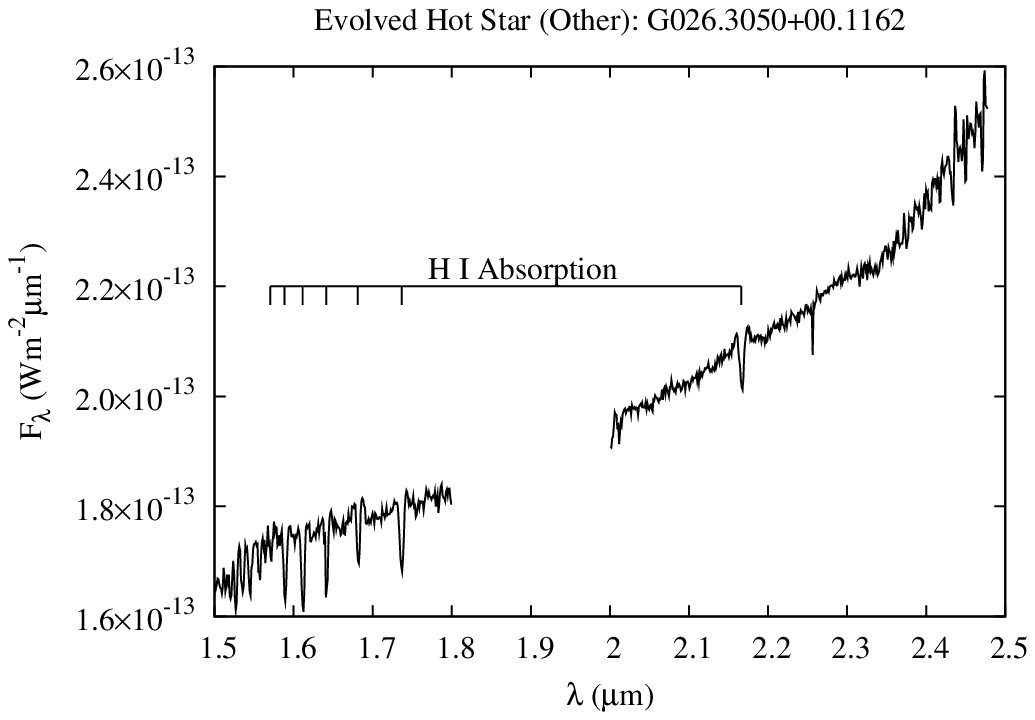}
\end{minipage}
\caption{Example spectra of each class of contaminants found within
the sample i.e. the non-YSOs. From
top-left to bottom-right: the H~II region G019.7540-00.1279; the
evolved star G049.5373-00.3929; the PPN G026.4897-00.5455; the PN
G030.2340-00.1392; the M-Giant G025.4018+00.0198A; and finally the
hot star G026.3050+00.1162, labelled as `Other' in the
data base and tables. Note the similarity of the PN spectrum to that
of the H~II region. This object required the RMS follow-up
campaign's mid-IR imaging data in order to be correctly classified.}
\label{fig2}
\end{figure*} 
 
Figure \ref{fig1} shows a comparison of catalogue with measured
magnitudes, for both H- and K-bands. The dotted lines represent
Akritas-Thiel-Sen (ATS) regressions. The ATS regression and
Kendall's $\tau$ were calculated using the {\sc cenken}
\citep{cenken} function in the NADA library \citep{NADA} from the
R-project statistics package, and compared with other regression methods and
correlation tests performed using the {\sc Asurv} statistics
package \citep{stats-methods,asurv1,asurv2}. These statistical
methods were used to take account of the many censored data present
(see also Section \ref{sec:accretion-disks}). Based on a $y=a+bx$
fit, the ATS fit coefficients for H are $a_{{\rm H}}=0.96$ and $b_{{\rm
H}}=0.92$. For K, the coefficients are $a_{{\rm K}}=-0.08$ and
$b_{{\rm K}}=1.02$. The Kendall's $\tau$ values for H and K are
0.48 and 0.70 respectively, both with $p$-values $\simeq0$; i.e. there is
essentially no possibility that a correlation does not exist. 

The lower $\tau$ and greater deviation in the fit from $m_{\rm obs}
= m_{\rm cat}$ for H-band magnitudes is primarily due to the fact
that many of the fainter objects have very noisy H-band spectra, and
some were not detected at shorter wavelengths. This means that there are more limits, 
and the errors at the high end of the magnitude range
are larger, both of which decrease the measured correlation
coefficient and pull the regression line away from $m_{\rm obs} =
m_{\rm cat}$. Nevertheless, the correlation is clearly there for
both H- and K-bands, and the regression coefficients for the K-band
are very close to $m_{\rm obs} = m_{\rm cat}$. This shows that, for
most objects, the photometry measured from the spectra is good
despite the frequently poor weather conditions. 
 
\begin{figure*}
\begin{minipage}{84mm}
\includegraphics[width=84mm]{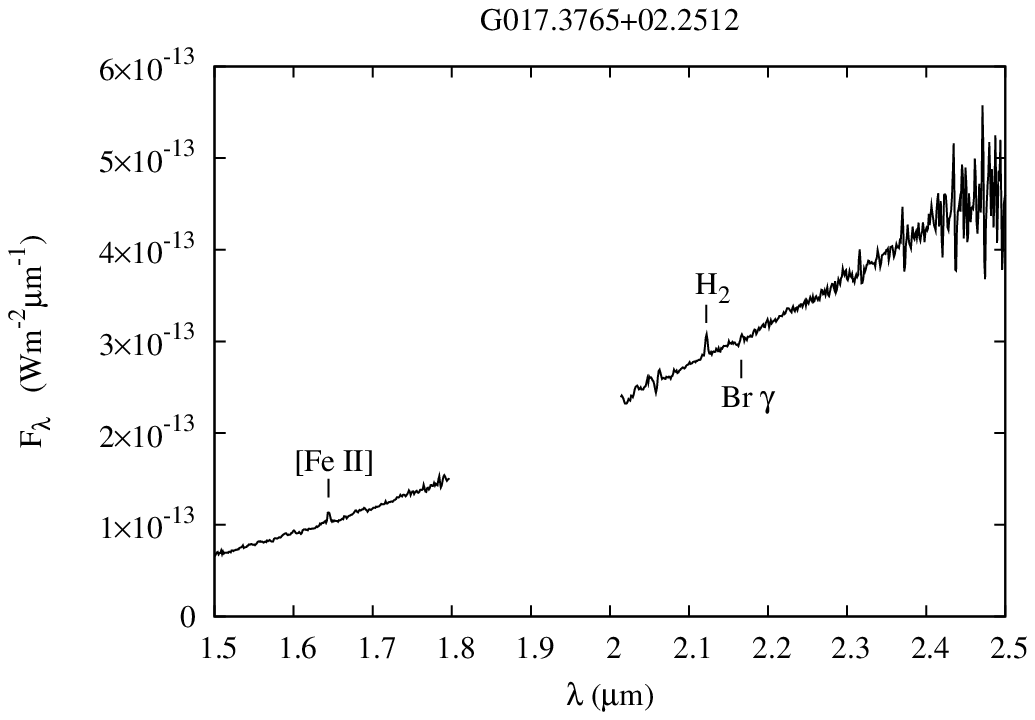}
\end{minipage}
\hspace{0.7cm}
\begin{minipage}{84mm}
\includegraphics[width=84mm]{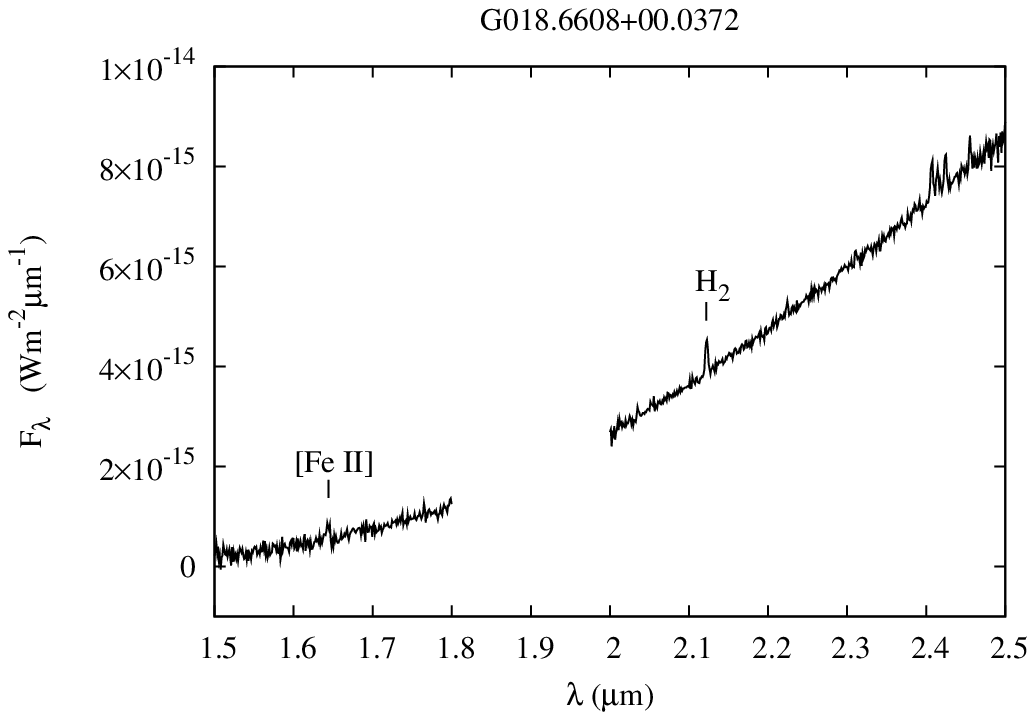}
\end{minipage}
\hspace{0.7cm}
\begin{minipage}{84mm}
\includegraphics[width=84mm]{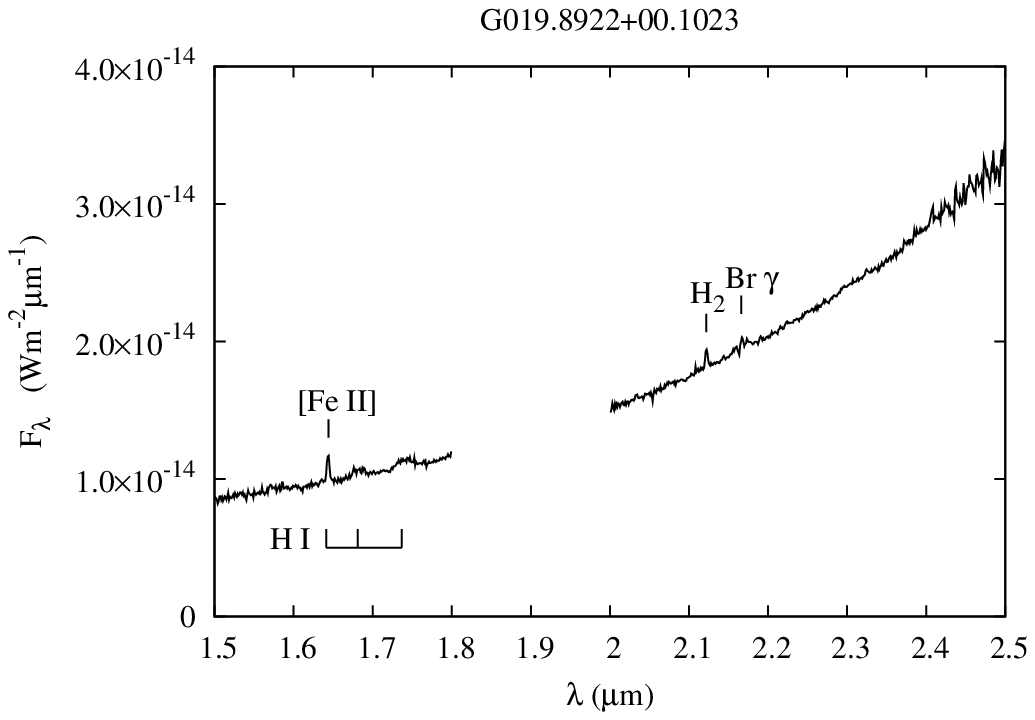}
\end{minipage}
\hspace{0.7cm}
\begin{minipage}{84mm}
\includegraphics[width=84mm]{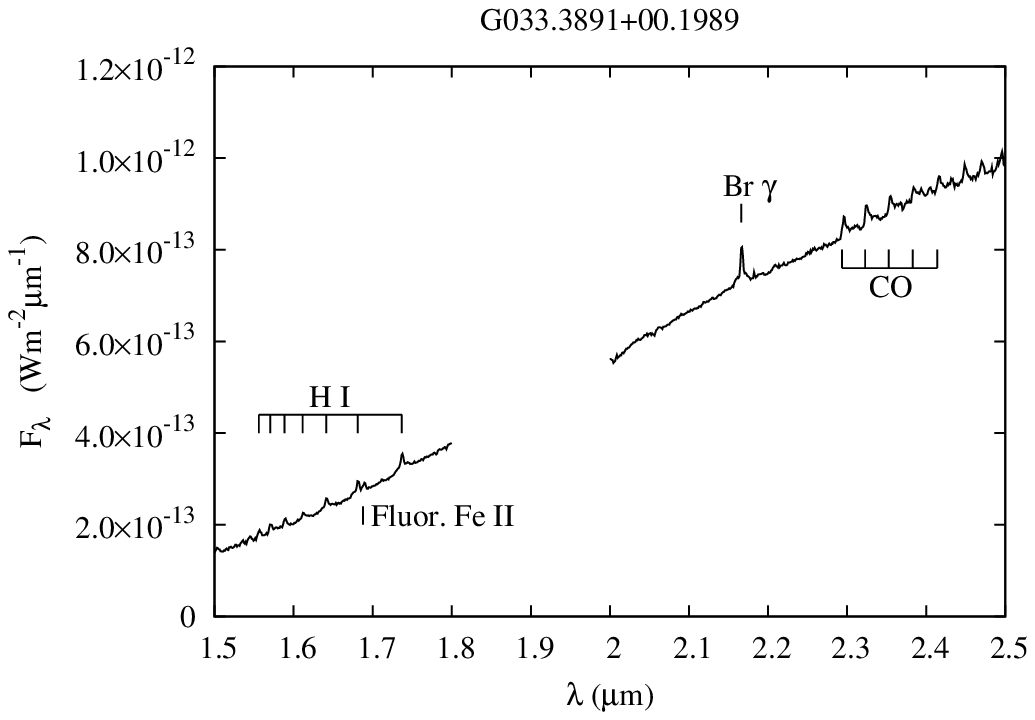}
\end{minipage}
\hspace{0.7cm}
\begin{minipage}{84mm}
\includegraphics[width=84mm]{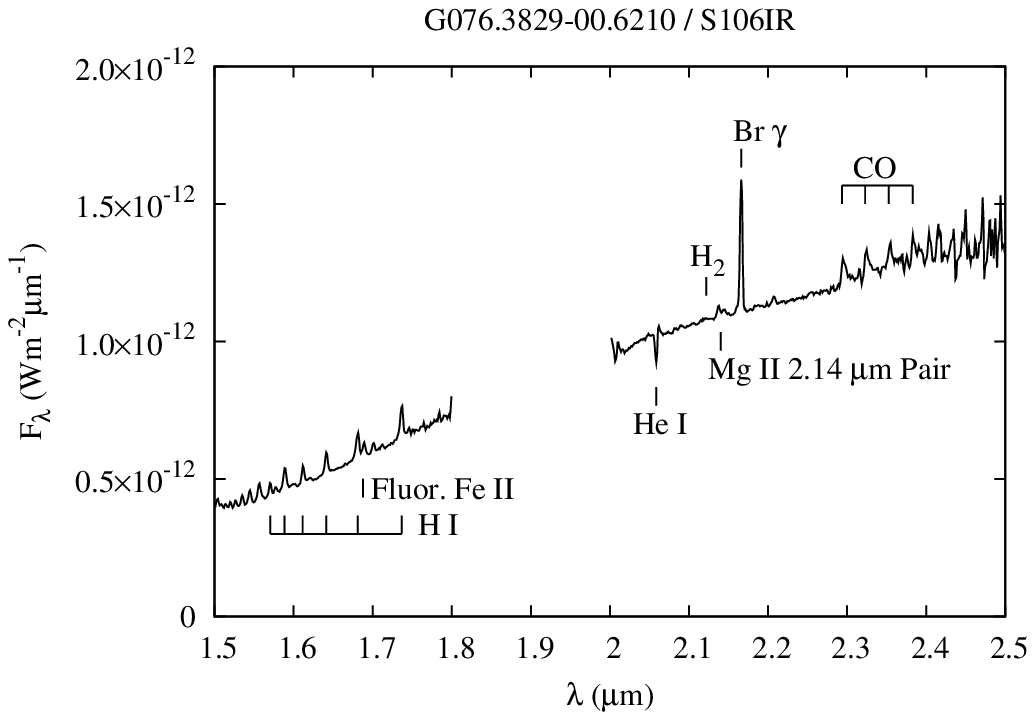}
\end{minipage}
\hspace{0.7cm}
\begin{minipage}{84mm}
\includegraphics[width=84mm]{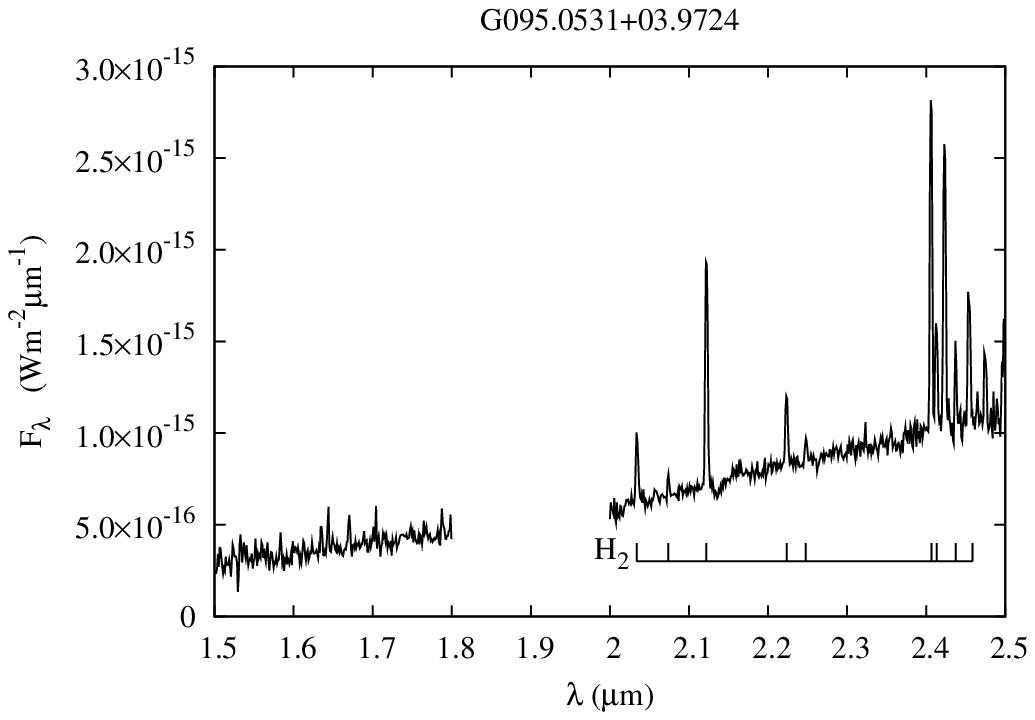}
\end{minipage}
\caption{Example spectra of the MYSOs. All have the steep rising
continuum and some emission lines, however there is variation in
which lines are present. This figure shows all the
major emission lines seen in MYSOs, but is not intended to be
representative in terms of the proportion of MYSOs that show each
emission line.} \label{fig3}
\end{figure*}
 
Those objects that deviate strongly from the $m_{\rm obs} = m_{\rm
cat}$ were all taken in very poor weather conditions and/or with
poor seeing. The conditions were highly time-variable for these
badly affected observations. The much longer exposure times for the
objects average out such variations whereas the shorter exposure
times for the standards do not. The raw standard star
data counts were noticeably very low for these observations in
particular. The ratio of object to standard then becomes exaggerated. 
Since there is only one standard star spectrum per
object, there was no way to mitigate against this. For consistency,
all the fluxes were normalised to the catalogue magnitudes. The
correction factor, $f_{\rm obs}/f_{\rm corr}=10^{(m_{\rm obs}-m_{\rm
cat})/-2.5}$, was calculated using the standard equation for
apparent magnitude, and is shown in Table A3. 

%%%%%%%%%%%%%%%%%%%%%%%%%%%%%%%%%%%%%%%%
\section{Classification}\label{sec:classification}
  
\begin{figure*}
\begin{minipage}{\textwidth}
\centering
\includegraphics[width=84mm]{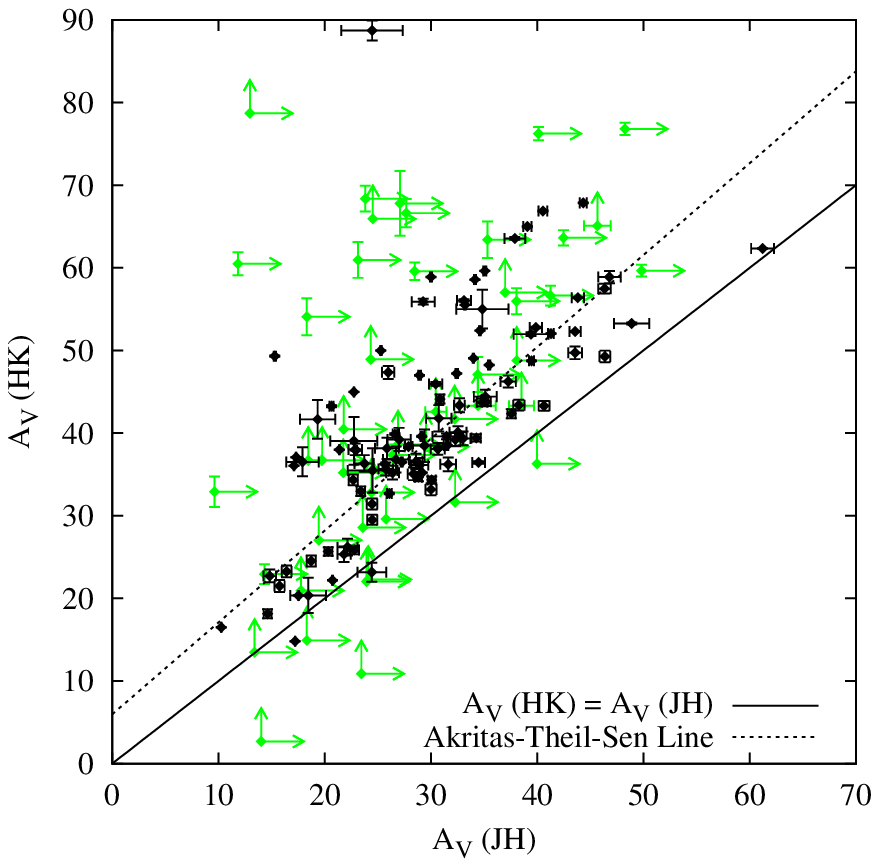}
\hspace{0.7cm}
\includegraphics[width=84mm]{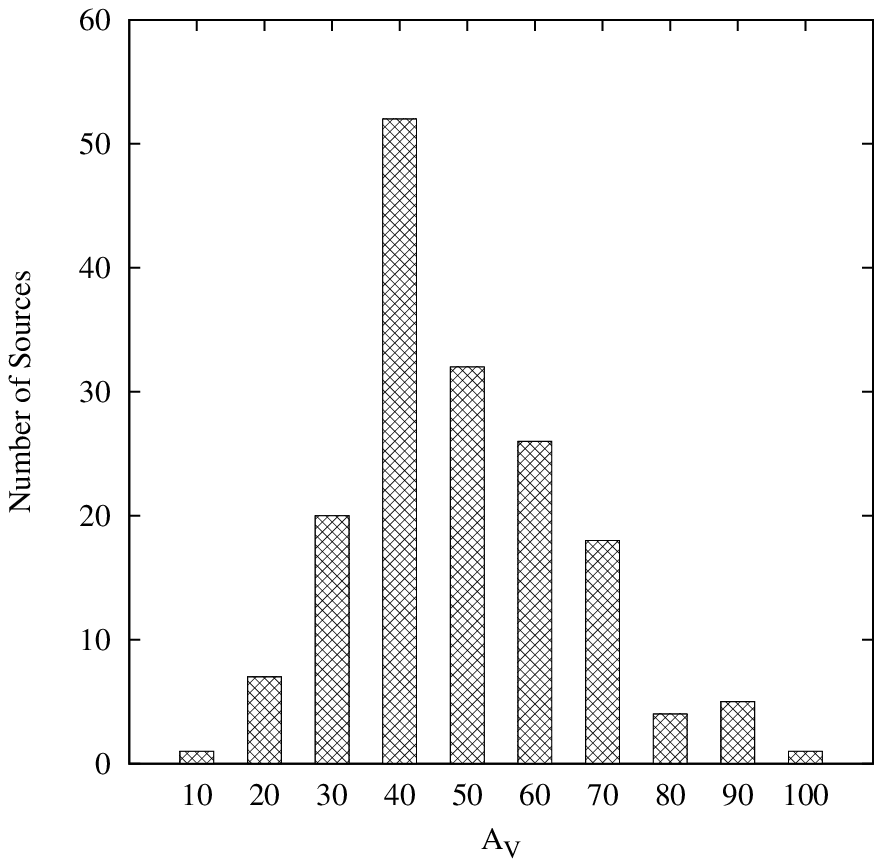}
\caption{\emph{Left}: Comparison of extinction calculated using J-
and H-band magnitudes ($A_{V}$ (JH)) and H- and K-band magnitudes
($A_{V}$ (HK)), for the YSOs. Detections are denoted by black
diamonds and limits by green arrows. The solid line represents
$A_{V}$ (HK) = $A_{V}$ (JH), and the dotted line is the ATS
regression line. \emph{Right}: Number distribution of $A_{V}$ for
the YSOs, using $A_{V}$ (HK) where available, or otherwise using
$A_{V}$ (JH).}\label{fig4}
\end{minipage}
\end{figure*}
 
\begin{figure}
\begin{minipage}{\columnwidth}
\centering
\includegraphics[width=84mm]{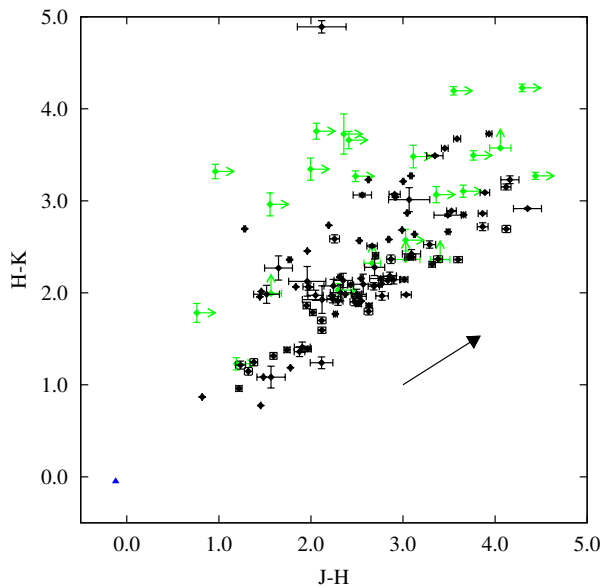}
\caption{Colour-colour diagram showing YSO $J-H$ and $H-K$ colours
for the RMS objects. Detections are denoted by black diamonds and
limits by green arrows. The black arrow shows the extinction vector,
of length $A_{V} = 10$, calculated using the
\citet{1989ESASP.290...93D} extinction law. The
blue triangle shows the intrinsic colours of a B0 star, taken from
\citet{1983A&A...128...84K}: $J-H=-0.12$ and
$H-K=-0.05$.}\label{fig5}
\end{minipage}
\end{figure}
 
The objects were primarily classified using the RMS follow-up
observations. These criteria are described in detail in
Lumsden et al. (in preparation). In summary, MYSOs conform to the colour cuts
defined in \citet{2002Lumsden}, are radio-quiet due to insufficient
ionized gas, show strong millimetre CO emission as expected in a
star formation region, and appear as point sources in mid-IR images
at 1\arcsec resolution. They show clustering in near-IR images in
contrast to more evolved stars, which appear as isolated point sources. This
left a sample of predominantly MYSOs, but also of some H~II regions,
proto-planetary nebulae, compact planetary nebulae and embedded
evolved stars. Near-IR spectroscopy was necessary to distinguish
between the MYSOs and contaminants.

The spectra were classified according to the shape of their
continuum and the presence or absence of emission and/or absorption
features. All the spectra show telluric water absorption in the
$1.8 - 2.0\ \umu$m range. Consequently, this range could not be used
in the analysis and was taken out of the figures. Example spectra of
the contaminants are shown in Figure \ref{fig2}. Figure \ref{fig3}
shows example spectra of MYSOs. Each major emission line is present
in at least one of the MYSO spectra shown in the example figures.
Spectra of all objects observed will be available on the RMS
data base.\footnote{See RMS data base
http://www.ast.leeds.ac.uk/RMS/,Lumsden et al. (in preparation)} 

MYSOs are still embedded in their natal clouds, so their spectra show
strong dust extinction in the form of a steep rising continuum, and
occasionally excess dust emission at $\lambda \ga 2.3\ \umu$m.
Almost all have emission lines, the most common being \brg\ and \h.
Some also have H-band Brackett series, shocked [Fe~II], fluorescent
Fe~II, or CO bandhead emission. Both \brg\ and He~I can be seen in
emission, absorption, or with a P-Cygni-type profile, though at this
low resolution it was not possible to obtain expansion velocities for
such profiles. 

Massive YSOs were differentiated from low-mass YSOs by their
luminosity: a loose cut of ${\rm L} > 10^{3}$~\lsol\ was applied.
Most of the RMS YSOs have luminosities characteristic of massive
stars: 131/195 ($\sim 70 \%$) of the RMS YSOs in this paper have
${\rm L} > 5~\times~10^{3}$~\lsol, which corresponds to ${\rm M} >
8$~\msol\ \citep{2011Mottram-lum-funcs}. However, their spectral
properties are very different from those of typical main sequence
stars \citep{2005ApJS..161..154H}.

H~II regions have comparatively flat continua, strong H~I emission,
and, often, He~I emission. The H~I emission conforms to case-B of
\citet{BakerMenzel1938} since it originates in an optically thin
ionized region. It is worth noting that it was occasionally
difficult to distinguish between MYSOs with strong H~I emission and
H~II regions by their spectra alone. Generally, this was achieved
using other data from the RMS Survey follow-up or in the literature.
In particular, radio observations and GLIMPSE images were used. In
most cases, H~II regions produce free-free radio emission owing to
the large amount of ionized gas, whereas H~I emission in MYSOs does
not produce bright free-free emission. In addition, H~II regions
appear diffuse in mid-IR GLIMPSE images, so even radio-quiet `weak'
H~II regions can usually be distinguished from MYSOs. Sometimes the
suspected central star of an H~II region has characteristics of a
YSO. These were classified as `YSO/HIIR'.
 
Evolved stars have much less dust extinction and therefore have
comparatively blue continua, or continua that peak in the middle of
the 1.5--2.5~$\umu$m range. They occasionally have the Brackett
series in emission or absorption, and frequently have CO bandhead
absorption. Proto-planetary nebulae (PPN) are very red and can have
strong CO absorption. Some of the PPN spectra closely resemble those
of FU-Ori type objects, so it is possible that some of the objects
we have classified as PPN are in fact FU-Ori type objects. The
spectra of planetary nebulae (PN) are flat or red and are often very
similar to YSO or H~II region spectra. In these cases, other data
from the RMS Survey follow-up and the literature were used to
classify the object. The final class found within our sample were
M-Giants. These have continua that peak in the middle of the
1.5--2.5~$\umu$m region, show molecular absorption lines owing to
their cool atmospheres, and have strong \water\ absorption. 

Of the 247 objects whose spectra were taken, there were 180 YSOs, 15
YSOs within H~II regions, 26 H~II regions, 7 evolved stars, 5 PPN, 4
PN, 5 M-type supergiants, and 5 evolved hot stars labelled as
`Other' in the table and on the RMS data base website due to unusual
characteristics in their data. Spectra were found to be a valuable
confirmation of classification in addition to our imaging data in
45\% of cases, and the only means of classifying the source in 5\%
of cases.

%%%%%%%%%%%%%%%%%%%%%%%%%%%%%%%%%%%%%%%%
\section{Extinction}

Two estimates of extinction to each object were calculated, using
the \citet{1984ApJ...285...89D} / \citet{1989ESASP.290...93D}
interstellar extinction model; one using $J-H$ colours and the other
using $H-K$ colours. The MYSO colours were normalised to
the intrinsic colours of a B0 star, taken from
\citet{1983A&A...128...84K}: $J-H=-0.12$ and $H-K=-0.05$. For these
calculations, the catalogue magnitudes in Table A2 were used. The
results are listed in Table A3 as `$A_{V}$ (JH)' and `$A_{V}$ (HK)'
for calculations using the $J-H$ and $H-K$ colours respectively.
The equation for
$A_{V}$ is: 
\begin{equation}
A_{V} = \frac{m_{1} - m_{2} + c_{int}}{0.55_{}^{1.75}
(\lambda_{1}^{-1.75} - \lambda_{2}^{-1.75})}
\end{equation}
where $m_{1}$ and $m_{2}$ are the magnitudes used to calculate the
extinction (either J- and H-band magnitudes, or H- and K-band magnitudes),
$\lambda_{1}$ and $\lambda_{2}$ are their reference wavelengths,
$c_{int}$ is the intrinsic $m_{1} - m_{2}$ from
\citet{1983A&A...128...84K}, and $1.75$ is the
\citet{1984ApJ...285...89D} / \citet{1989ESASP.290...93D} reddening
constant. The
errors were calculated in quadrature as follows: 
\begin{equation} 
\delta A_{V} = \frac{\sqrt{\delta
m_{1}^{2} + \delta m_{2}^{2}}}{0.55_{}^{1.75} (\lambda_{1}^{-1.75}
- \lambda_{2}^{-1.75})}
\end{equation}
These are shown in Table A3. Where one or more magnitudes are
unavailable, the extinctions cannot be calculated, and
`--' is marked in the table. Where the 2MASS magnitude is a
lower-limit, the $A_{V}$ given is also a lower-limit. For 35 sources
(27 of them YSOs), there was not enough magnitude data to calculate
an extinction. 

The left panel of Figure \ref{fig4} shows $A_{V}$ (HK) against
$A_{V}$ (JH) for the YSOs. The solid line represents $A_{V}$ (HK) =
$A_{V}$ (JH), and the dotted line is the ATS regression line. The
Kendall's $\tau$ is 0.36, with a $p$-value of $3.0~\times~10^{-11}$,
indicating a good correlation between the two.
The equation for the fit is 
\begin{equation}
A_{V}\ {\rm (HK)}\ = 6.0 + 1.1 A_{V}\ {\rm (JH)}
\end{equation} 
The regression line falls above the $A_{V}$ (HK) = $A_{V}$ (JH)
line. This was explained by \citet{PorterDrewLumsden1998}: the $H-K$
colours appear redder in young objects due to dust excess, and the
$J-H$ colours appear bluer due to scattering at short wavelengths.
This artificially increases $A_{V}$ (HK) and decreases $A_{V}$ (JH),
meaning the true value for extinction lies somewhere between the
two. As the amount of dust increases, the effect becomes more
pronounced and the regression line deviates further from the $A_{V}$
(HK) = $A_{V}$ (JH) line. 

Figure \ref{fig5} shows a colour-colour diagram demonstrating
this point. Some of the objects do fall very close to the extinction
line, i.e. the line that runs parallel to the extinction vector and
starts at the intrinsic $J-H$ and $H-K$ colours. However, most fall
above this line, showing evidence of the dust excess emission and/or
scattering described above. 

For extinction correction, $A_{V}$ (HK)
was used where available, or $A_{V}$ (JH) when $A_{V}$ (HK) was not
available, or was a lower limit. The number distribution of these
extinctions is shown in the right panel of Figure \ref{fig4}. The
median is $A_{V} = 42$, and the extinctions range from $A_{V}=2.7$
to $A_{V}=114$, demonstrating that this study covers the minimally
obscured objects right through to very red, dusty sources.
  
\begin{figure*}
\begin{minipage}{84mm}
\includegraphics[width=84mm]{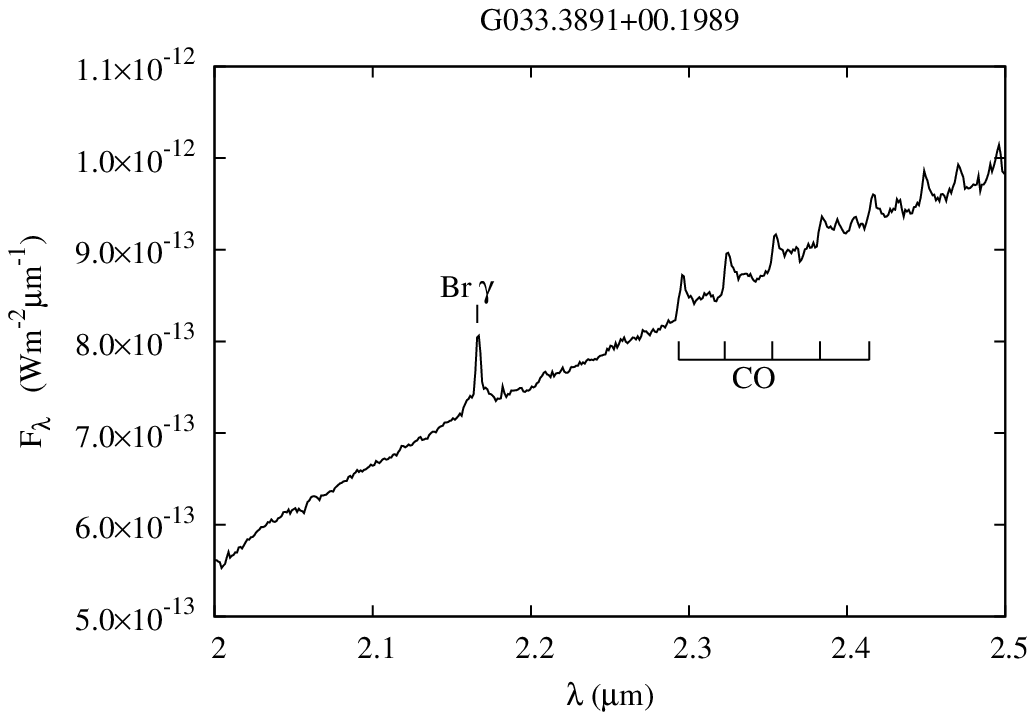}
\caption{The K-band section of the spectrum of the YSO
G033.3891+00.1989. This shows a very clear example of the saw-tooth
profile of the CO bandhead. It also has \brg\ emission.
\newline
\newline}
\label{fig6}
\end{minipage}
\hspace{0.7cm}
\begin{minipage}{84mm}
\includegraphics[width=84mm]{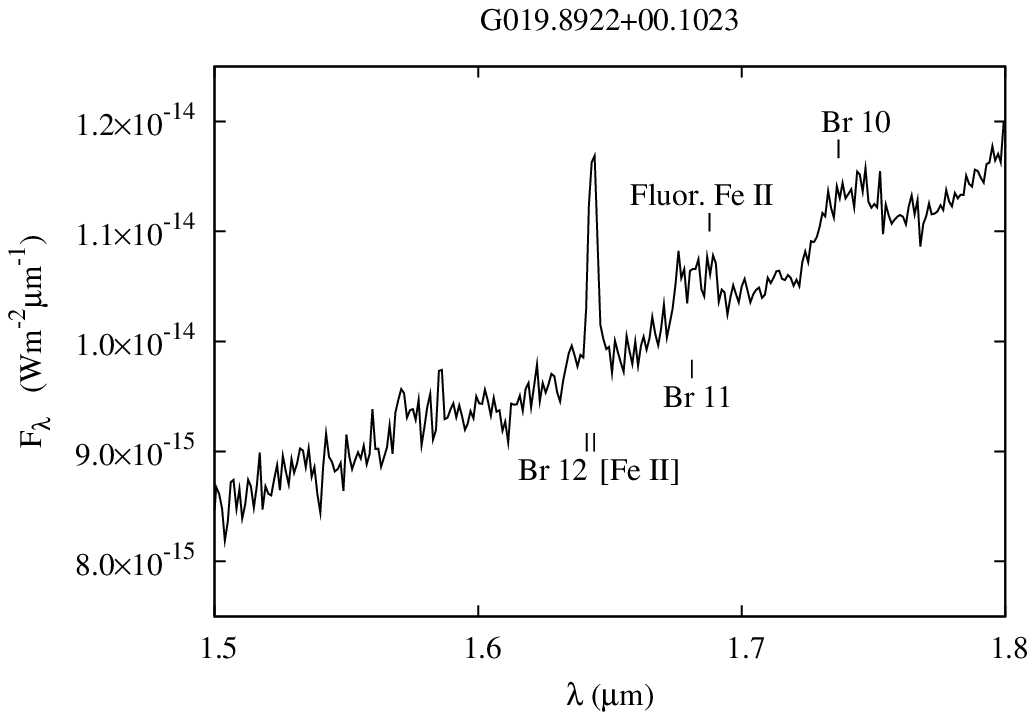}
\caption{The H-band section of the spectrum of the YSO 
G019.8922+00.1023. This object has both the Brackett series and the
[Fe~II] 1.64402~$\umu$m line, as well as the fluorescent Fe~II 1.6878~$\umu$m line. Note that the Br~12 / [Fe~II] line at $\sim 1.64\
\umu$m is much stronger than the other Brackett series lines.} 
\label{fig7}
\end{minipage}
\end{figure*}
 
%%%%%%%%%%%%%%%%%%%%%%%%%%%%%%%%%%%%%%%%
\section{Young Stellar Objects}\label{sec:YSOs}
\subsection{Spectral Features}\label{sec:YSO-features}

In addition to the strong dust extinction and corresponding steep
rising continuum, most YSOs have emission lines in their spectra.
However, not all of them show the same lines. One of the objectives
of the RMS survey is to identify why there are differences. The main
emission lines observed in YSOs in our sample were \brg, Br~10, Br
11, Br~12, [Fe~II] 1.64~$\umu$m line, \h\ 2.1218~$\umu$m, \h\ 2.2477
$\umu$m, fluorescent Fe~II 1.6878~$\umu$m, and CO bandhead. Some
objects also exhibited the Pfund series, or other molecular \h\
lines. He~I 2.0587~$\umu$m was seen in both emission and absorption.
Both \brg\ and He~I were occasionally seen with P-Cygni-type
profiles: 13 MYSOs had \brg\ with a P-Cygni profile and 15 had He~I
with a P-Cygni profile, of
which 8 had P-Cygni profiles for both \brg\ and He~I. These objects
are listed in Table B3. 
However, the He~I 2.11~$\umu$m doublet was not seen in the YSOs,
only in H~II regions. Detection frequencies are shown in Table
\ref{tab:detection-frequency}. 

\begin{table}
\begin{minipage}{\columnwidth}
\caption{Detection frequency of spectral features in YSO spectra.
All detections are at the 3$\sigma$ level or better. The percentages
are based on
totals of 195 objects for the K-band, 188 for Br~10 and fluorescent
Fe~II, and 186	for the shorter wavelength H-band lines. This takes
account of the fact that some of the objects were not detected for part
or all of the H-band meaning determining the presence or absence of
the lines in that region was impossible.
}\label{tab:detection-frequency}
\centering
\begin{tabular}{lcc}
\hline
Emission Feature & Number of Sources & Frequency (\%) \\
\hline
\brg\ & 147 & 75 \\
Br~10 & 85 & 45 \\
Br~11 & 69 & 37 \\
Br~12 Un-blended & 35 & 19 \\
Br~12/ [Fe~II] Blend & 46 & 25 \\
{[}Fe~II] Un-blended & 20 & 11 \\
\h\ 2.1218~$\umu$m & 109 & 56 \\
\h\ 2.2477~$\umu$m & 18 & 9 \\
Fluor. Fe~II & 48 & 26 \\
He~I 2.0587~$\umu$m & 30 & 15 \\
Pfund Series & 13 & 7 \\
CO 2--0 Bandhead & 34 & 17 \\
\hline
\end{tabular}
\end{minipage}
\end{table}

\subsection{Emission and Absorption Line Fluxes}
The fluxes of the YSO emission and absorption lines were measured
using the Starlink package {\sc Dipso}. The continuum was fitted by
a 3$^{rd}$ order polynomial function, then subtracted. In many cases,
the steepness of the continuum obscured the presence of emission
lines that became visible when the continuum was subtracted. For all spectral features
except the CO 2--0 bandhead, the line profiles were fitted with a
Gaussian function, which {\sc Dipso} then integrated to find the
total flux of the line. The majority of the spectral lines are
unresolved so a Gaussian function is a good assumption. 

Detections were defined as having $f_{\lambda} > 3\ \delta
f_{\lambda}$, where $f_{\lambda}$ is the line flux and $\delta
f_{\lambda}$ is the error. For the unresolved lines, {\sc Dipso}
gave a good estimation of the error. Where the lines had broad wings
or interference in the profile shape from noise, {\sc Dipso} could
still fit the line with a Gaussian, but could not accurately
estimate the error. When this was the case, errors were calculated
using the equation for a Gaussian with the height equal to the
continuum r.m.s. noise,
i.e. \begin{equation}
\delta f_{\lambda} = h \left(\frac{W}{2\sqrt{2\ln{2}}}\right) 
\end{equation}
where $h$ is height, and $W$ is the FWHM of an unresolved peak at
that wavelength.

The continuum r.m.s. noise was measured using {\sc Figaro}'s {\sc
istat} function, taking wavelength ranges with no spectral lines,
close enough to the pertinent spectral line that extinction would
not distort the result. The r.m.s. noise was
measured for two ranges per line, one either side of the line, and
the mean value of these was taken. The upper limits for
non-detections were calculated using the same equation, with $h=3\
~\times$ continuum r.m.s. 
    
\begin{figure*}
\begin{minipage}{\textwidth}
\centering
\includegraphics[width=84mm]{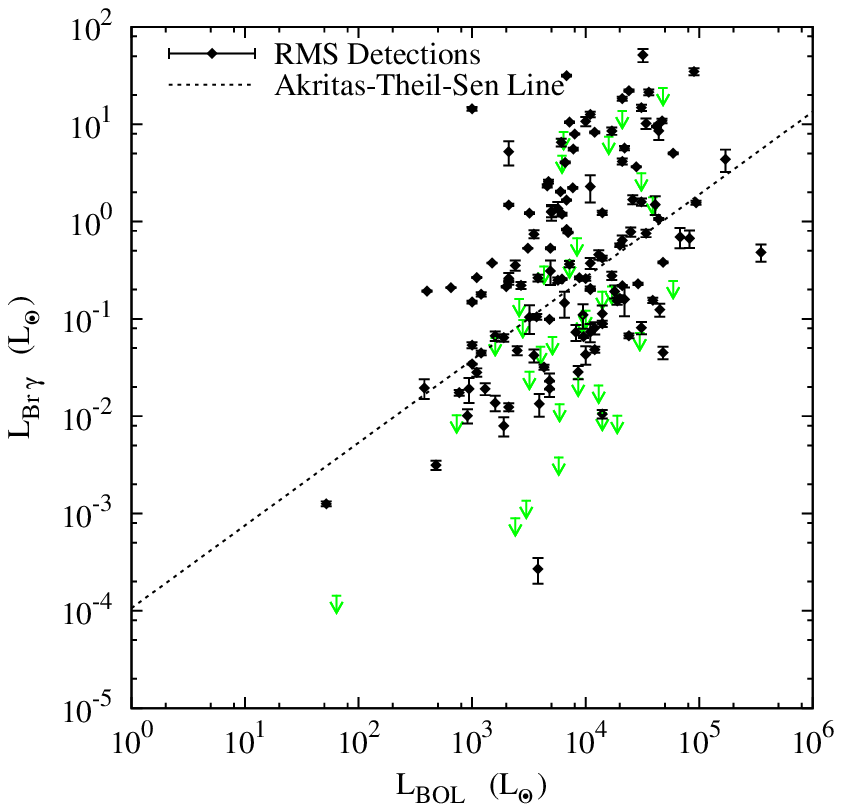}
\hspace{0.7cm}
\includegraphics[width=84mm]{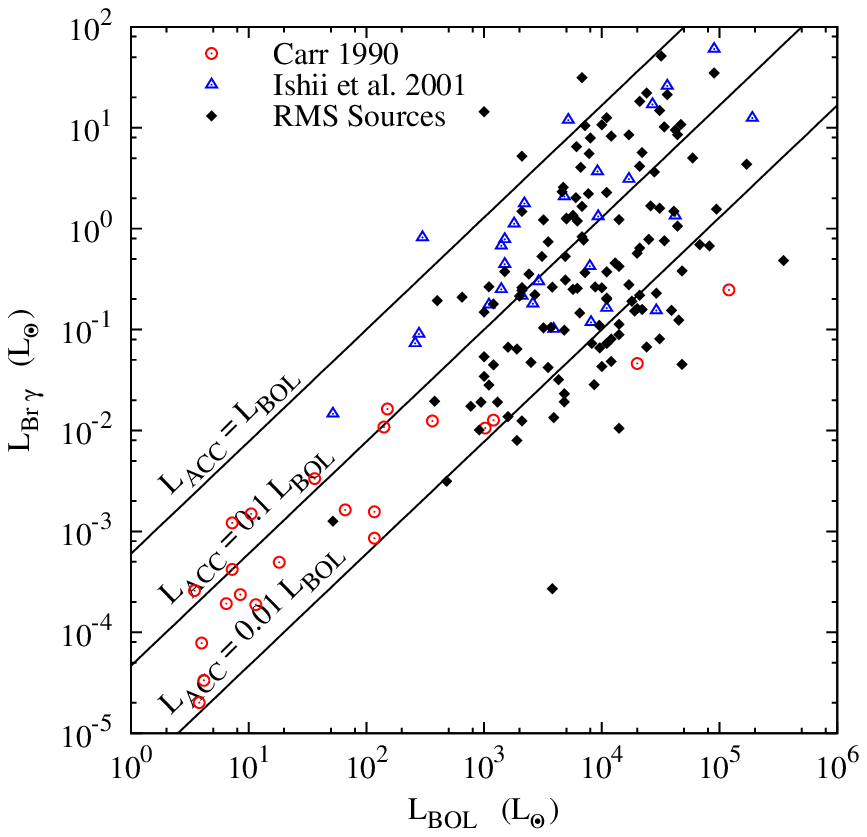}
\caption{\emph{Left}: \brg\ luminosity against bolometric luminosity for RMS sources. Detections
are denoted by black diamonds and limits by green arrows. Errors on
detections are shown, but are frequently too small to be easily
distinguishable on this scale. The dotted line represents the ATS
regression. 
\emph{Right}: Comparison of RMS data (diamonds) with those of
\citet{carr1990} (circles) and \citet{ishii2001} (triangles). Only
detections above the 3$\sigma$ limit are shown. The solid lines
represent the loci of ${\rm{L}}_{{\rm ACC}} = {\rm L}_{{\rm BOL}}$,
${\rm{L}}_{{\rm ACC}} = 0.1\ {\rm L}_{{\rm BOL}}$, and
${\rm{L}}_{{\rm ACC}} = 0.01\ {\rm L}_{{\rm BOL}}$ using the
\citet{calvet2004} relationship between ${\rm{L}}_{{\rm ACC}}$ and
${\rm{L}}_{{\rm Br}\gamma}$. In both plots, all the line
luminosities have been corrected for extinction, and the RMS data
have been corrected for the difference between measured and
catalogue magnitudes.}\label{fig8} 
\end{minipage}
\end{figure*}
 
The CO bandhead has a saw-tooth shaped profile (see Figure
\ref{fig6}), so different
treatment was required. The equivalent width (EW) was measured from
2.289--2.313~$\umu$m, again using {\sc Dipso}. The mean continuum flux
 density was measured across the 2.2825--2.2875~$\umu$m range
just to the blue end of the bandhead, using {\sc Figaro}'s {\sc
istat} function. The EWs were then converted into fluxes. This is
consistent with the method used by \citet{ishii2001}. Detections
were again defined as ${\rm EW} > 3\ \delta {\rm EW}$, where $\delta
{\rm EW}$ is the error given by {\sc Dipso}. Limits were taken as
${\rm EW}_{{\rm limit}}=3\ \delta {\rm EW}$.

The fluxes of the \brg, Br~10, Br~11, Br~12 / [Fe~II] 1.64~$\umu$m
line, \h\ 2.1218~$\umu$m, \h\ 2.2477~$\umu$m, fluorescent Fe~II
1.6878~$\umu$m, and He~I 2.0587~$\umu$m lines are listed in Table
B1, with the presence of the Pfund series denoted by `P' in the
final column. The equivalent widths of the CO 2--0
bandhead are listed in Table B2. The errors in the tables are at the
1$\sigma$ level. The fluxes and equivalent widths in the tables are
the raw values as measured directly from the spectra, and are not
corrected for extinction or normalised to the catalogue magnitude
(see Section \ref{sec:magnitudes}). 

Br~12 and [Fe~II] 1.64402~$\umu$m are close to each other and form a
blend when both are present (see Figure \ref{fig7}). To
determine whether the line seen at 1.64~$\umu$m is Br~12, [Fe~II], or
a blend, one has to compare this line with the rest of the Brackett series.
For this reason, the fluxes of Br~12 and [Fe~II] are listed in Table
B1 in the same column. Un-blended [Fe~II] corresponds to
the presence of a line at 1.64~$\umu$m when Br~10 and Br~11 are
absent. When the rest of the Brackett series is present, the presence of
[Fe~II] can only be inferred if the relative strength of the 1.64
$\umu$m line compared to Br~11 is higher than expected. Assuming
\citet{BakerMenzel1938} case-B recombination with the
\citet{StoreyHummer1995} line ratios for ${\rm T}_{{\rm e}}=7500{\rm\
K}$ and ${\rm N}_{{\rm e}}=10^6\ {\rm cm}^{-3}$, the 1.64~$\umu$m line was defined
as a blend if $f(1.64\ \umu {\rm m}) > 0.788~\times~f({\rm Br\ 11})$. 
See Section \ref{sec:hilines} for a full discussion of the
properties of H~I lines.

This cut relies on the emission lines being optically thin.
If the lines are not optically thin, the flux of the Br~12 line may
fall above this limit but not have a contribution from [Fe~II].
This means that there are probably some objects that are considered
to have a blend of Br~12 and [Fe~II] under this definition that
actually only have Br~12. However, the low resolution of the spectra
meant that de-convolving the lines was not possible, so this was
considered to be the best available separation method for the purposes of this
paper. 

In 9 of the YSO spectra, dust extinction rendered part or all of the
H-band too noisy to identify emission lines in this region with any
confidence. This caused difficulties in the analysis of these
spectra. However, given that there are 195 objects classified as YSO
or YSO/HIIR, this represents only $\sim 5$ \% of the objects.
Excluding these objects from the H-Band data analysis still leaves a
large enough sample to make statistically sound conclusions.

\subsection{H~I Emission}\label{sec:hilines}
The detection rate of \brg\ emission amongst the YSOs is 75 \%, and
many of these also have other H~I lines. This detection frequency is
lower than the 97\% found by \citet{ishii2001} for intermediate-mass
YSOs and Herbig Ae/Be stars, but comparable with the 71\% found by
\citet{carr1990} for low-mass YSOs. \brg\ was also seen with a
P-Cygni profile in the spectra of 13 objects (see Table B3),
indicating that for those objects, \brg\ originated in an outflowing
wind. 

\begin{table}
\begin{minipage}{\columnwidth}
\caption{Summary of the results of the correlation tests for the
emission line luminosities against ${\rm L}_{{\rm BOL}}$. These fits
were performed only on the RMS data; the low- and intermediate-mass
stars were not included in the correlation and regression
calculations.
}\label{tab:correlations}
\centering
\begin{tabular}{lcc}
\hline
Emission Line & Kendall's $\tau$ & $p$-value \\ \hline
\brg & 0.26 & $6.1~\times~10^{-7}$ \\ 
\h & 0.19 & 0.0003 \\ 
CO Bandhead -- all data & 0.04 & 0.4 \\ 
CO Bandhead -- detections only & 0.48 & 0.0006 \\ 
Fluor. Fe~II -- all data & 0.07 & 0.2 \\ 
Fluor. Fe~II -- detections only & 0.36 & 0.0003 \\ 
He~I -- all data & 0.05 & 0.4 \\ 
He~I 2.0587~$\umu$m -- detections only & 0.45 & 0.003 \\ \hline
\end{tabular}
\end{minipage}
\end{table}

The left panel of Figure \ref{fig8} shows a comparison of
\brg\ luminosity with bolometric luminosity (the latter from
\citealt{2011Mottram-luminosity}). Correlation and regression tests
were performed on the line vs. bolometric luminosity data for the RMS
objects only. The ATS regression is represented in the figure by
the dotted line. The equation of the line is:
\begin{equation}
{\rm L}_{{\rm Br}\gamma} = 10^{-4.0}~\times~({\rm L}_{{\rm BOL}})^{0.85}
\end{equation}
A summary of the results of the correlation tests for all the lines
is shown in Table \ref{tab:correlations}. The Kendall's $\tau$
coefficient for \brg\ vs. ${\rm L}_{{\rm BOL}}$ shows that there is a strong correlation between the
\brg\ luminosity and the bolometric luminosity of the star. This
correlation was also seen in studies of low- and
intermediate-mass YSOs and Herbig Ae/Be stars \citep{carr1990,
ishii2001, ConnelleyGreene2010}.

\brg\ emission is known to be related to accretion in low-mass stars
\citep{calvet2004}, and has been scaled up to Herbig Ae/Be stars by
\citet{Garcia-Lopez2006}. The right panel of Figure
\ref{fig8} compares detections in the RMS sources with those
of \citet{carr1990} and \citet{ishii2001}. The solid lines
represent the loci of ${\rm{L}}_{{\rm ACC}} = {\rm L}_{{\rm
BOL}}$, ${\rm{L}}_{{\rm ACC}} = 0.1\ {\rm L}_{{\rm BOL}}$, and ${\rm{L}}_{{\rm
ACC}} = 0.01\ {\rm L}_{{\rm BOL}}$ using the empirical relationship from \citet{calvet2004}:
\begin{equation}
\log_{10}\left({\rm{L}}_{{\rm ACC}} \right) = -0.7 +0.9\left(\log_{10}\left({\rm{L}}_{{\rm Br}\gamma} \right) +4 \right) 
\end{equation}
The majority of the objects have $0.01\ {\rm L}_{{\rm BOL}} <
{\rm{L}}_{{\rm ACC}} < {\rm L}_{{\rm BOL}}$. This appears to be true
for most RMS sources, implying that there is a common mechanism
producing \brg\ throughout the mass range covered by
\citet{carr1990}, \citet{ishii2001} and the RMS sources. This trend
is seen with other emission lines, suggesting that high-mass YSOs
resemble scaled-up versions of low- or intermediate-mass YSOs. 

One of the considerations when distinguishing MYSOs with Brackett
series emission from H~II regions is whether the recombination
conforms to case-B of \citet{BakerMenzel1938}; recombination
emission in typical compact H~II regions conforms to case-B, whereas this is
not always true for MYSOs. \citet{StoreyHummer1995} tabulated the
case-B line ratios for a range of temperatures and densities. 

The temperature of the gas in which the Br series lines originate
must lie between that of a UCHII region (${\rm T}_{{\rm
e}}\sim7500\ {\rm K}$) at the lower end, to an OB star (${\rm
T}_{{\rm e}}\la 30000\ {\rm K}$) at the upper end. Similarly, the
density of a UCHII region imposes a lower limit on the density of
${\rm N}_{{\rm e}}\ga 10^{4}\ {\rm cm}^{-3}$. A realistic maximum
density for the stellar wind would be ${\rm N}_{{\rm e}}\la 10^{11}\
{\rm cm}^{-3}$. For ${\rm T}_{{\rm e}}$ and ${\rm N}_{{\rm e}}$ in
this range, the \citet{StoreyHummer1995} Br~10/\brg\ ratio varies
from 0.327 to 0.490, i.e. around a 30\% variation. Similarly, the Br
12/Br~11 ratio varies from 0.756 to 0.834, i.e. about 10 \%.
 
\begin{figure}
\begin{minipage}{\columnwidth}
\centering
\includegraphics[width=84mm]{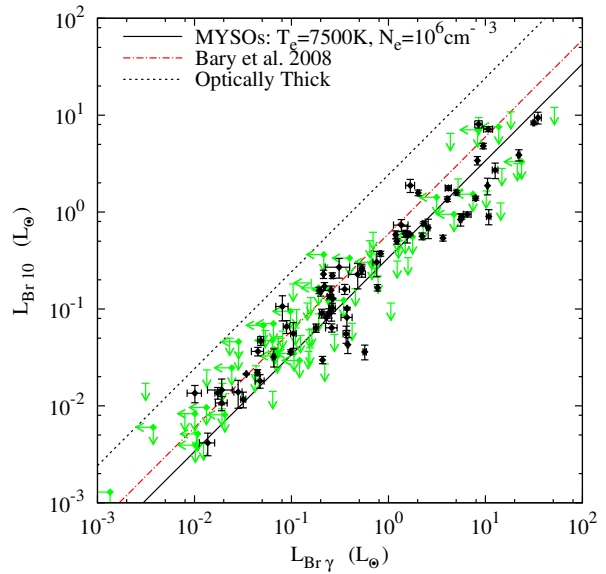}
\caption{Br~10 luminosity against \brg\ luminosity for RMS sources.
Detections are denoted by black diamonds and limits by green arrows.
The black solid line shows the case-B \citet{StoreyHummer1995}
Br~10/\brg\ ratio using values of ${\rm T}_{\rm e}$ and ${\rm N}_{\rm e}$
found from observations of MYSOs; the red dash-dot line represents
the Br~10/\brg\ ratio using ${\rm T}_{{\rm e}}$
and ${\rm N}_{{\rm e}}$ values for T-Tauri stars taken from
\citet{2008ApJ...687..376B}; and the black dotted line represents the
case in which Br~10 and \brg\ are both optically thick (i.e.
black-body emission). The line luminosities have been corrected for
extinction, and for the difference between measured and catalogue
magnitudes.}
\label{fig9}
\end{minipage}
\end{figure}
 
We adopt the case-B values for MYSOs based on observational evidence
(e.g. \citealt{PorterDrewLumsden1998}). Though the values for ${\rm
T}_{{\rm e}}$ and ${\rm N}_{{\rm e}}$ differ from those found for
T-Tauri stars, the corresponding Br~12/Br~11 line ratios are
comparable. For example, \citet{1991ApJ...382..617H} (and
subsequently \citealt{1998ApJ...492..743M})
found values of ${\rm T}_{{\rm e}}\sim10000{\rm\ K}$ and ${\rm
N}_{{\rm e}}=10^{14}\ {\rm cm}^{-3}$ when modelling production in
infalling
magnetospheric gas, whereas \citet{2008ApJ...687..376B} observed
Paschen and Brackett series H~I recombination line ratios consistent
with temperatures ${\rm T}_{{\rm e}}\leq 2000{\rm\ K}$ and densities
$10^9<{\rm N}_{{\rm e}}<10^{10}\ {\rm cm}^{-3}$, inconsistent with
production in accreting material. 
The \citet{1991ApJ...382..617H} and \citet{2008ApJ...687..376B}
${\rm T}_{{\rm e}}$ and ${\rm N}_{{\rm e}}$ values correspond to
\citet{StoreyHummer1995} Br~12/Br~11 ratios of 0.757 and 0.764
respectively, which are not vastly different from the 0.788 ratio
given by using the observational values for ${\rm T}_{{\rm e}}$ and
${\rm N}_{{\rm e}}$, given a typical flux error for the 1.64~$\umu$m
line of $\sim 10 \%$. However, while the Br~12/Br~11 ratio is very
similar whichever set of ${\rm T}_{{\rm e}}$ and ${\rm N}_{{\rm e}}$
values are used, this is not the case for the Br~10/\brg\ ratios:
The observed ${\rm T}_{{\rm e}}$ and ${\rm N}_{{\rm e}}$ and the
\citet{1991ApJ...382..617H} ${\rm T}_{{\rm e}}$ and ${\rm N}_{{\rm
e}}$ produce line ratios of 0.338 and 0.301 respectively, whereas
the \citet{2008ApJ...687..376B} ratio is 0.597. 

Figure \ref{fig9} shows Br~10 luminosity against \brg\
luminosity. Shown on the plot are 3 lines: one showing the
\citet{StoreyHummer1995} Br~10/\brg\ ratio using observed ${\rm
T}_{{\rm e}}$ and ${\rm N}_{{\rm e}}$ values for MYSOs; one showing
the Br~10/\brg\ ratio for T-Tauri stars using ${\rm T}_{{\rm e}}$
and ${\rm N}_{{\rm e}}$ from \citet{2008ApJ...687..376B}; and
finally one showing the theoretical ratio for optically thick
emission. The \citet{1991ApJ...382..617H} ratio for T-Tauri stars
falls right next to the MYSO line, so is not shown on the
plot.
 
\begin{figure*}
\begin{minipage}{\textwidth}
\centering
\includegraphics[width=84mm]{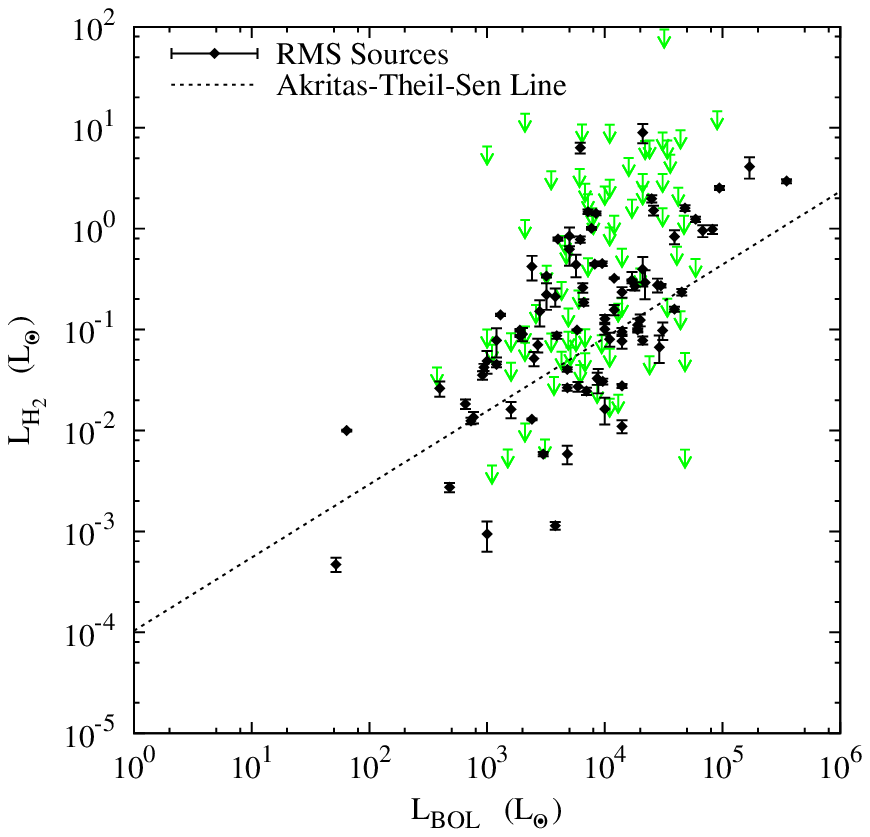}
\hspace{0.7cm}
\includegraphics[width=84mm]{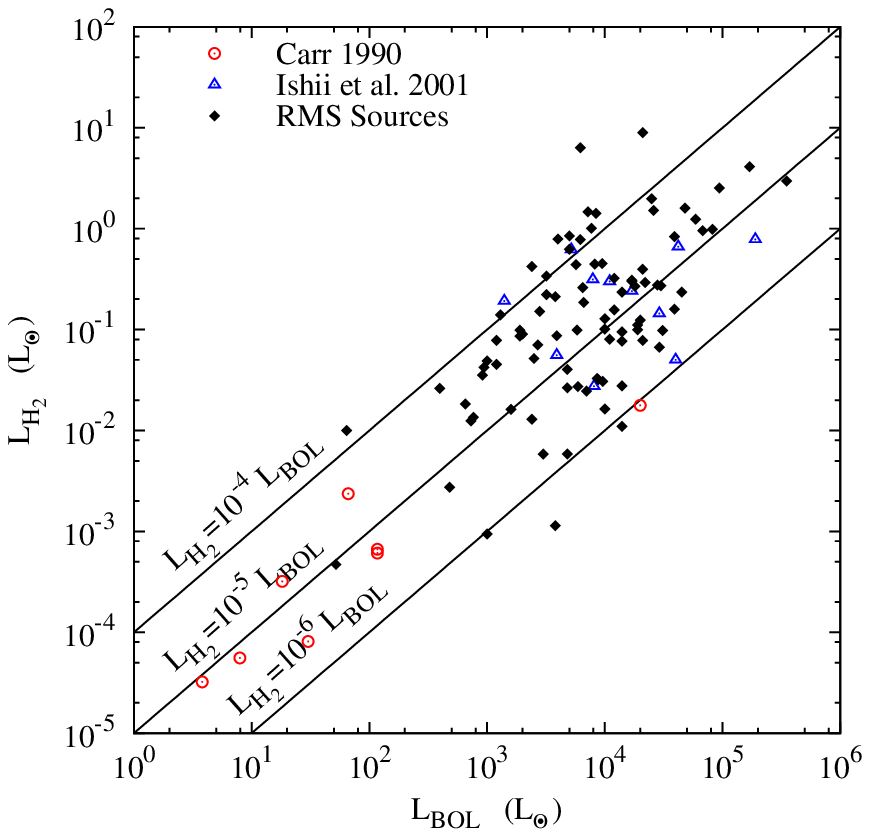}
\caption{\emph{Left}: \h\ luminosity against bolometric luminosity for RMS sources. Detections
are denoted by black diamonds and limits by green arrows. The dotted
line represents the ATS regression. Errors on detections are shown,
but are frequently too small to be easily distinguishable on this
scale. \emph{Right}: Comparison of RMS data (diamonds) with those
of \citet{carr1990} (circles) and \citet{ishii2001} (triangles).
Only detections above the 3$\sigma$ limit are shown. The solid
lines represent the loci of
${\rm L}_{{\rm H}_{2}} = 10^{-4}\ {\rm L}_{{\rm BOL}}$,
${\rm L}_{{\rm H}_{2}} = 10^{-5}\ {\rm L}_{{\rm BOL}}$, and
${\rm L}_{{\rm H}_{2}} = 10^{-6}\ {\rm L}_{{\rm BOL}}$.
In both plots, all the line luminosities have been corrected for
extinction, and the RMS data have been corrected for the difference
between measured and catalogue magnitudes. } \label{fig10}
\end{minipage}
\end{figure*}
 
Most of the MYSOs fall very close to, or on the ${\rm T}_{{\rm
e}}=7500{\rm\ K}$ and ${\rm N}_{{\rm e}}=10^6\ {\rm cm}^{-3}$
line. However, for a minority of sources,
the Br~10/\brg\ ratio is much greater than that predicted by case-B
recombination for the MYSO values of ${\rm T}_{{\rm e}}$ and ${\rm
N}_{{\rm e}}$ -- closer to the \citet{2008ApJ...687..376B} T-Tauri
line. This could imply that for these YSOs \brg\ is
optically thick and therefore that the recombination does not
conform to case-B. This is in contrast to the optically thin H~II
regions in which recombination line ratios do follow the case-B
values. Alternatively, the ${\rm T}_{{\rm e}}$
and ${\rm N}_{{\rm e}}$ values for these stars could be closer to
those found by \citet{2008ApJ...687..376B} for T-Tauri stars than
the previously observed values for MYSOs. Very few of
the objects have Br~10/\brg\ ratios very close to the line for which
both Br~10 and \brg\ are assumed to be optically thick. This
suggests that though \brg\ may be optically thick for some sources,
Br~10 is usually optically thin. 

\subsection{Shock Emission: Molecular \h\ and Forbidden [Fe~II]}
The detection rate of the molecular \h\ 2.1218~$\umu$m line is 56\%, and
9\% have additional molecular \h\ lines. This is a higher detection rate
than found by \citet{ishii2001} for intermediate-mass YSOs (34\%),
and much higher than the rate found by \citet{carr1990} for low-mass
YSOs (25\%). \citet{ishii2001} interpret the presence of strong \h\
emission as emission produced by the shock where an outflow impacts
the surrounding circumstellar material, indicating that at least
56\% of the YSOs in our sample have a strong outflow. There may be
more YSOs with outflows in our sample, with either not enough
circumstellar material to form a shock, or aligned at such an angle
that we cannot see the light from the shock. 

The left panel of Figure \ref{fig10} shows \h\ luminosity
against bolometric luminosity. The equation of the regression line is
\begin{equation}
{\rm L}_{{\rm H}_{2}} = 10^{-4.0}~\times~({\rm L}_{{\rm BOL}})^{0.73}
\end{equation}
The Kendall's $\tau$ value (see Table \ref{tab:correlations}) shows that there is a correlation between
the \h\ luminosity and the bolometric luminosity of the
star. 

The right panel of Figure \ref{fig10} compares detections in
the RMS sources with those of \citet{carr1990} and
\citet{ishii2001}. The solid lines represent the loci of 
${\rm L}_{{\rm H}_{2}} = 10^{-4}\ {\rm L}_{{\rm BOL}}$, 
${\rm L}_{{\rm H}_{2}} = 10^{-5}\ {\rm L}_{{\rm BOL}}$, and 
${\rm L}_{{\rm H}_{2}} = 10^{-6}\ {\rm L}_{{\rm BOL}}$.
The \citet{carr1990} and \citet{ishii2001} sources all have
$10^{-6}\ {\rm L}_{{\rm BOL}} < {\rm L}_{{\rm H}_{2}} < 10^{-4}\
{\rm L}_{{\rm BOL}}$, as do most of the RMS sources. This implies a
common production mechanism for \h\ for low-mass and high-mass YSOs,
as with \brg, adding further evidence to the idea that high-mass
YSOs are very similar to low-mass YSOs.

[Fe~II] 1.64402~$\umu$m emission is also shock excited
\citep{1994MNRAS.268..821L, PorterDrewLumsden1998}. As described in
Section \ref{sec:YSO-features}, its close proximity to Br~12 makes
analysis of this line difficult. 20 objects (i.e. 11\%) were found
to have [Fe~II] emission without the rest of the Br series. An
estimated 46 further objects (i.e. 25\%) have [Fe~II] blended with
Br~12. This figure is probably an upper limit, since the
assumption that H~I emission in YSOs conform to case-B of
\citet{BakerMenzel1938} is probably not valid for all YSOs (see
Section \ref{sec:hilines} and Figure \ref{fig9}). 
 
\begin{figure*}
\begin{minipage}{\textwidth}
\centering
\includegraphics[width=84mm]{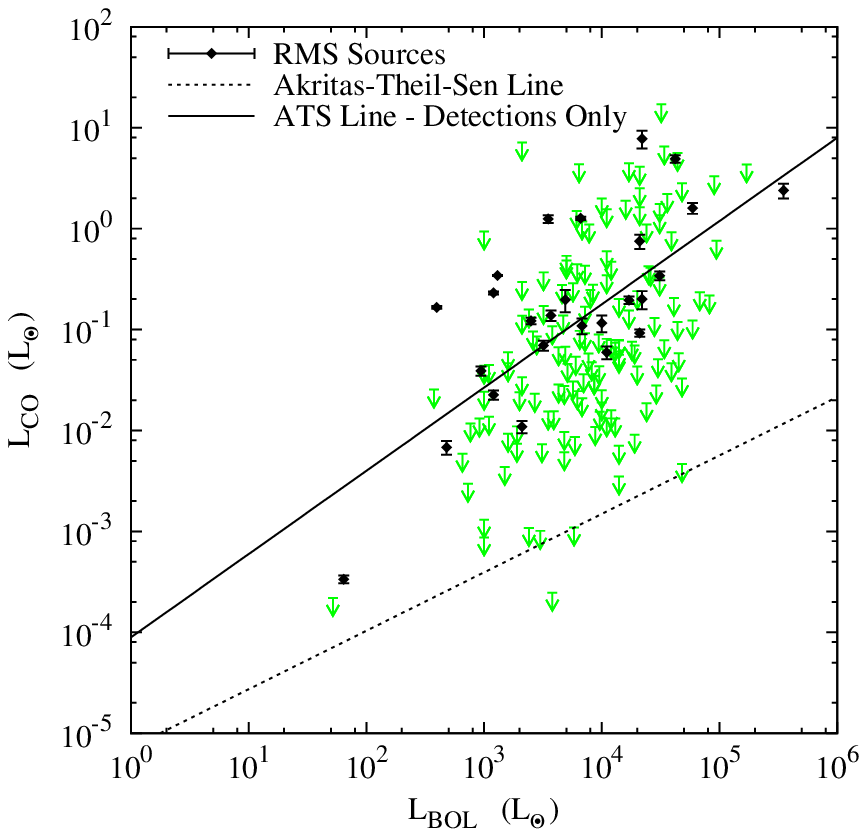}
\hspace{0.7cm}
\includegraphics[width=84mm]{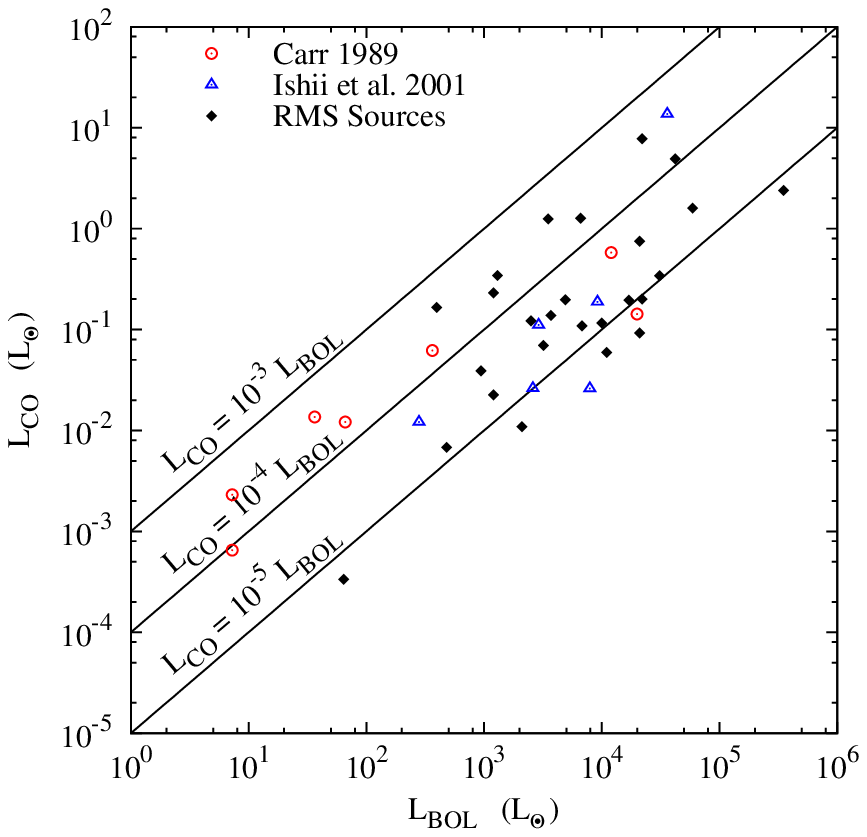}
\caption{\emph{Left}: CO 2--0 bandhead luminosity against
bolometric luminosity for RMS sources.
Detections are denoted by black diamonds and limits by green arrows.
The lines represent the ATS regressions. Two lines are shown, one
for regression with detections only (solid line), and one for all
the data including non-detections (dotted line). Errors on
detections are shown, but are frequently too small to be easily
distinguishable on this scale. \emph{Right}: Comparison of RMS data
(diamonds) with those of \citet{carr1989} (circles) and
\citet{ishii2001} (triangles). Only detections above the 3$\sigma$
limit are shown. The solid lines represent the loci of ${\rm
L}_{{\rm CO}} = 10^{-3}\ {\rm L}_{{\rm BOL}}$,
${\rm L}_{{\rm CO}} = 10^{-4}\ {\rm L}_{{\rm BOL}}$, and
${\rm L}_{{\rm CO}} = 10^{-5}\ {\rm L}_{{\rm BOL}}$.
In both plots, all the line luminosities have been corrected for
extinction, and the RMS data have been corrected for the difference
between measured and catalogue magnitudes.} \label{fig11}
\end{minipage}
\end{figure*}
 
\subsection{Accretion Discs: CO Bandhead and Fluorescent Fe~II
Emission}\label{sec:accretion-disks}
The production of CO bandhead emission requires high densities and
temperatures which can only be found in a circumstellar disc
(e.g. \citealt{carr1989, 1995ApJ...446..793C, 2004A&A...427L..13B,
2010MNRAS.402.1504D, 2010MNRAS.408.1840W}). Fluorescent Fe~II
1.6878~$\umu$m emission in the MYSOs probably also originates in
discs: it is known to originate in a disc in both AGN (e.g.
\citealt{2004ApJ...615..610B}) and classical Be stars (e.g.
\citealt{2007A&A...470..239Z}), and has previously been observed in
high-mass YSOs by \citet{PorterDrewLumsden1998} and
\citet{Lumsden-feii}, and in low-mass YSOs
(though less frequently) by \citet{1992ApJS...82..247H}. 

The CO bandhead was seen in 17\% of the YSO spectra. This detection
rate is lower than both the \citet{carr1989} rate for low-mass YSOs
(20\%) and the \citet{ishii2001} rate for intermediate-mass YSOs
(22\%). Fe~II emission was seen in 26\% of our sources with visible
H-band, usually in those objects with strong H-band H~I emission. 7
objects were not visible at this wavelength, and more were very
noisy so this detection rate should be considered as a lower limit.
There is seldom overlap between the two lines -- only 6 YSOs have
both fluorescent Fe~II and CO bandhead emission. In total, 76 RMS
objects (i.e. 39\%) have evidence of a circumstellar disc. 

\citet{ConnelleyGreene2010} note in their paper on low-mass Class I
YSOs that all objects with CO bandhead emission also show \brg\ emission.
This trend is seen, though not as strongly, in the high-mass RMS
YSOs: 75\% of all YSOs have \brg\ emission, and 85\% of YSOs with CO
bandhead emission have \brg. This is a higher \brg\ detection rate
for YSOs with CO bandhead. However, 5 out of 36 (i.e. 14\%) RMS objects with CO bandhead emission do not
also have \brg\ emission. 
 
The left panel of Figure \ref{fig11} shows CO 2--0 bandhead
luminosity against bolometric luminosity. The lines represent the
ATS regressions. Due to the number of censored data points (i.e.
limits), a least-squares fit was not appropriate; the least-squares
method either ignores the censored points or considers them with
equal weight to the uncensored points, either of which gives a bias
to the results of a fit. This is why regression was performed using
the Akritas-Thiel-Sen (ATS) non-parametric algorithm rather than the
usual least-squares method, and correlation tests were performed
using the Kendall's $\tau$ method \citep{cenken}.
These methods take into account the effect of censored data points
on the correlation and regression, without giving them equal weight
to the detections. For data with few censored points, the ATS
line is comparable with the least squares method, so can handle a
wide variety of data sets. For consistency and to allow like-for-like
comparison with the other lines, the ATS method was used for all
regressions and correlation tests. 
 
\begin{figure*} 
\begin{minipage}{84mm}
\includegraphics[width=84mm]{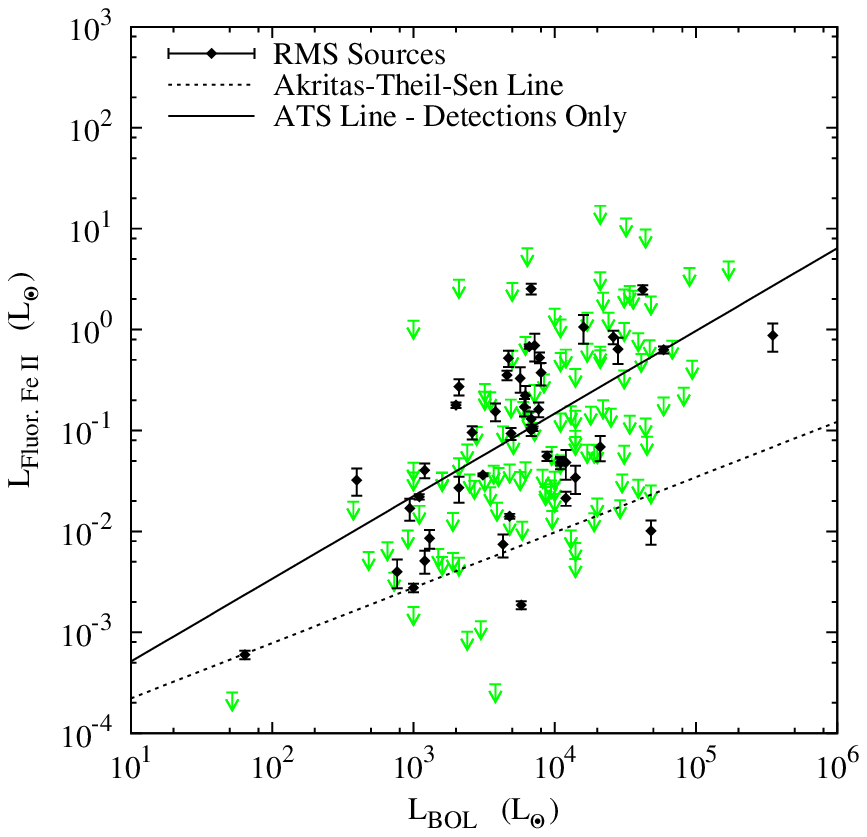}
\caption{ Fe~II 1.6878~$\umu$m luminosity against bolometric luminosity for RMS sources. Detections
are denoted by black diamonds and limits by green arrows. The lines
represent the ATS regressions. Two lines are shown, one for
regression with all the data including non-detections (dotted line),
and one for regression with detections only (solid line). Line
luminosities have been corrected for extinction and for the
difference between measured and catalogue magnitudes. 
} \label{fig12}
\end{minipage}
\hspace{0.7cm}
\begin{minipage}{84mm}
\includegraphics[width=84mm]{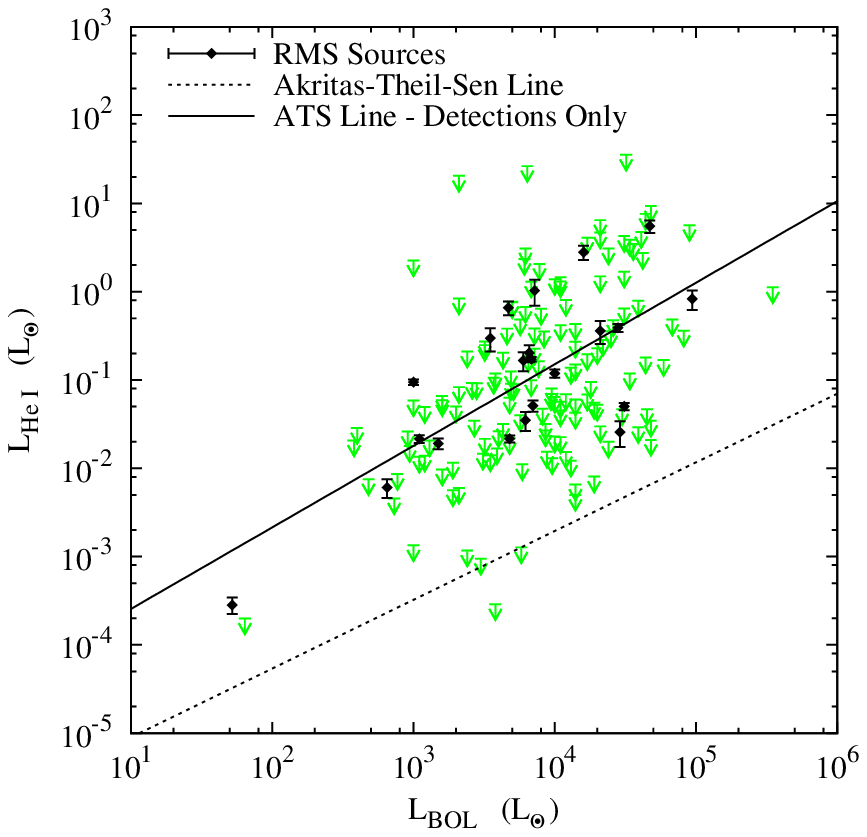}
\caption{He~I 2.0587~$\umu$m luminosity against bolometric luminosity for RMS sources. Detections
are denoted by black diamonds and limits by green arrows. The lines
represent the ATS regressions. Two lines are shown, one for
regression with all the data including non-detections (dotted line),
and one for regression with detections only (solid line). Line
luminosities have been corrected for extinction and for the
difference between measured and catalogue magnitudes. 
} \label{fig13}
\end{minipage}
\end{figure*}
 
Initially, the fit and correlation tests were performed on all the
data. The results show that there is no overall correlation between
CO bandhead luminosity and bolometric luminosity (see Table
\ref{tab:correlations}). However, when non-detections are excluded,
a correlation between the CO bandhead and bolometric luminosity is
found. This difference is reflected in the regression lines: The
equation for all the data including non-detections is
\begin{equation}
{\rm L}_{{\rm CO}} = 10^{-5.1}~\times~({\rm L}_{{\rm BOL}})^{0.58}
\end{equation}
The equation for detections only is 
\begin{equation}
{\rm L}_{{\rm CO}} = 10^{-4.1}~\times~({\rm L}_{{\rm BOL}})^{0.83}
\end{equation} 
The non-detections `push' the line down to well below where it would
be if only detections were taken into account. This demonstrates the
effect that censored data points have on a large sample, and shows
that it is important to include the censored points in the analysis
when trying to understand the overall trends in the data-set. This
regression method was employed for all the data for consistency,
even when the censored points have only a minimal effect on the
results (e.g. \brg\ emission). 

For the CO bandhead emission, the reason for this difference may be
that the ability to detect CO bandhead is unreliable. In some
objects the broad CO bandhead emission may simply be obscured by the
spectral noise, or it may be swamped in a steeply rising continuum
caused by excess dust emission. This effect probably predominates
for the sources with the higher limits in Figure \ref{fig11}. It is
known that CO bandhead emission is variable in some MYSOs (e.g.
\citealt{1997ApJ...491..359B, 2006A&A...457..183C}). If this is true
for other MYSOs in our sample, we may not have observed the object
when CO bandhead emission is present. There may be an inclination
effect, though we have no way of testing this suggestion at this low
resolution. Alternatively, some objects may be genuinely CO weak,
particularly those objects with very low limits in Figure
\ref{fig11}. This may indicate the lack of a disk with suitable
neutral material for these objects, suggestive of very low accretion
rates.

The right panel of Figure \ref{fig11} compares detections in
the RMS sources with those of \citet{carr1989} and
\citet{ishii2001}. The solid lines represent the loci of
${\rm L}_{{\rm CO}} = 10^{-3}\ {\rm L}_{{\rm BOL}}$,
${\rm L}_{{\rm CO}} = 10^{-4}\ {\rm L}_{{\rm BOL}}$, and
${\rm L}_{{\rm CO}} = 10^{-5}\ {\rm L}_{{\rm BOL}}$.
Most sources from all 3 surveys have $10^{-5}\ {\rm L}_{{\rm BOL}} <
{\rm L}_{{\rm CO}} < 10^{-3}\ {\rm L}_{{\rm BOL}}$. This shows
that, once again, there is a trend through the low-, intermediate-
and high-mass YSOs, implying a common mechanism throughout the mass
range for sources in which CO is detected.

Fluorescent Fe~II 1.6878~$\umu$m emission is also thought to
originate in a disk \citep{2004ApJ...615..610B,
2007A&A...470..239Z}. Figure \ref{fig12} shows fluorescent
Fe~II 1.6878~$\umu$m luminosity against bolometric luminosity. The
lines represent the ATS regressions. Again, due to the large number
of limits, two regressions were performed. No overall correlation
was found, but there was a correlation between ${\rm L}_{{\rm Fe\
II}}$ and ${\rm L}_{{\rm BOL}}$ when non-detections were excluded
(see Table \ref{tab:correlations}). As for CO bandhead, this
difference is reflected in the regression lines: 
The equation for all the data including non-detections is
\begin{equation}
{\rm L}_{{\rm Fe\ II}} = 10^{-4.2}~\times~({\rm L}_{{\rm BOL}})^{0.54}
\end{equation}
The equation for detections only is
\begin{equation}
{\rm L}_{{\rm Fe\ II}} = 10^{-4.1}~\times~({\rm L}_{{\rm BOL}})^{0.82}
\end{equation} 
Spectral noise may have an effect, and is probably the cause of the
higher upper limits in Figure \ref{fig12}. However, we
believe that the main effect differentiating between detections only
and all data including non-detections is probably that there are
genuine differences between sources. These differences will be
discussed in more detail in a future paper.

\subsection{He~I}
The He~I 2.0587~$\umu$m line was seen in emission in 15\% of YSOs.
Due to the requirement of hard UV photons to ionize helium, it was
naively expected that there should be a threshold luminosity of
$\sim 10^{4}$~\lsol\ or more. This is not seen in the data; He~I
emission is seen across the whole luminosity range covered by the
RMS data. This implies that either luminosity is not necessarily a
good indicator of temperature for high mass YSOs, or temperature
does not drive He~I production. Collisionally excited He~I emission
is frequently seen in low-mass T-Tauri stars, originating in a
stellar wind \citep{2006ApJ...646..319E, 2007IAUS..243..171E,
2009AIPC.1094...29E}. The He~I emission in the RMS sources with
luminosity ${\rm L} < 10^{4}$~\lsol\ is probably also collisionally
excited, originating in a wind, rather than ionized as in H~II
regions. This may also be the emission mechanism for some (or all)
of the stars with ${\rm L} > 10^{4}$~\lsol.

In addition, He~I was seen in absorption in 5 spectra (e.g.
G076.3829-00.6210 / S106IR, see Figure \ref{fig3}), or with
a P-Cygni type profile in 15 spectra (see Table B3). Unfortunately, at this low
resolution it was not possible to obtain expansion velocities for
these lines. He~I 2.0587~$\umu$m absorption has been studied in
detail with high resolution spectroscopy for G076.3829-00.6210 by
\citet{1993MNRAS.265...12D}. They concluded that to be seen in
absorption, this line requires high kinetic temperatures (${\rm
T}\ga 10^{4}$ K) and electron densities ($n_{e}\ga 10^{8}\ {\rm
cm}^{-3}$) in order for the transition to be preferentially
populated from below rather than above. This suggests that the He~I
absorption in G076.3829-00.6210 is produced in an outflow
originating very close to the star. It is likely that He~I
absorption in other high-mass YSOs also originates in an outflow. 
 
Figure \ref{fig13} shows He~I 2.0587~$\umu$m luminosity
against bolometric luminosity, for sources with He~I in
emission. He~I absorption is not included in this plot. The lines
represent the ATS regressions. Again, two regressions and
correlation tests were performed, finding that a correlation between
He~I luminosity and bolometric luminosity is not an overall trend in
the MYSO population, but does exist where He~I is detected (see Table
\ref{tab:correlations}). 
The equation for all the data including non-detections is
\begin{equation}
{\rm L}_{{\rm He\ I}} = 10^{-5.8}~\times~({\rm L}_{{\rm BOL}})^{0.78}
\end{equation}
The equation for detections only is
\begin{equation}
{\rm L}_{{\rm He\ I}} = 10^{-4.5}~\times~({\rm L}_{{\rm BOL}})^{0.92}
\end{equation} 
In this case, spectral noise is less likely to be a causal factor
for the differences in the analysis; the spectrum at 2.0587~$\umu$m
is not as noisy as at $\sim 2.3\ \umu$m or the H-band. It is
possible that the low resolution of the spectra mean that
self-absorption could prevent us from detecting He~I if the line is
sufficiently narrow. At higher resolutions we may see He~I in more
objects when the profile could be resolved. However, it is more
likely that the results of these correlation tests reflect real
differences between the sources. 

%%%%%%%%%%%%%%%%%%%%%%%%%%%%%%%%%%%%%%%%
\section{Conclusions}
Near-IR spectroscopic data were obtained for 247 objects selected
from the RMS survey. The spectra were classified, finding 180 YSOs,
15 YSOs within H~II regions, 26 H~II regions, 7 evolved stars, 5
PPN, 4 PN, 5 M-type supergiants, and 5 evolved hot stars. Spectra
were found to be a valuable confirmation of classification
in addition to our imaging data in 45\% of cases, and the only means of
classifying the source in 5\%. The properties of the YSOs are as follows:
\begin{itemize} 
\item The YSOs have luminosities ranging from 52~\lsol\ to $3.5~\times~10^{5}$~\lsol, and extinctions ranging from $A_{V}=2.7$ to
$A_{V}=114$.
\item Most YSOs were found to have emission lines, though not all
had the same lines. The detection rates of \brg, \h, CO bandhead,
[Fe~II] 1.64~$\umu$m, fluorescent Fe~II 1.6878~$\umu$m, and He~I
2.0587~$\umu$m were 75\%, 56\%, 17\%, 11\%, 26\%, and 15\%
respectively. He~I 2.0587~$\umu$m was seen in absorption in 5
spectra. \brg\ and He~I 2.0587~$\umu$m were seen with P-Cygni type
profiles in 13 and 15 objects respectively, of which 8 objects had
both lines with P-Cygni profiles. This indicates the presence
of an outflowing wind.
\item In total, 39\% of the YSOs were found to have either
fluorescent Fe~II 1.6878~$\umu$m or CO bandhead emission (or both),
both of which are evidence of a circumstellar disc. 56\% had \h\
emission, indicating the presence of an outflow.
\item The \brg, \h, and CO bandhead data were compared with low-mass
YSOs from \citet{carr1989, carr1990} and intermediate-mass YSOs from
\citet{ishii2001}. The \brg\ detection rate was lower than that for
the intermediate-mass YSOs, but comparable with the detection rate
for low-mass YSOs. The \h\ detection rate was higher than the
detection rates for both the low- and intermediate-mass YSOs.
Conversely, the CO bandhead detection rate was lower for the RMS
YSOs than for both the low- and intermediate-mass YSOs. The
differences in detection rates between the 3 data sets may be due to
differences in obscuration and evolution. We will discuss these
differences in more detail in Cooper et al. (in preparation).
\item Strong correlations were found between ${\rm L}_{{\rm
Br}\gamma}$ and ${\rm L}_{{\rm BOL}}$ and between ${\rm L}_{{\rm
H}_{2}}$ and ${\rm L}_{{\rm BOL}}$.
\item For CO bandhead, no overall correlation was found with ${\rm
L}_{{\rm BOL}}$. However, when non-detections were excluded, there
was a correlation. This may be due to the difficulty in
distinguishing broad emission lines from a steeply rising continuum,
variability in the emission, or because some objects genuinely do
not have CO bandhead emission (perhaps indicating an evolutionary
sequence).
\item The same effect was seen for both Fe~II 1.6878~$\umu$m and He~I 2.0587~$\umu$m emission: no overall correlation with ${\rm
L}_{{\rm BOL}}$ was found but there were correlations found when
only detections above $3\sigma$ were considered. For Fe~II, spectral
noise is likely to be an issue for some sources, but we believe the
major contribution is that there are real differences between the
sources. For He~I, spectral noise is unlikely to be the issue: the
continuum at around $\sim 2.06\ \umu$m is not as noisy as at $\sim
2.3\ \umu$m or in the H-band. However, self-absorption might cause
the emission line not to be seen at this low resolution. Again, it
may also be explained by genuine differences between the sources. 
\item The line luminosity data for \brg, \h, and CO bandhead
detections were
compared with the low-mass YSOs from \citet{carr1989, carr1990} and
intermediate-mass YSOs from \citet{ishii2001}. For all three lines,
a trend was seen across the whole mass range, from luminosities
spanning five orders of magnitude. This suggests that the
production mechanism for each line is the same for low-,
intermediate-, and high-mass stars, i.e. high-mass YSOs appear to
resemble scaled-up versions of low-mass YSOs. This is consistent
with current theories in which all stars form in a similar fashion,
e.g. \citet{Krumholz2009}.
\end{itemize}

\section*{Acknowledgements}
This publication makes use of data products from the Two Micron All
Sky Survey, which is a joint project of the University of
Massachusetts and the Infrared Processing and Analysis
Center/California Institute of Technology, funded by the National
Aeronautics and Space Administration and the National Science
Foundation. We also make use of data obtained at the United Kingdom
Infrared Telescope, which is operated by the Joint Astronomy Centre
on behalf of the Science and Technology Facilities Council of the
U.K., and the SIMBAD database, operated at CDS, Strasbourg, France.

\bsp

%%%%%%%%%%% APPENDIX %%%%%%%%
\appendix
\section{Target Object Data Tables}  
\section{Emission Line Fluxes and Equivalent Widths} 

%%%%% TABLE-OBJECTS (6 LANDSCAPE PAGES) %%%%%
\clearpage
\begin{minipage}{\textwidth}
\begin{center}
\includegraphics[height=\textheight]{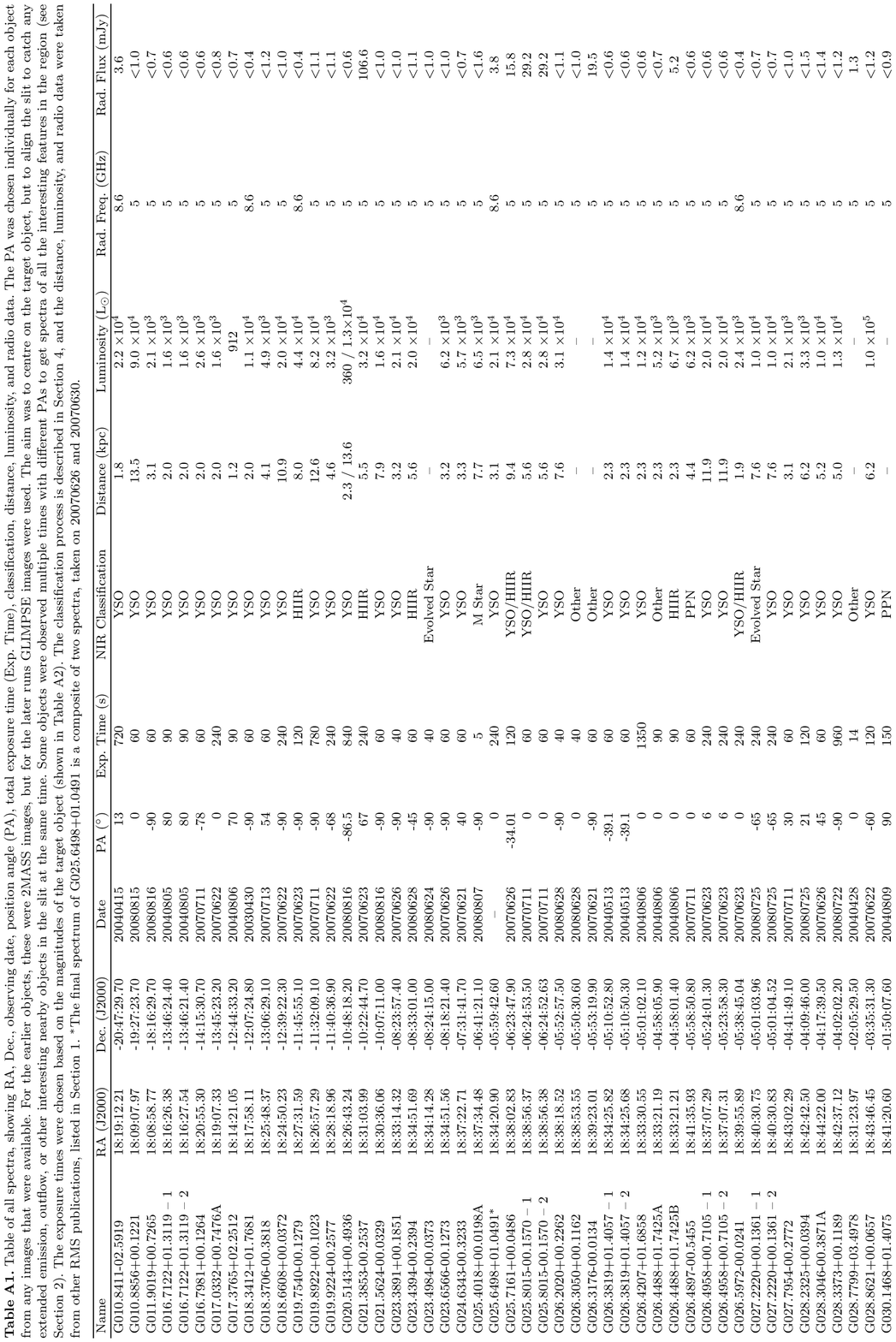}
\end{center}
\end{minipage}

\clearpage
\begin{minipage}{\textwidth}
\begin{center}
\includegraphics[height=\textheight]{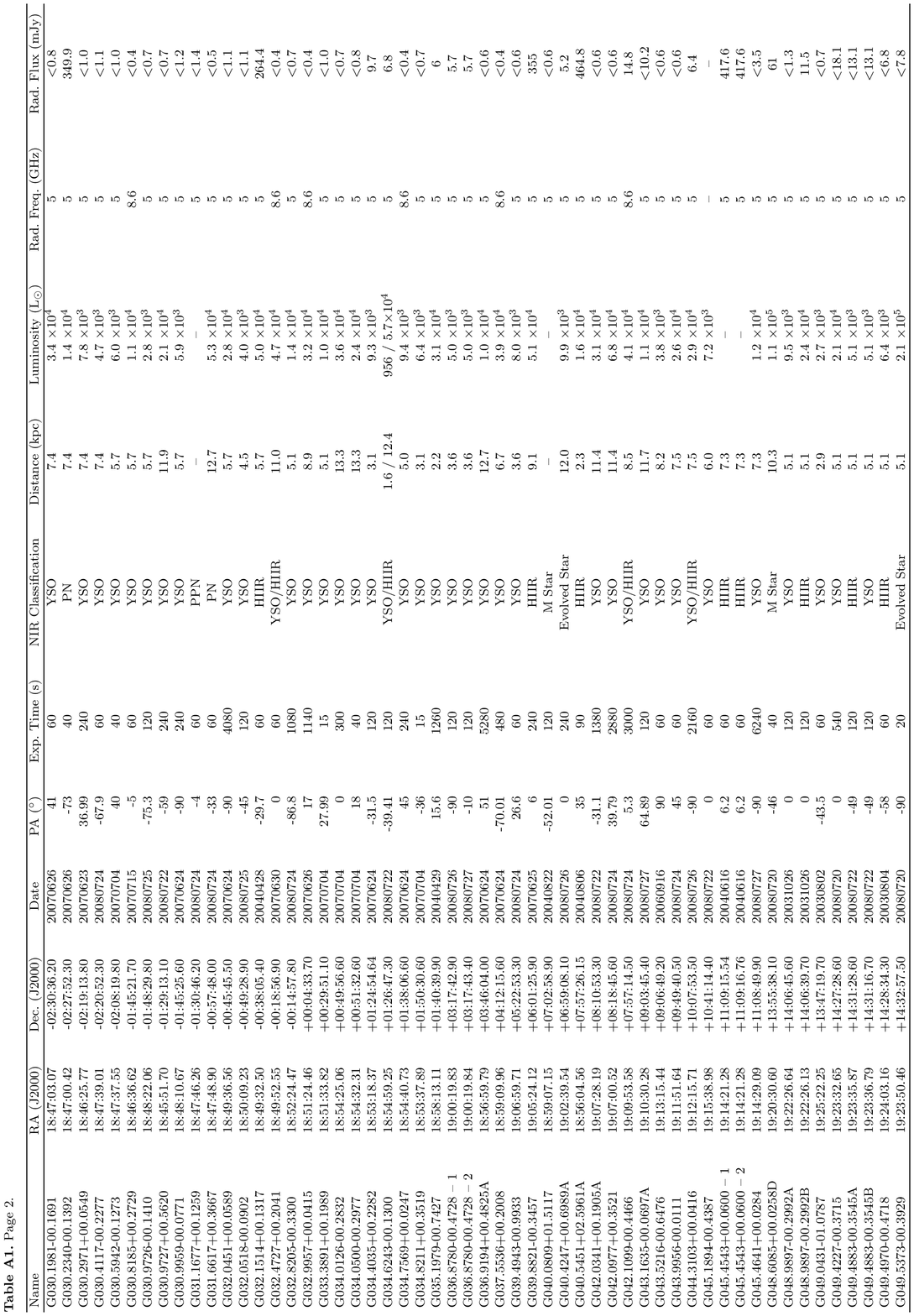}
\end{center}
\end{minipage}

\clearpage
\begin{minipage}{\textwidth}
\begin{center}
\includegraphics[height=\textheight]{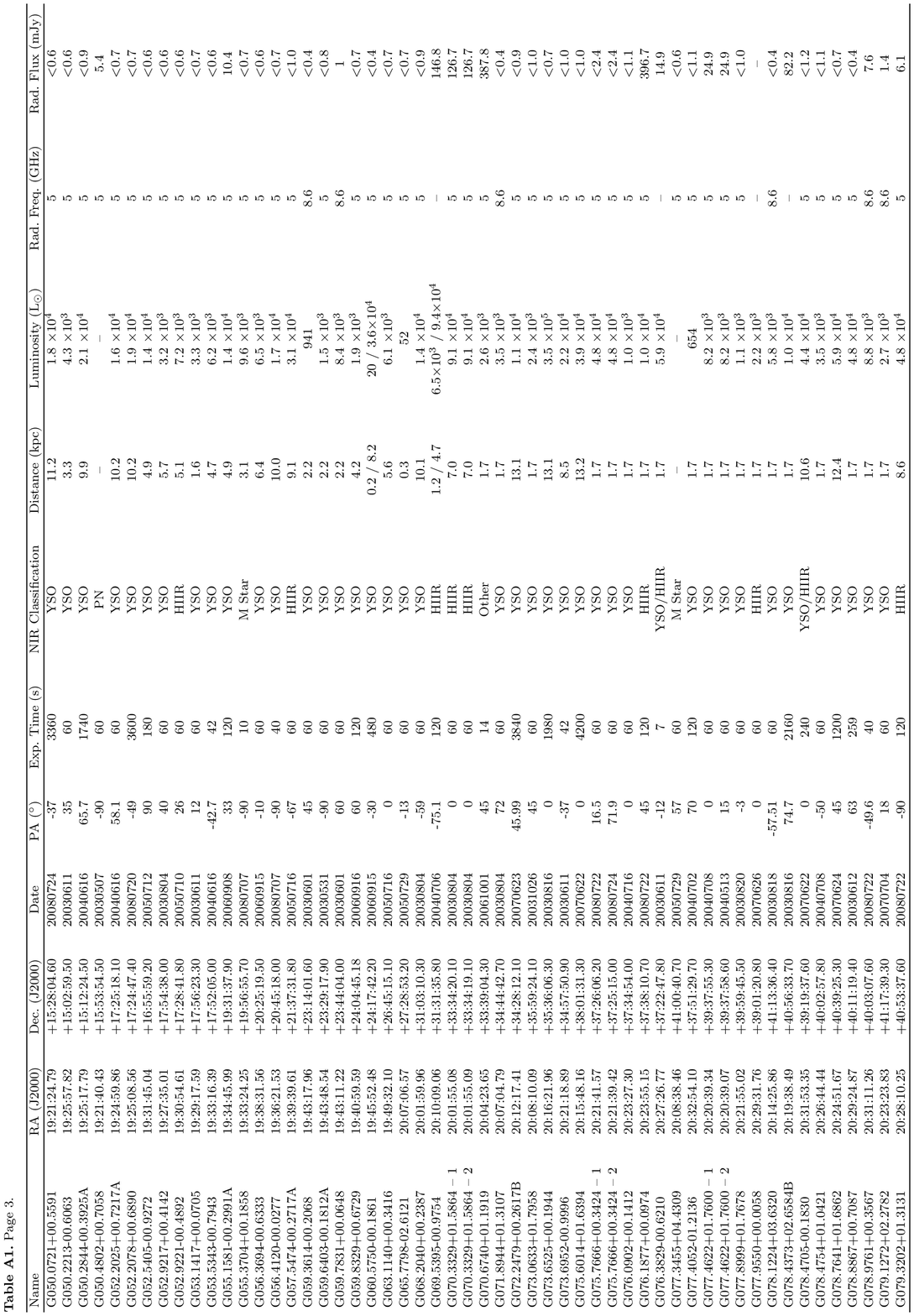}
\end{center}
\end{minipage}

\clearpage
\begin{minipage}{\textwidth}
\begin{center}
\includegraphics[height=\textheight]{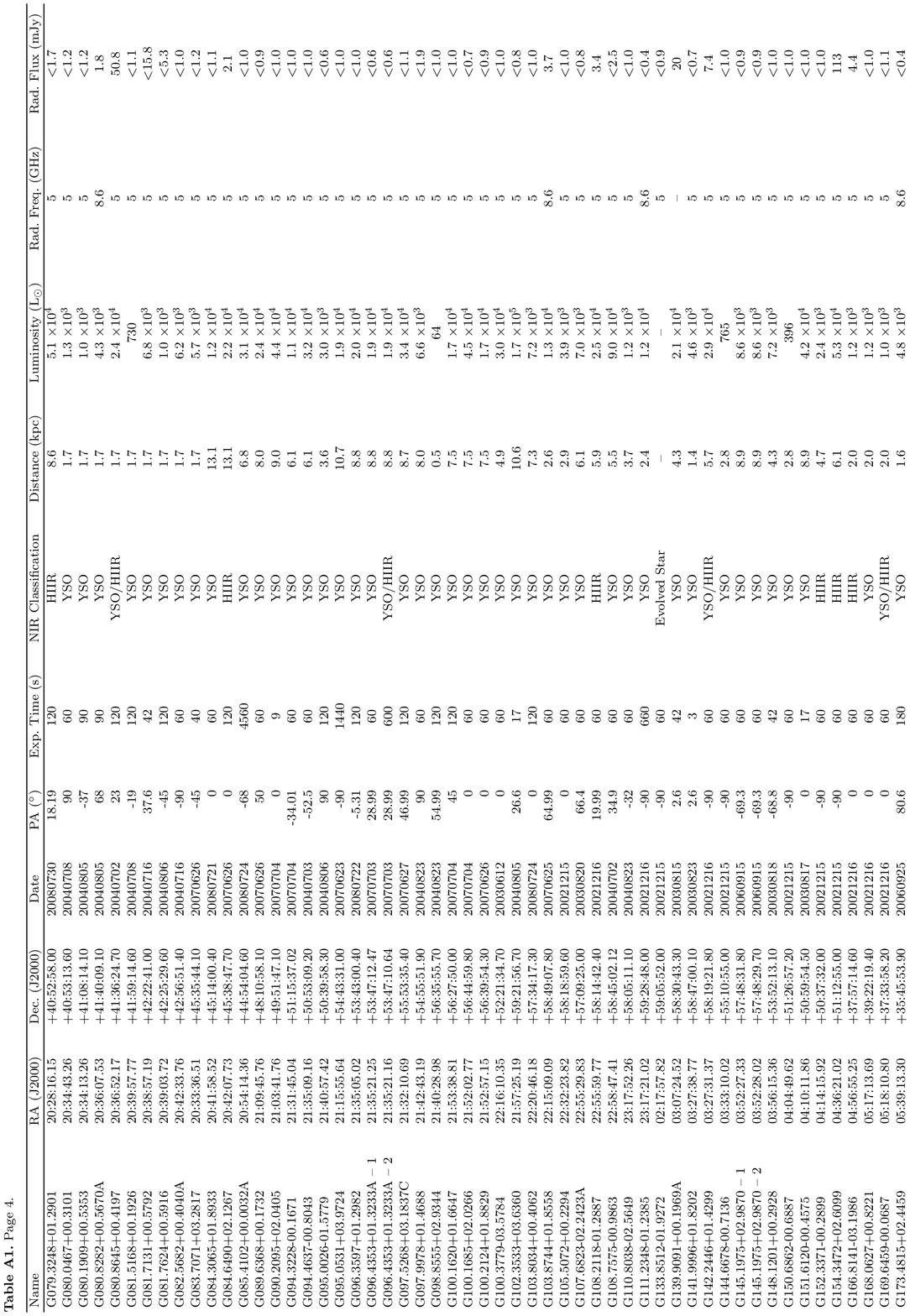}
\end{center}
\end{minipage}

\clearpage
\begin{minipage}{\textwidth}
\begin{center}
\includegraphics[height=\textheight]{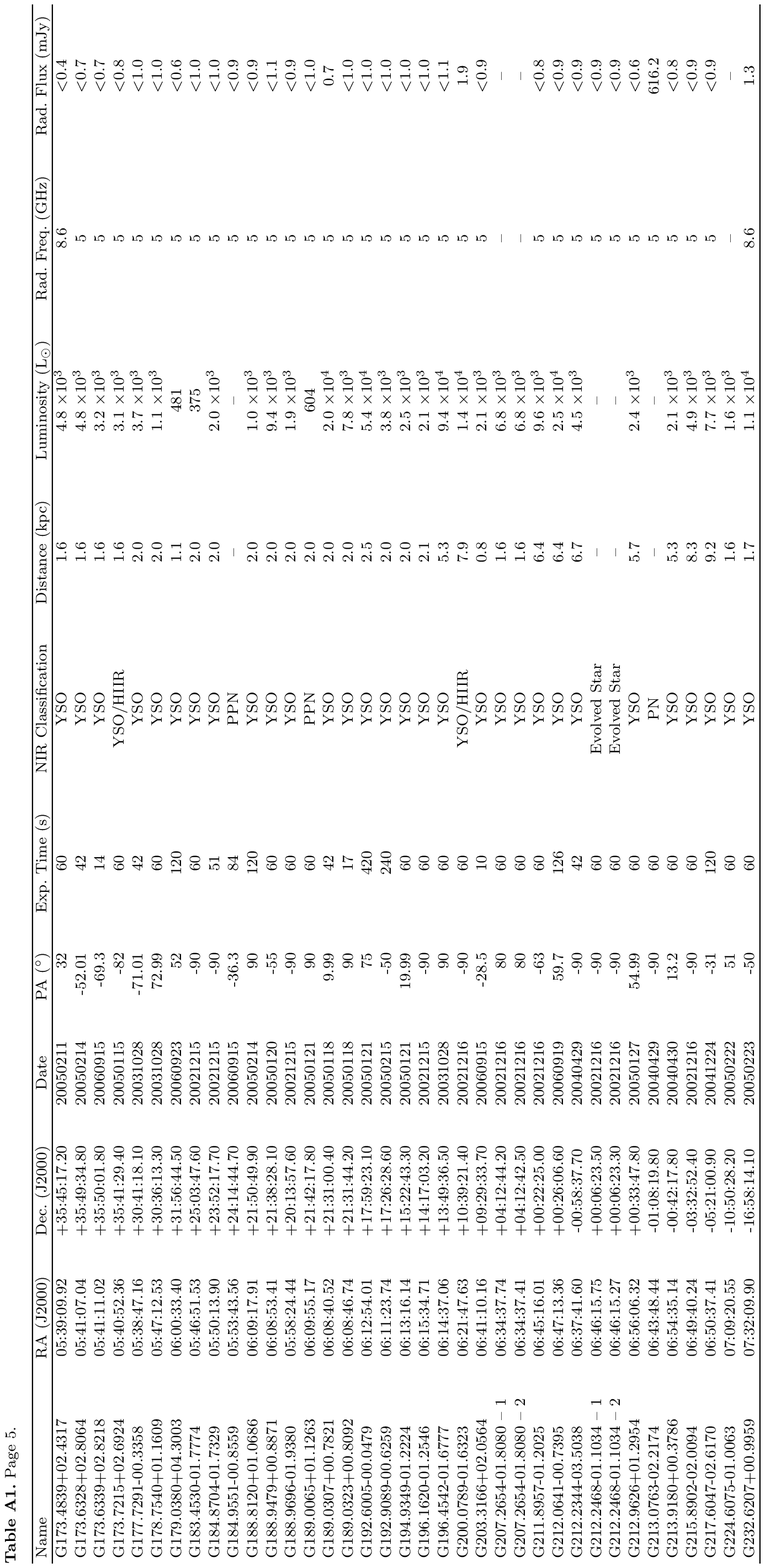}
\end{center}
\end{minipage}

%%%%%% TABLE-MAGNITUDES (6 LANDSCAPE PAGES) %%%%%% 
\clearpage
\begin{minipage}{\textwidth}
\begin{center}
\includegraphics[height=\textheight]{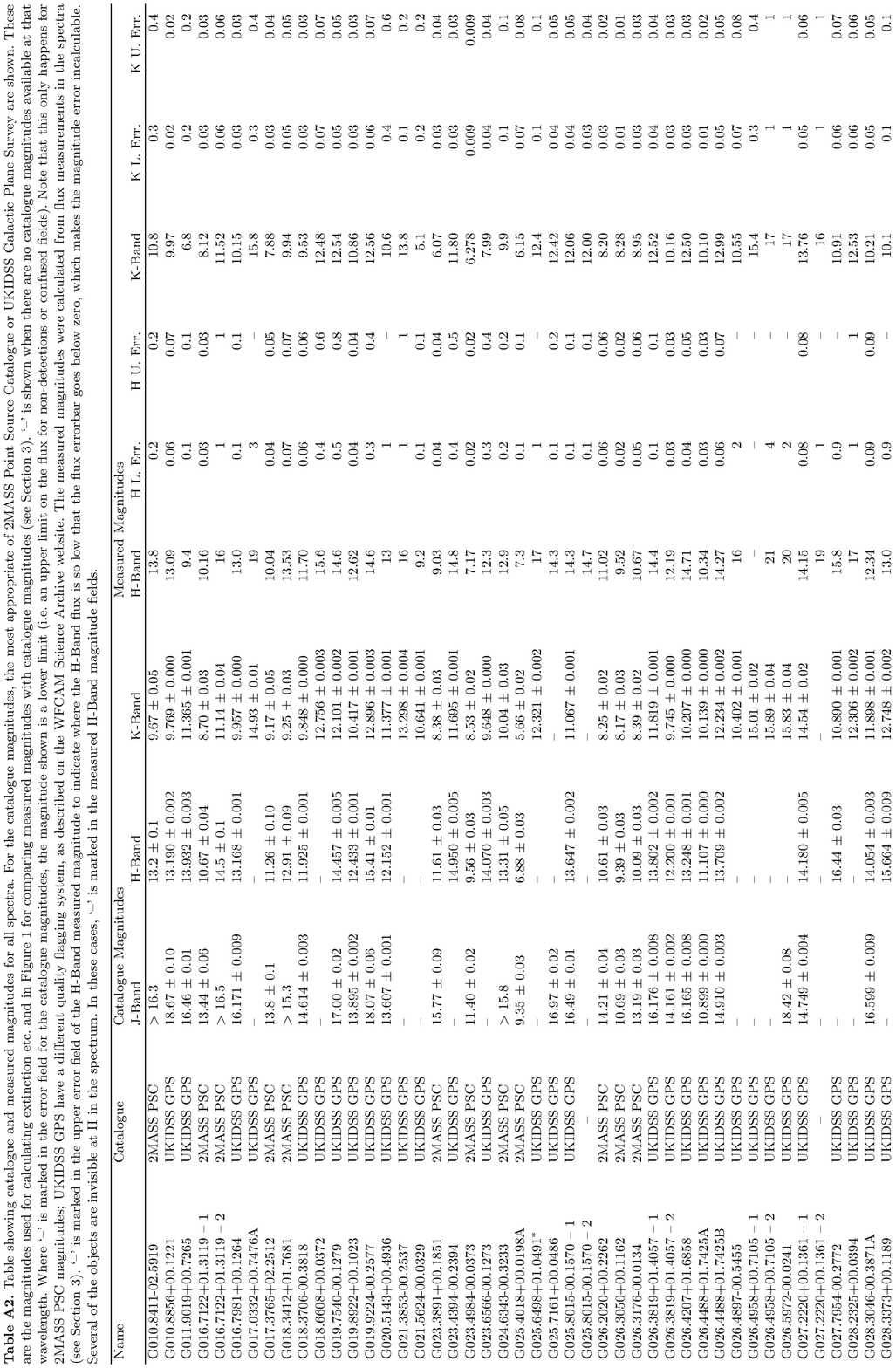}
\end{center}
\end{minipage}

\clearpage
\begin{minipage}{\textwidth}
\begin{center}
\includegraphics[height=\textheight]{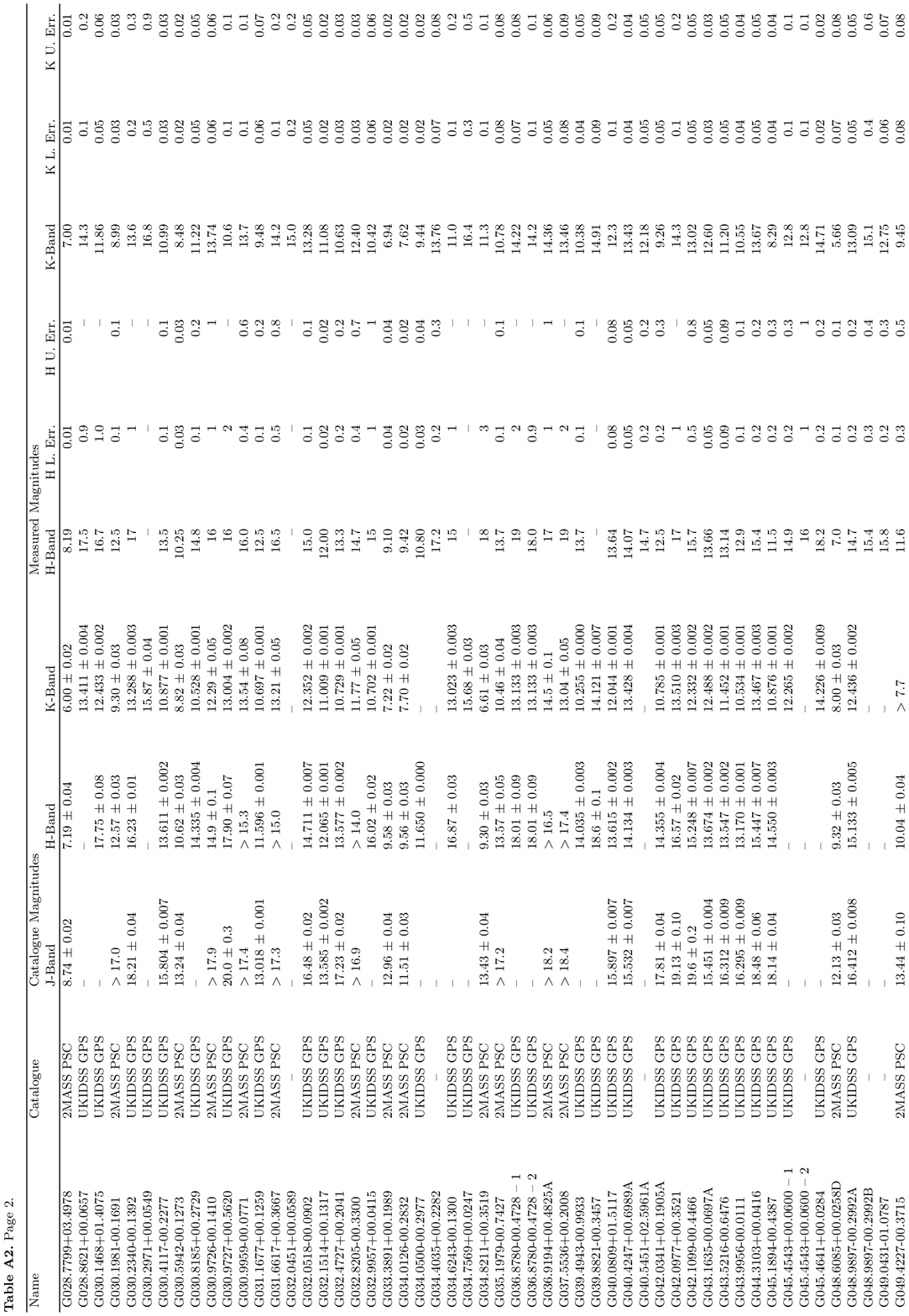}
\end{center}
\end{minipage}

\clearpage
\begin{minipage}{\textwidth}
\begin{center}
\includegraphics[height=\textheight]{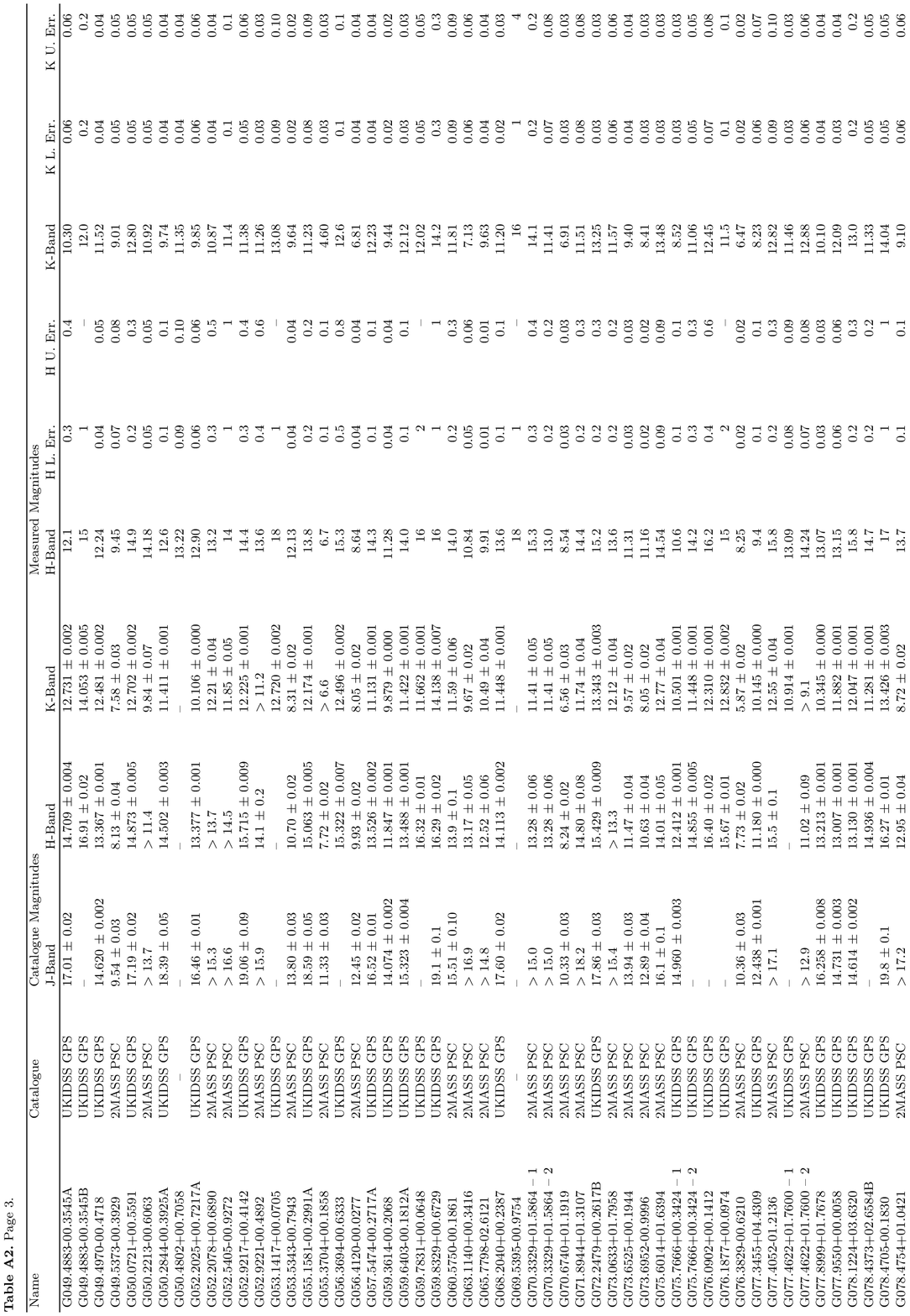}
\end{center}
\end{minipage}

\clearpage
\begin{minipage}{\textwidth}
\begin{center}
\includegraphics[height=\textheight]{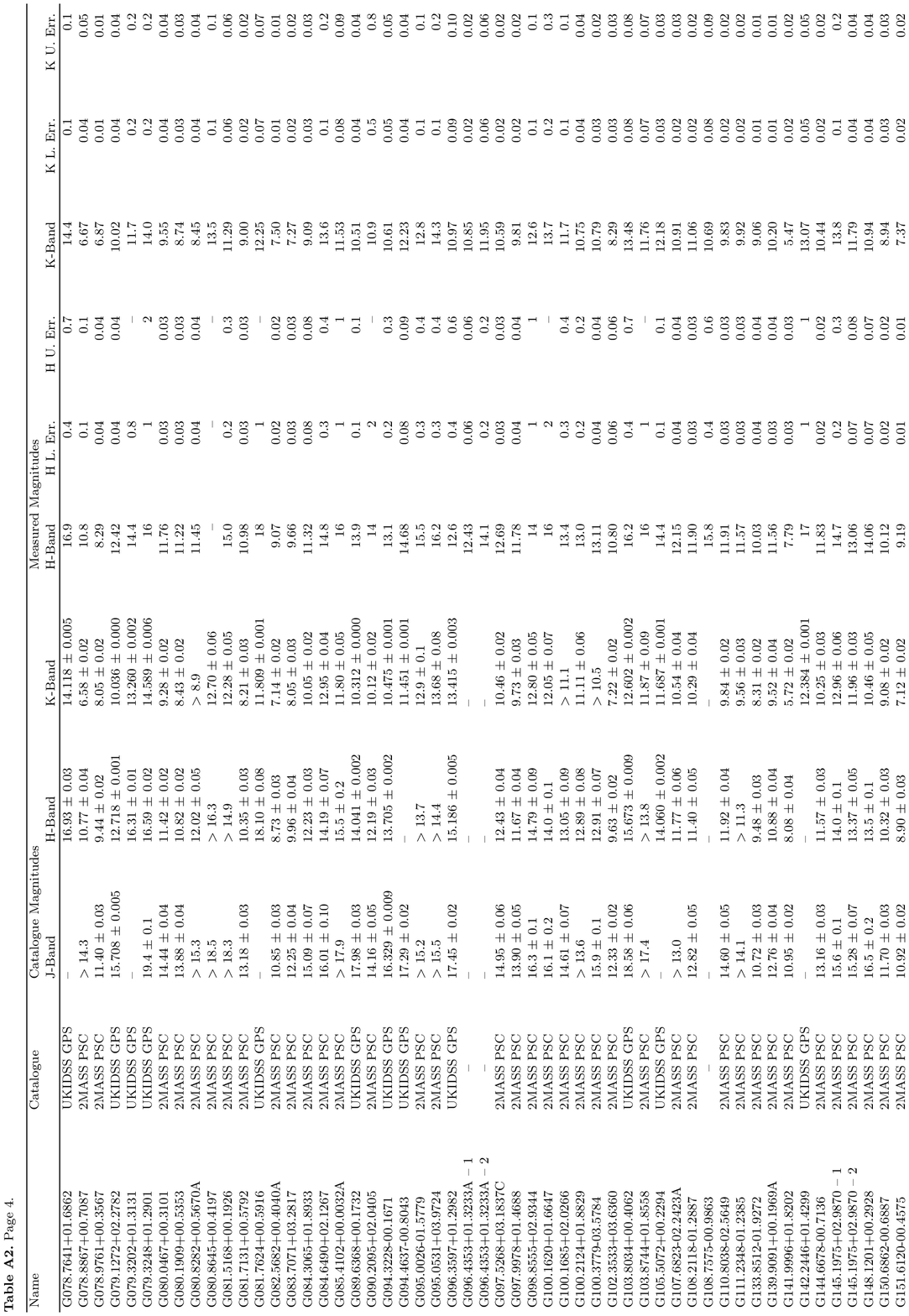}
\end{center}
\end{minipage}

\clearpage
\begin{minipage}{\textwidth}
\begin{center}
\includegraphics[height=\textheight]{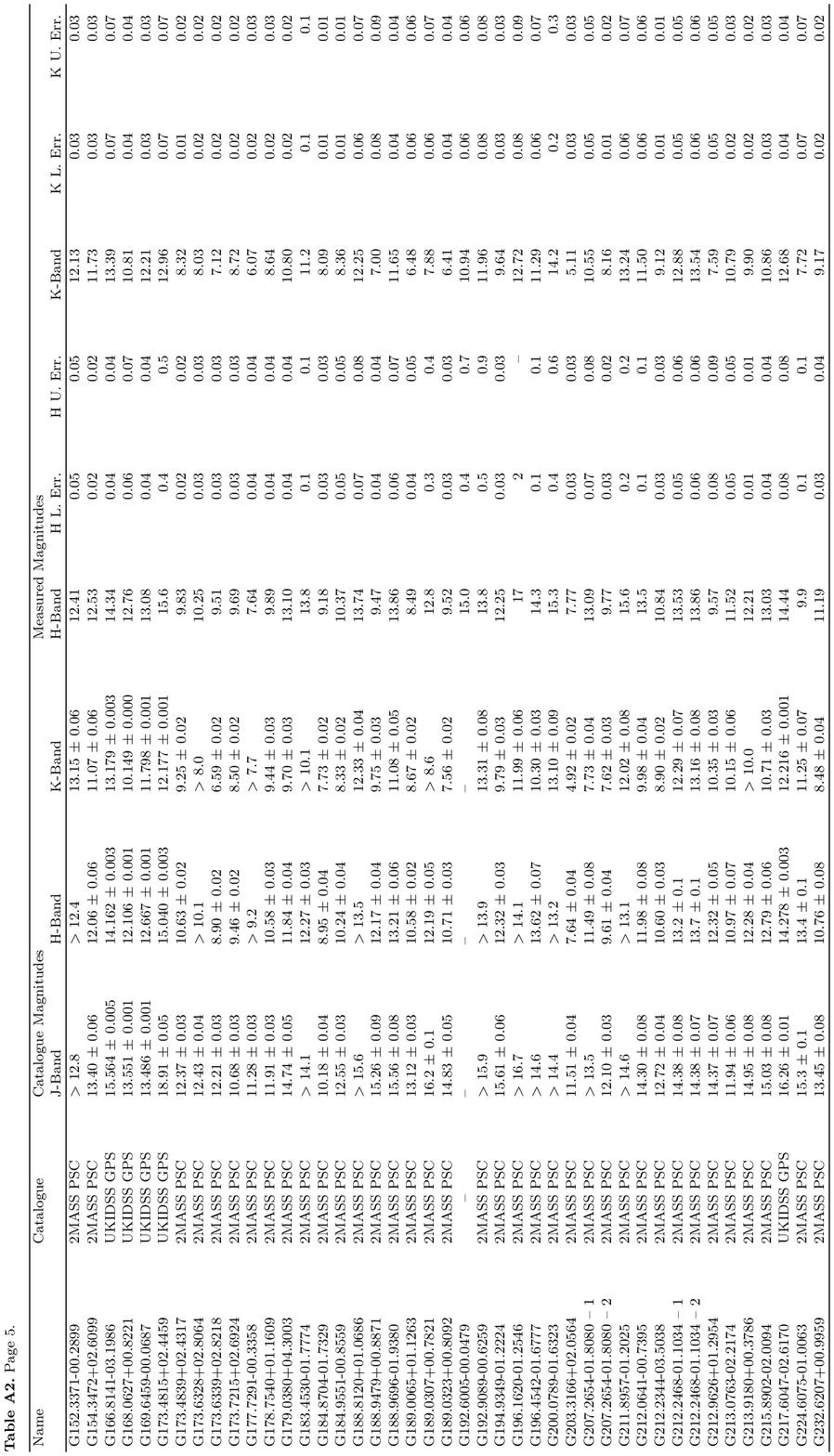}
\end{center}
\end{minipage}

%%%%%% TABLE-EXTICTION (2 PAGES) %%%%%%
\clearpage
\begin{minipage}{\textwidth}
\begin{center}
\includegraphics[height=\textheight]{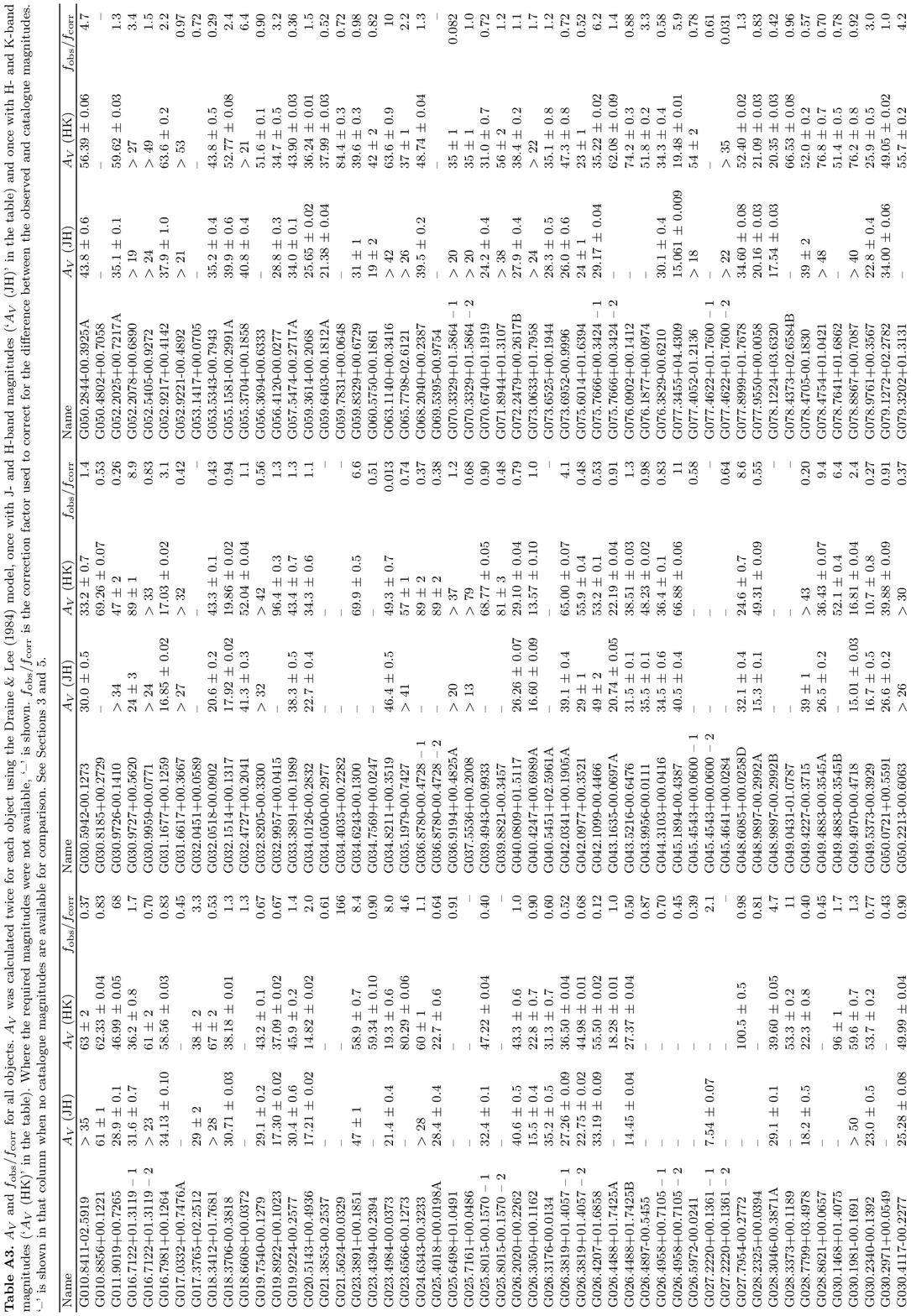}
\end{center}
\end{minipage}

\clearpage
\begin{minipage}{\textwidth}
\begin{center}
\includegraphics[height=\textheight]{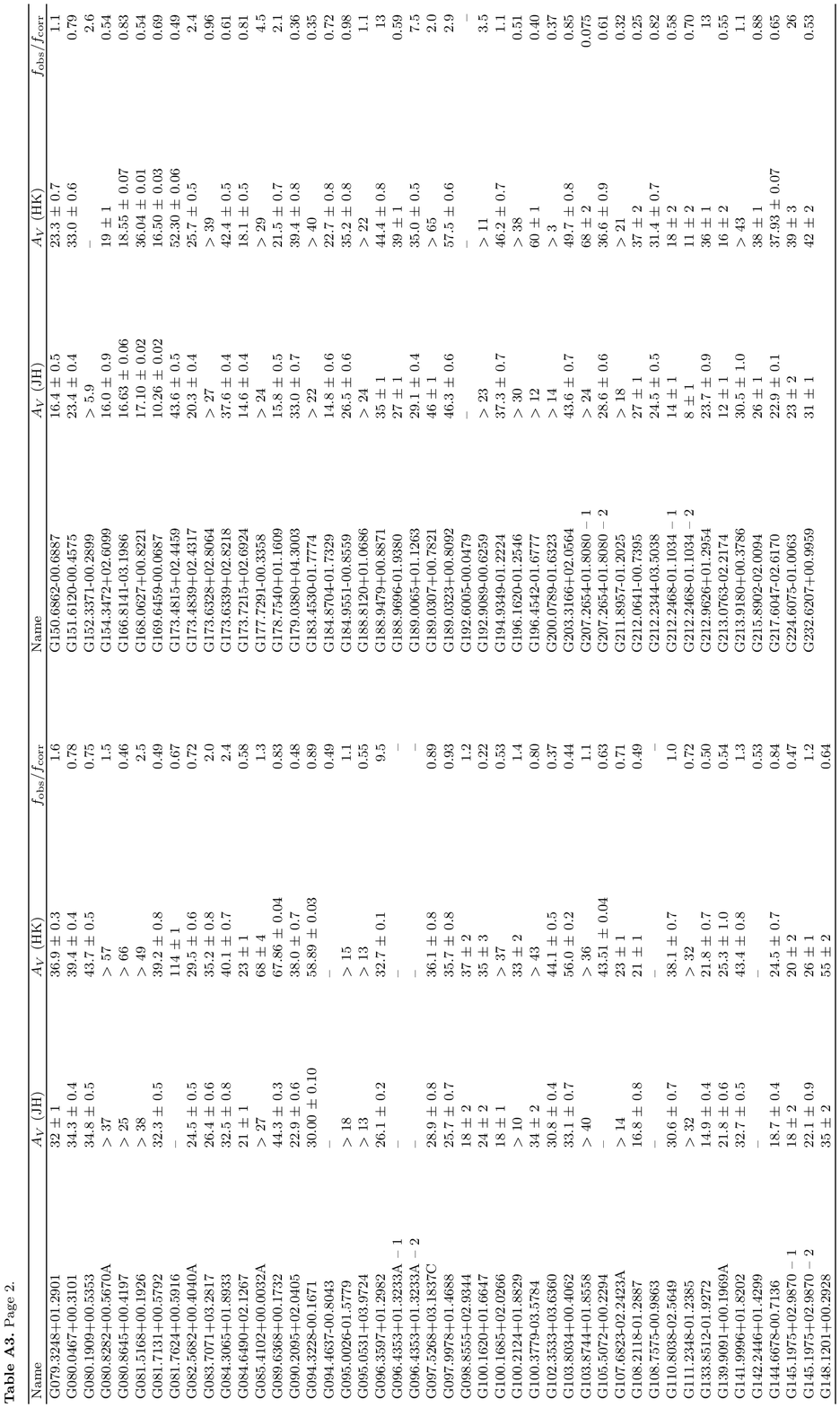}
\end{center}
\end{minipage}

%%%%%% FLUXES & EWS %%%%%%
\clearpage
\begin{minipage}{\textwidth} 
\begin{center}
\includegraphics[height=\textheight]{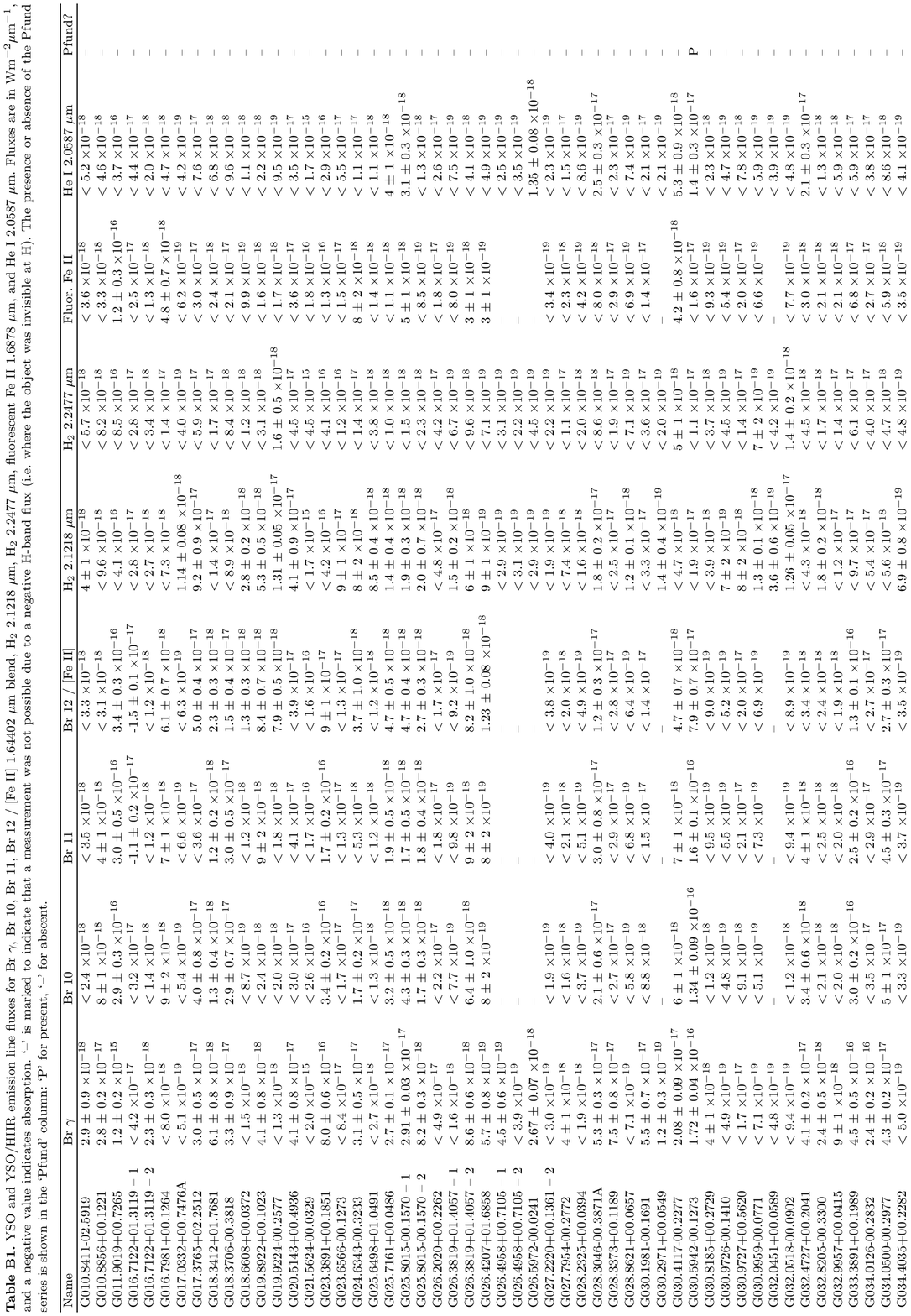}
\end{center}
\end{minipage}

\clearpage
\begin{minipage}{\textwidth}
\begin{center}
\includegraphics[height=\textheight]{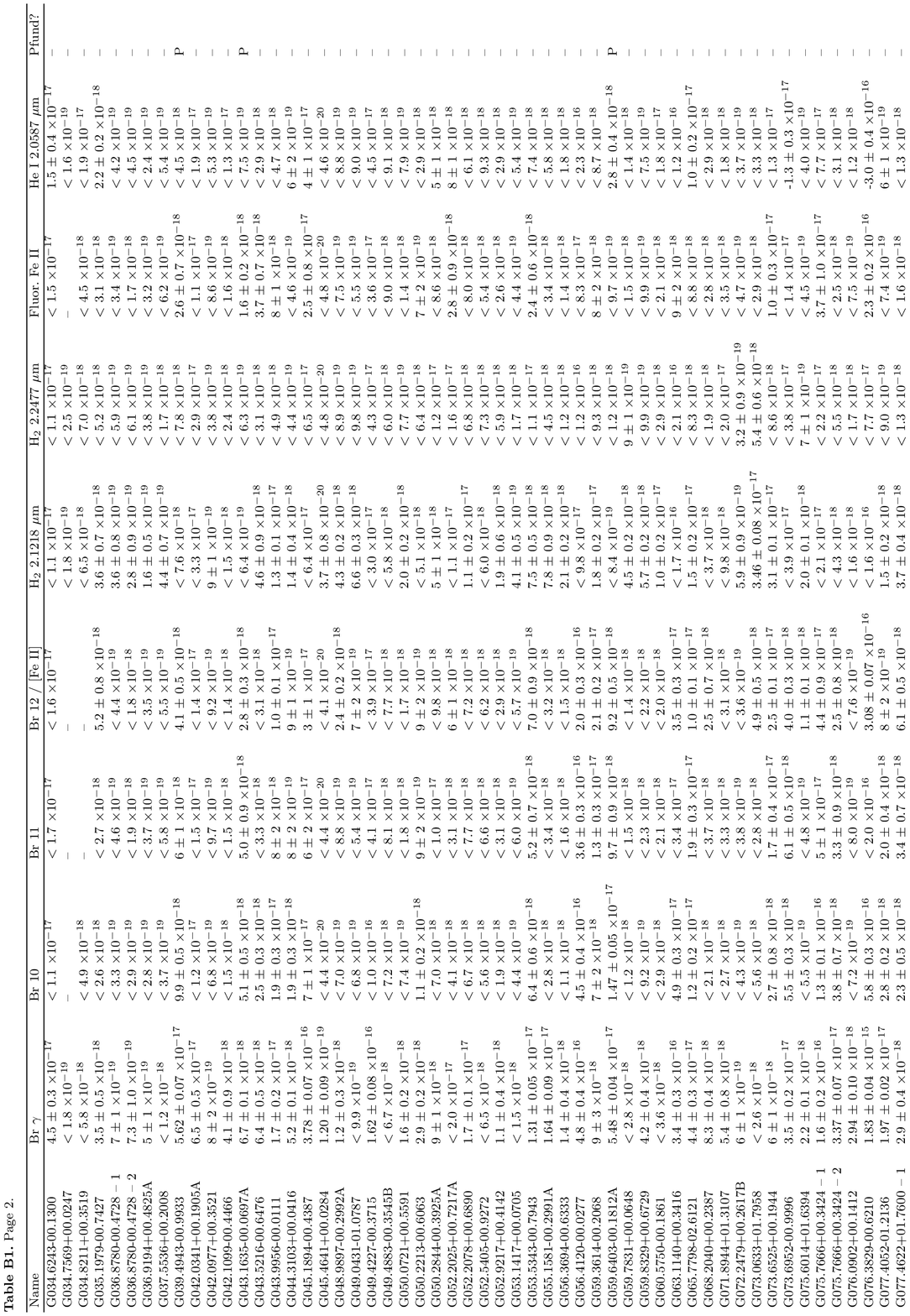}
\end{center}
\end{minipage}

\clearpage
\begin{minipage}{\textwidth}
\begin{center}
\includegraphics[height=\textheight]{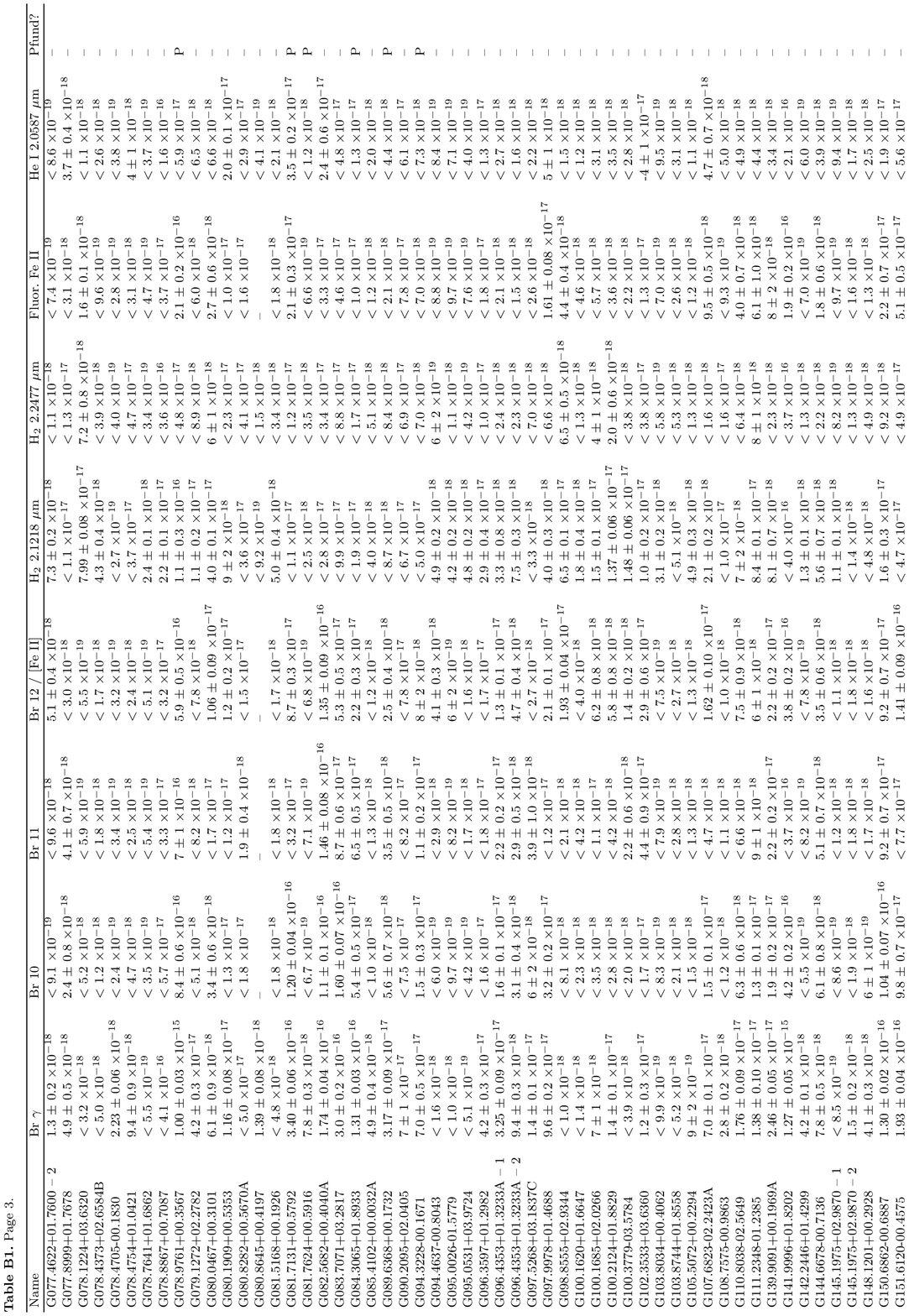}
\end{center}
\end{minipage}

\clearpage
\begin{minipage}{\textwidth}
\begin{center}
\includegraphics[height=\textheight]{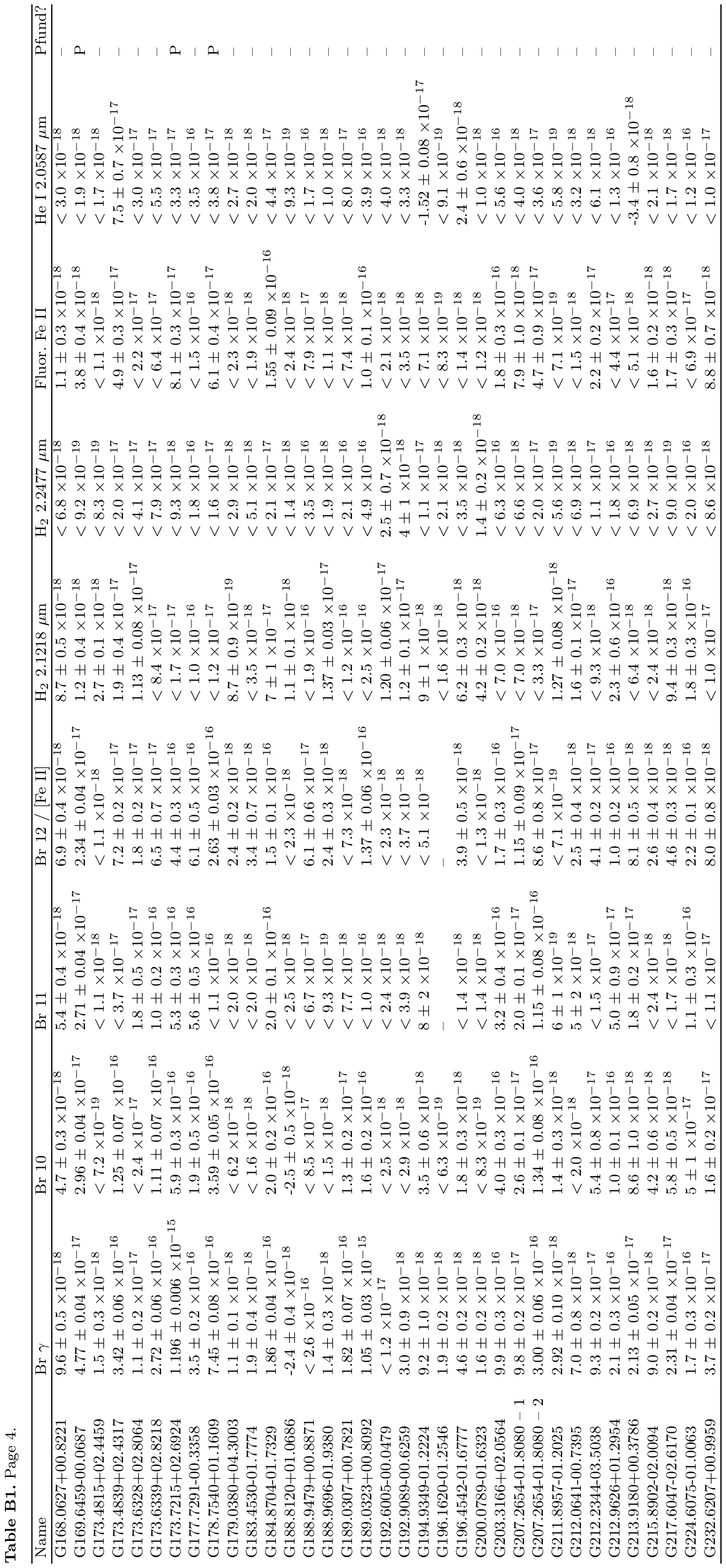}
\end{center}
\end{minipage}

\clearpage
\begin{minipage}{\textwidth}
\begin{center}
\includegraphics[height=\textheight]{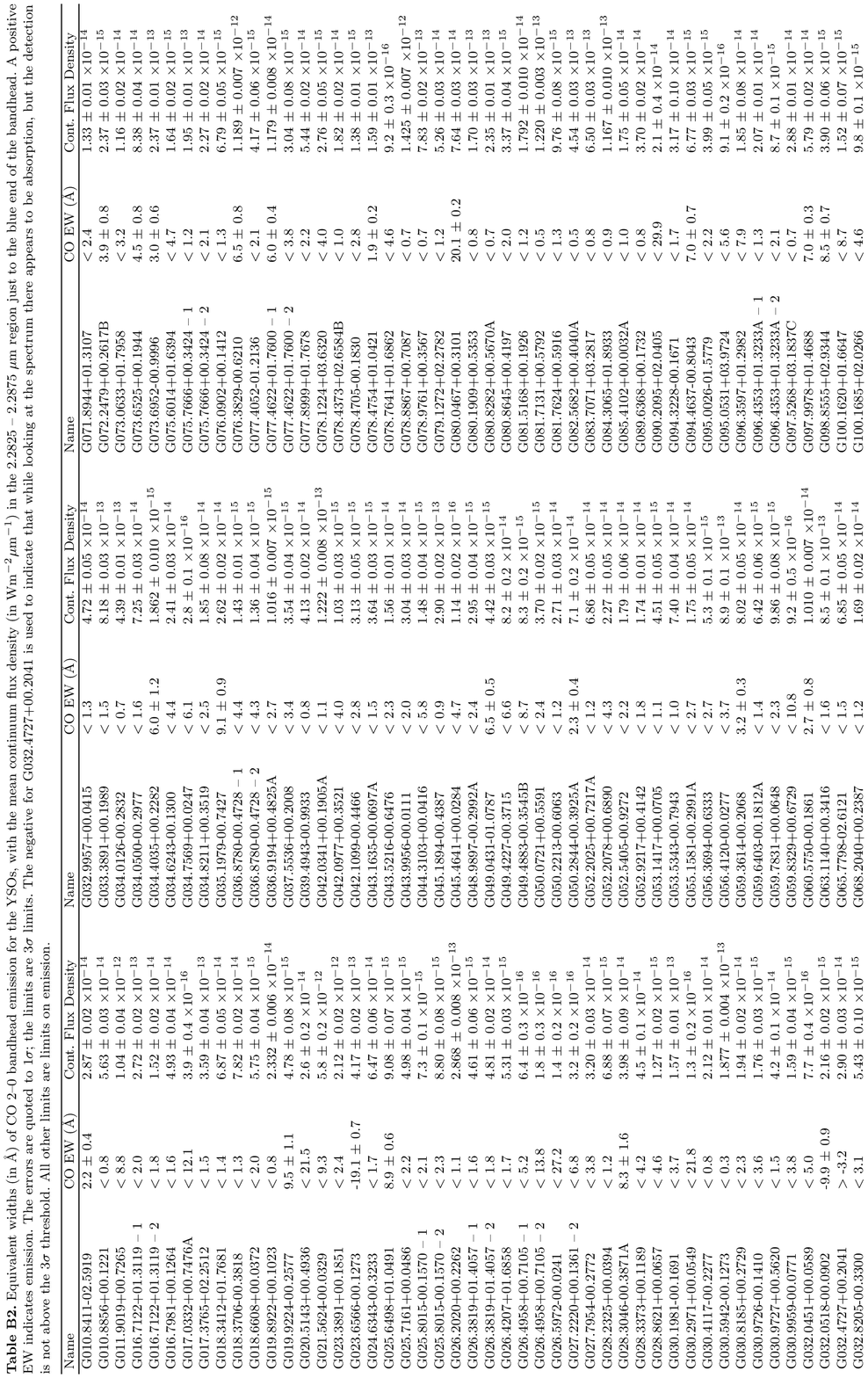}
\end{center}
\end{minipage}

\clearpage
\begin{minipage}{\textwidth}
\begin{center}
\includegraphics[height=\textheight]{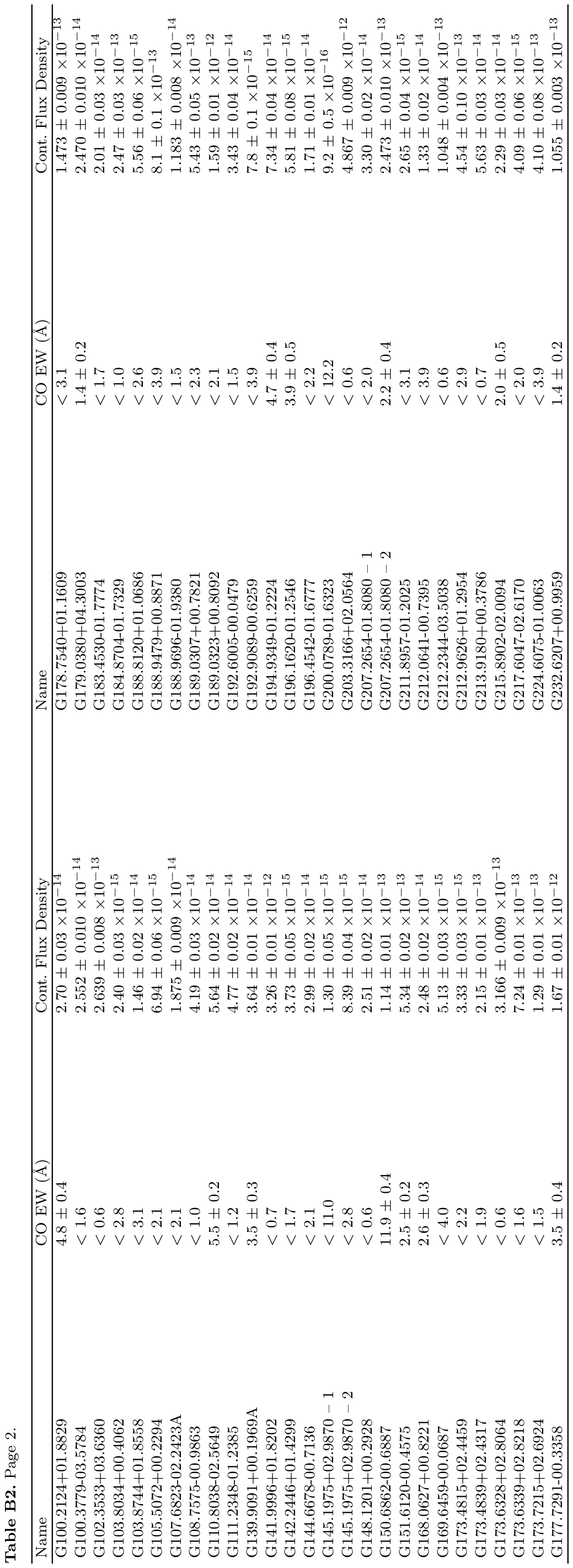}
\end{center}
\end{minipage}

%%%%% P-CYGNI PROFILES %%%%%
\setcounter{section}{2}
\setcounter{table}{2}
\begin{table*}
\raggedright
\begin{minipage}{84mm}
\caption{List of objects for which either or both of the \brg\ or He~I
2.0587~$\umu$m emission lines have P-Cygni profiles.}
\begin{tabular*}{\textwidth}{@{\extracolsep{\fill} }ll}
\hline
Object	&	P-Cygni Profile	\\	\hline
G017.3765+02.2512 	&	Both \brg\ and He I	\\	
G018.3412+01.7681 	&	\brg\ only	\\	
G023.3891+00.1851 	&	He I only	\\	
G027.7954-00.2772 	&	\brg\ only	\\	
G056.4120-00.0277 	&	Both \brg\ and He I	\\	
G073.6525+00.1944 	&	\brg\ only	\\	
G073.6952-00.9996 	&	He I only	\\	
G076.3829-00.6210 	&	He I only	\\	
G077.8999+01.7678 	&	\brg\ only	\\	
G102.3533+03.6360 	&	Both \brg\ and He I	\\	
G141.9996+01.8202 	&	Both \brg\ and He I	\\	
G151.6120-00.4575 	&	He I only	\\	
G168.0627+00.8221 	&	He I only	\\	
G173.6328+02.8064 	&	He I only	\\	
G177.7291-00.3358 	&	Both \brg\ and He I	\\	
G189.0307+00.7821 	&	Both \brg\ and He I	\\	
G189.0323+00.8092 	&	Both \brg\ and He I	\\	
G194.9349-01.2224 	&	He I only	\\	
G203.3166+02.0564	&	\brg\ only	\\	
G213.9180+00.3786 	&	Both \brg\ and He I	\\	
\hline
\end{tabular*}
\end{minipage}
\end{table*}

\label{lastpage}

\end{document}

%%%%%%%%%%%%%%%%%%%%%%%%%%%%%%%%%%%%%%%%%%%%%%%%
\thispagestyle{empty}
\appendix
\setcounter{section}{1}
\setcounter{table}{0}
\begin{table*}
\begin{minipage}{\textwidth}
%\vspace{0.5cm}
\caption{Table of all spectra, showing RA, Dec., observing date,
position angle (PA), total exposure time (Exp. Time), classification,
distance, luminosity, and radio data. The PA was chosen individually
for each object from any images that were available. For the earlier
objects, these were 2MASS images, but for the later runs GLIMPSE
images were used. The aim was to centre on the target object, but
to align the slit to catch any extended emission, outflow, or other
interesting nearby objects in the slit at the same time. Some objects
were observed multiple times with different PAs to get spectra of all
the interesting features in the region (see Section 2). The exposure times were chosen
based on the magnitudes of the target object (shown in Table A2).
The classification process is described in Section 4, and the distance,
luminosity, and radio data were taken from other RMS publications,
listed in Section 1. $^{*}$The final spectrum of G025.6498+01.0491 is a
composite of two spectra, taken on 20070626 and 20070630. 
}
\begin{center}
\begin{tabular*}{\textwidth}{@{\extracolsep{\fill} }lcccrcccccc}
\hline
Name & RA (J2000) & Dec. (J2000) &  Date  & PA (\degr) & Exp. Time (s) & NIR Classification & Distance (kpc) & Luminosity (\lsol)  & Rad. Freq. (GHz) & Rad. Flux (mJy) \\ \hline
G010.8411-02.5919 & 18:19:12.21 & -20:47:29.70 & 20040415 & 13 & 720 & YSO & 1.8 & 2.2 $\times 10^{4}$ & 8.6 & 3.6 \\
G010.8856+00.1221 & 18:09:07.97 & -19:27:23.70 & 20080815 & 0 & 60 & YSO & 13.5 & 9.0 $\times 10^{4}$ & 5 & $<$1.0 \\
G011.9019+00.7265 & 18:08:58.77 & -18:16:29.70 & 20080816 & -90 & 60 & YSO & 3.1 & 2.1 $\times 10^{3}$ & 5 & $<$0.7 \\
G016.7122+01.3119 -- 1 & 18:16:26.38 & -13:46:24.40 & 20040805 & 80 & 90 & YSO & 2.0 & 1.6 $\times 10^{3}$ & 5 & $<$0.6 \\
G016.7122+01.3119 -- 2 & 18:16:27.54 & -13:46:21.40 & 20040805 & 80 & 90 & YSO & 2.0 & 1.6 $\times 10^{3}$ & 5 & $<$0.6 \\
G016.7981+00.1264 & 18:20:55.30 & -14:15:30.70 & 20070711 & -78 & 60 & YSO & 2.0 & 2.6 $\times 10^{3}$ & 5 & $<$0.6 \\
G017.0332+00.7476A & 18:19:07.33 & -13:45:23.20 & 20070622 & 0 & 240 & YSO & 2.0 & 1.6 $\times 10^{3}$ & 5 & $<$0.8  \\
G017.3765+02.2512 & 18:14:21.05 & -12:44:33.20 & 20040806 & 70 & 90 & YSO & 1.2 & 912  & 5 & $<$0.7 \\
G018.3412+01.7681 & 18:17:58.11 & -12:07:24.80 & 20030430 & -90 & 60 & YSO & 2.0 & 1.1 $\times 10^{4}$ & 8.6 & $<$0.4 \\
G018.3706-00.3818 & 18:25:48.37 & -13:06:29.10 & 20070713 & 54 & 60 & YSO & 4.1 & 4.9 $\times 10^{3}$ & 5 & $<$1.2 \\
G018.6608+00.0372 & 18:24:50.23 & -12:39:22.30 & 20070622 & -90 & 240 & YSO & 10.9 & 2.0 $\times 10^{4}$ & 5 & $<$1.0 \\
G019.7540-00.1279 & 18:27:31.59 & -11:45:55.10 & 20070623 & -90 & 120 & HIIR & 8.0 & 4.4 $\times 10^{4}$ & 8.6 & $<$0.4 \\
G019.8922+00.1023 & 18:26:57.29 & -11:32:09.10 & 20070711 & -90 & 780 & YSO & 12.6 & 8.2 $\times 10^{4}$ & 5 & $<$1.1 \\
G019.9224-00.2577 & 18:28:18.96 & -11:40:36.90 & 20070622 & -68 & 240 & YSO & 4.6 & 3.2 $\times 10^{3}$ & 5 & $<$1.1 \\
G020.5143+00.4936 & 18:26:43.24 & -10:48:18.20 & 20080816 & -86.5 & 840 & YSO & 2.3 / 13.6 & 360 / 1.3$\times10^{4}$  & 5 & $<$0.6  \\
G021.3853-00.2537 & 18:31:03.99 & -10:22:44.70 & 20070623 & 67 & 240 & HIIR & 5.5 & 3.2 $\times 10^{4}$ & 5 & 106.6 \\
G021.5624-00.0329 & 18:30:36.06 & -10:07:11.00 & 20080816 & -90 & 60 & YSO & 7.9 & 1.6 $\times 10^{4}$ & 5 & $<$1.0 \\
G023.3891+00.1851 & 18:33:14.32 & -08:23:57.40 & 20070626 & -90 & 40 & YSO & 3.2 & 2.1 $\times 10^{4}$ & 5 & $<$1.0 \\
G023.4394-00.2394 & 18:34:51.69 & -08:33:01.00 & 20080628 & -45 & 60 & HIIR & 5.6 & 2.0 $\times 10^{4}$ & 5 & $<$1.1 \\
G023.4984-00.0373 & 18:34:14.28 & -08:24:15.00 & 20080624 & -90 & 40 & Evolved Star & -- & --  & 5 & $<$1.0 \\
G023.6566-00.1273 & 18:34:51.56 & -08:18:21.40 & 20070626 & -90 & 60 & YSO & 3.2 & 6.2 $\times 10^{3}$ & 5 & $<$1.0 \\
G024.6343-00.3233 & 18:37:22.71 & -07:31:41.70 & 20070621 & 40 & 60 & YSO & 3.3 & 5.7 $\times 10^{3}$ & 5 & $<$0.7 \\
G025.4018+00.0198A & 18:37:34.48 & -06:41:21.10 & 20080807 & -90 & 5 & M Star & 7.7 & 6.5 $\times 10^{3}$ & 5 & $<$1.6 \\
G025.6498+01.0491$^{*}$ & 18:34:20.90 & -05:59:42.60 & -- & 0 & 240 & YSO & 3.1 & 2.1 $\times 10^{4}$ & 8.6 &  3.8 \\
G025.7161+00.0486 & 18:38:02.83 & -06:23:47.90 & 20070626 & -34.01 & 120 & YSO/HIIR & 9.4 & 7.3 $\times 10^{4}$ & 5 & 15.8 \\
G025.8015-00.1570 -- 1 & 18:38:56.37 & -06:24:53.50 & 20070711 & 0 & 60 & YSO/HIIR & 5.6 & 2.8 $\times 10^{4}$ & 5 & 29.2 \\
G025.8015-00.1570 -- 2 & 18:38:56.38 & -06:24:52.63 & 20070711 & 0 & 60 & YSO & 5.6 & 2.8 $\times 10^{4}$ & 5 & 29.2 \\
G026.2020+00.2262 & 18:38:18.52 & -05:52:57.50 & 20080628 & -90 & 40 & YSO & 7.6 & 3.1 $\times 10^{4}$ & 5 & $<$1.1 \\
G026.3050+00.1162 & 18:38:53.55 & -05:50:30.60 & 20080628 & 0 & 40 & Other & -- & --  & 5 & $<$1.0 \\
G026.3176-00.0134 & 18:39:23.01 & -05:53:19.90 & 20070621 & -90 & 60 & Other & -- & --  & 5 & 19.5 \\
G026.3819+01.4057 -- 1 & 18:34:25.82 & -05:10:52.80 & 20040513 & -39.1 & 60 & YSO & 2.3 & 1.4 $\times 10^{4}$ & 5 & $<$0.6 \\
G026.3819+01.4057 -- 2 & 18:34:25.68 & -05:10:50.30 & 20040513 & -39.1 & 60 & YSO & 2.3 & 1.4 $\times 10^{4}$ & 5 & $<$0.6 \\
G026.4207+01.6858 & 18:33:30.55 & -05:01:02.10 & 20040806 & 0 & 1350 & YSO & 2.3 & 1.2 $\times 10^{4}$ & 5 & $<$0.6 \\
G026.4488+01.7425A & 18:33:21.19 & -04:58:05.90 & 20040806 & 0 & 90 & Other & 2.3 & 5.2 $\times 10^{3}$ & 5 & $<$0.7 \\
G026.4488+01.7425B & 18:33:21.21 & -04:58:01.40 & 20040806 & 0 & 90 & HIIR & 2.3 & 6.7 $\times 10^{3}$ & 5 & 5.2 \\
G026.4897-00.5455 & 18:41:35.93 & -05:58:50.80 & 20070711 & 0 & 60 & PPN & 4.4 & 6.2 $\times 10^{3}$ & 5 & $<$0.6 \\
G026.4958+00.7105 -- 1 & 18:37:07.29 & -05:24:01.30 & 20070623 & 6 & 240 & YSO & 11.9 & 2.0 $\times 10^{4}$ & 5 & $<$0.6 \\
G026.4958+00.7105 -- 2 & 18:37:07.31 & -05:23:58.30 & 20070623 & 6 & 240 & YSO & 11.9 & 2.0 $\times 10^{4}$ & 5 & $<$0.6 \\
G026.5972-00.0241 & 18:39:55.89 & -05:38:45.04 & 20070623 & 0 & 240 & YSO/HIIR & 1.9 & 2.4 $\times 10^{3}$ & 8.6 & $<$0.4 \\
G027.2220+00.1361 -- 1 & 18:40:30.75 & -05:01:03.96 & 20080725 & -65 & 240 & Evolved Star & 7.6 & 1.0 $\times 10^{4}$ & 5 & $<$0.7  \\
G027.2220+00.1361 -- 2 & 18:40:30.83 & -05:01:04.52 & 20080725 & -65 & 240 & YSO & 7.6 & 1.0 $\times 10^{4}$ & 5 & $<$0.7 \\
G027.7954-00.2772 & 18:43:02.29 & -04:41:49.10 & 20070711 & 30 & 60 & YSO & 3.1 & 2.1 $\times 10^{3}$ & 5 & $<$1.0 \\
G028.2325+00.0394 & 18:42:42.50 & -04:09:46.00 & 20080725 & 21 & 120 & YSO & 6.2 & 3.3 $\times 10^{3}$ & 5 & $<$1.5 \\
G028.3046-00.3871A & 18:44:22.00 & -04:17:39.50 & 20070626 & 45 & 60 & YSO & 5.2 & 1.0 $\times 10^{4}$ & 5 & $<$1.4  \\
G028.3373+00.1189 & 18:42:37.12 & -04:02:02.20 & 20080722 & -90 & 960 & YSO & 5.0 & 1.3 $\times 10^{4}$ & 5 & $<$1.2 \\
G028.7799+03.4978 & 18:31:23.97 & -02:05:29.50 & 20040428 & 0 & 14 & Other & -- & --  & 5 & 1.3 \\
G028.8621+00.0657 & 18:43:46.45 & -03:35:31.30 & 20070622 & -60 & 120 & YSO & 6.2 & 1.0 $\times 10^{5}$ & 5 & $<$1.2 \\
G030.1468+01.4075 & 18:41:20.60 & -01:50:07.60 & 20040809 & 90 & 150 & PPN & -- & --  & 5 & $<$0.9 \\ 

\end{tabular*}
\end{center}
\end{minipage}
\end{table*}

%%%%%%%%%%%%%%%%%%%%%%%%%%%%%%%%%%%%%%%%%%%%%%%%
\thispagestyle{empty}
\appendix
\setcounter{section}{1}
\setcounter{table}{1}
\begin{table*}

\begin{minipage}{\textwidth}
\caption{Table showing catalogue and
measured magnitudes for all spectra. For the catalogue magnitudes, the
most appropriate of 2MASS Point Source Catalogue or  UKIDSS Galactic
Plane Survey are shown. These are the magnitudes used for calculating
extinction etc. and in Figure 1 for comparing measured magnitudes
with catalogue magnitudes (see Section 3). `--' is shown when there are no catalogue
magnitudes available at that wavelength. Where `--' is marked in the
error field for the catalogue magnitudes, the magnitude shown is a
lower limit (i.e. an upper limit on the flux for non-detections or
confused fields). Note that this only happens for 2MASS PSC
magnitudes; UKIDSS GPS have a different quality flagging system, as
described on the WFCAM Science Archive website. The
measured magnitudes were calculated from flux measurements in the
spectra (see Section 3). `--' is marked in the upper error field
of the H-Band measured magnitude to indicate where the H-Band flux is
so low that the flux errorbar goes below zero, which makes the
magnitude error incalculable.  Several of the objects are invisible at
H in the spectrum. In these cases, `--' is marked in the measured
H-Band magnitude fields. 
}  
\begin{center}

\begin{tabular*}{\textwidth}{@{\extracolsep{\fill} }lllllrrrrrr}
\hline
Name & Catalogue & \multicolumn{3}{l}{Catalogue Magnitudes}           & \multicolumn{6}{l}{Measured Magnitudes}           \\
 &  & J-Band    & H-Band  & K-Band  & H-Band & H L. Err. & H U. Err. & K-Band & K L. Err. & K U. Err. \\ \hline
G010.8411-02.5919 & 2MASS PSC &  $>$ 16.3 & 13.2 $\pm$ 0.1 & 9.67 $\pm$ 0.05 & 13.8 & 0.2 & 0.2 & 10.8 & 0.3 & 0.4 \\
G010.8856+00.1221 & UKIDSS GPS & 18.67 $\pm$ 0.10 & 13.190 $\pm$ 0.002 & 9.769 $\pm$ 0.000 & 13.09 & 0.06 & 0.07 & 9.97 & 0.02 & 0.02 \\
G011.9019+00.7265 & UKIDSS GPS & 16.46 $\pm$ 0.01 & 13.932 $\pm$ 0.003 & 11.365 $\pm$ 0.001 & 9.4 & 0.1 & 0.1 & 6.8 & 0.2 & 0.2 \\
G016.7122+01.3119 -- 1 & 2MASS PSC & 13.44 $\pm$ 0.06 & 10.67 $\pm$ 0.04 & 8.70 $\pm$ 0.03 & 10.16 & 0.03 & 0.03 & 8.12 & 0.03 & 0.03 \\
G016.7122+01.3119 -- 2 & 2MASS PSC &  $>$ 16.5 & 14.5 $\pm$ 0.1 & 11.14 $\pm$ 0.04 & 16 & 1 & 1 & 11.52 & 0.06 & 0.06 \\
G016.7981+00.1264 & UKIDSS GPS & 16.171 $\pm$ 0.009 & 13.168 $\pm$ 0.001 & 9.957 $\pm$ 0.000 & 13.0 & 0.1 & 0.1 & 10.15 & 0.03 & 0.03 \\
G017.0332+00.7476A & UKIDSS GPS & --   & --   & 14.93 $\pm$ 0.01 & 19 & 3 & -- & 15.8 & 0.3 & 0.4 \\
G017.3765+02.2512 & 2MASS PSC & 13.8 $\pm$ 0.1 & 11.26 $\pm$ 0.10 & 9.17 $\pm$ 0.05 & 10.04 & 0.04 & 0.05 & 7.88 & 0.03 & 0.04 \\
G018.3412+01.7681 & 2MASS PSC &  $>$ 15.3 & 12.91 $\pm$ 0.09 & 9.25 $\pm$ 0.03 & 13.53 & 0.07 & 0.07 & 9.94 & 0.05 & 0.05 \\
G018.3706-00.3818 & UKIDSS GPS & 14.614 $\pm$ 0.003 & 11.925 $\pm$ 0.001 & 9.848 $\pm$ 0.000 & 11.70 & 0.06 & 0.06 & 9.53 & 0.03 & 0.03 \\
G018.6608+00.0372 & UKIDSS GPS & --   & --   & 12.756 $\pm$ 0.003 & 15.6 & 0.4 & 0.6 & 12.48 & 0.07 & 0.07 \\
G019.7540-00.1279 & UKIDSS GPS & 17.00 $\pm$ 0.02 & 14.457 $\pm$ 0.005 & 12.101 $\pm$ 0.002 & 14.6 & 0.5 & 0.8 & 12.54 & 0.05 & 0.05 \\
G019.8922+00.1023 & UKIDSS GPS & 13.895 $\pm$ 0.002 & 12.433 $\pm$ 0.001 & 10.417 $\pm$ 0.001 & 12.62 & 0.04 & 0.04 & 10.86 & 0.03 & 0.03 \\
G019.9224-00.2577 & UKIDSS GPS & 18.07 $\pm$ 0.06 & 15.41 $\pm$ 0.01 & 12.896 $\pm$ 0.003 & 14.6 & 0.3 & 0.4 & 12.56 & 0.06 & 0.07 \\
G020.5143+00.4936 & UKIDSS GPS & 13.607 $\pm$ 0.001 & 12.152 $\pm$ 0.001 & 11.377 $\pm$ 0.001 & 13 & 1 & -- & 10.6 & 0.4 & 0.6 \\
G021.3853-00.2537 & UKIDSS GPS & --   & --   & 13.298 $\pm$ 0.004 & 16 & 1 & 1 & 13.8 & 0.1 & 0.2 \\
G021.5624-00.0329 & UKIDSS GPS & --   & --   & 10.641 $\pm$ 0.001 & 9.2 & 0.1 & 0.1 & 5.1 & 0.2 & 0.2 \\
G023.3891+00.1851 & 2MASS PSC & 15.77 $\pm$ 0.09 & 11.61 $\pm$ 0.03 & 8.38 $\pm$ 0.03 & 9.03 & 0.04 & 0.04 & 6.07 & 0.03 & 0.04 \\
G023.4394-00.2394 & UKIDSS GPS & --   & 14.950 $\pm$ 0.005 & 11.695 $\pm$ 0.001 & 14.8 & 0.4 & 0.5 & 11.80 & 0.03 & 0.03 \\
G023.4984-00.0373 & 2MASS PSC & 11.40 $\pm$ 0.02 & 9.56 $\pm$ 0.03 & 8.53 $\pm$ 0.02 & 7.17 & 0.02 & 0.02 & 6.278 & 0.009 & 0.009 \\
G023.6566-00.1273 & UKIDSS GPS & --   & 14.070 $\pm$ 0.003 & 9.648 $\pm$ 0.000 & 12.3 & 0.3 & 0.4 & 7.99 & 0.04 & 0.04 \\
G024.6343-00.3233 & 2MASS PSC &  $>$ 15.8 & 13.31 $\pm$ 0.05 & 10.04 $\pm$ 0.03 & 12.9 & 0.2 & 0.2 & 9.9 & 0.1 & 0.1 \\
G025.4018+00.0198A & 2MASS PSC & 9.35 $\pm$ 0.03 & 6.88 $\pm$ 0.03 & 5.66 $\pm$ 0.02 & 7.3 & 0.1 & 0.1 & 6.15 & 0.07 & 0.08 \\
G025.6498+01.0491$^{*}$ & UKIDSS GPS & --   & --   & 12.321 $\pm$ 0.002 & 17 & 1 & -- & 12.4 & 0.1 & 0.1 \\
G025.7161+00.0486 & UKIDSS GPS & 16.97 $\pm$ 0.02 & --   & --   & 14.3 & 0.1 & 0.2 & 12.42 & 0.04 & 0.05 \\
G025.8015-00.1570 -- 1 & UKIDSS GPS & 16.49 $\pm$ 0.01 & 13.647 $\pm$ 0.002 & 11.067 $\pm$ 0.001 & 14.3 & 0.1 & 0.1 & 12.06 & 0.04 & 0.05 \\
G025.8015-00.1570 -- 2 & \multicolumn{1}{c}{--} & --   & --   & --   & 14.7 & 0.1 & 0.1 & 12.00 & 0.03 & 0.04 \\
G026.2020+00.2262 & 2MASS PSC & 14.21 $\pm$ 0.04 & 10.61 $\pm$ 0.03 & 8.25 $\pm$ 0.02 & 11.02 & 0.06 & 0.06 & 8.20 & 0.03 & 0.02 \\
G026.3050+00.1162 & 2MASS PSC & 10.69 $\pm$ 0.03 & 9.39 $\pm$ 0.03 & 8.17 $\pm$ 0.03 & 9.52 & 0.02 & 0.02 & 8.28 & 0.01 & 0.01 \\
G026.3176-00.0134 & 2MASS PSC & 13.19 $\pm$ 0.03 & 10.09 $\pm$ 0.03 & 8.39 $\pm$ 0.02 & 10.67 & 0.05 & 0.06 & 8.95 & 0.03 & 0.03 \\
G026.3819+01.4057 -- 1 & UKIDSS GPS & 16.176 $\pm$ 0.008 & 13.802 $\pm$ 0.002 & 11.819 $\pm$ 0.001 & 14.4 & 0.1 & 0.1 & 12.52 & 0.04 & 0.04 \\
G026.3819+01.4057 -- 2 & UKIDSS GPS & 14.161 $\pm$ 0.002 & 12.200 $\pm$ 0.001 & 9.745 $\pm$ 0.000 & 12.19 & 0.03 & 0.03 & 10.16 & 0.03 & 0.03 \\
G026.4207+01.6858 & UKIDSS GPS & 16.165 $\pm$ 0.008 & 13.248 $\pm$ 0.001 & 10.207 $\pm$ 0.000 & 14.71 & 0.04 & 0.05 & 12.50 & 0.03 & 0.03 \\
G026.4488+01.7425A & UKIDSS GPS & 10.899 $\pm$ 0.000 & 11.107 $\pm$ 0.000 & 10.139 $\pm$ 0.000 & 10.34 & 0.03 & 0.03 & 10.10 & 0.01 & 0.02 \\
G026.4488+01.7425B & UKIDSS GPS & 14.910 $\pm$ 0.003 & 13.709 $\pm$ 0.002 & 12.234 $\pm$ 0.002 & 14.27 & 0.06 & 0.07 & 12.99 & 0.05 & 0.05 \\
G026.4897-00.5455 & UKIDSS GPS & --   & --   & 10.402 $\pm$ 0.001 & 16 & 2 & -- & 10.55 & 0.07 & 0.08 \\
G026.4958+00.7105 -- 1 & UKIDSS GPS & --   & --   & 15.01 $\pm$ 0.02 & -- & -- & -- & 15.4 & 0.3 & 0.4 \\
G026.4958+00.7105 -- 2 & UKIDSS GPS & --   & --   & 15.89 $\pm$ 0.04 & 21 & 4 & -- & 17 & 1 & 1 \\
G026.5972-00.0241 & UKIDSS GPS & 18.42 $\pm$ 0.08 & --   & 15.83 $\pm$ 0.04 & 20 & 2 & -- & 17 & 1 & 1 \\
G027.2220+00.1361 -- 1 & UKIDSS GPS & 14.749 $\pm$ 0.004 & 14.180 $\pm$ 0.005 & 14.54 $\pm$ 0.02 & 14.15 & 0.08 & 0.08 & 13.76 & 0.05 & 0.06 \\
G027.2220+00.1361 -- 2 & \multicolumn{1}{c}{--} & --   & --   & --   & 19 & 1 & -- & 16 & 1 & 1 \\
G027.7954-00.2772 & UKIDSS GPS & --   & 16.44 $\pm$ 0.03 & 10.890 $\pm$ 0.001 & 15.8 & 0.9 & -- & 10.91 & 0.06 & 0.07 \\
G028.2325+00.0394 & UKIDSS GPS & --   & --   & 12.306 $\pm$ 0.002 & 17 & 1 & 1 & 12.53 & 0.06 & 0.06 \\
G028.3046-00.3871A & UKIDSS GPS & 16.599 $\pm$ 0.009 & 14.054 $\pm$ 0.003 & 11.898 $\pm$ 0.001 & 12.34 & 0.09 & 0.09 & 10.21 & 0.05 & 0.05 \\
G028.3373+00.1189 & UKIDSS GPS & --   & 15.664 $\pm$ 0.009 & 12.748 $\pm$ 0.002 & 13.0 & 0.9 & -- & 10.1 & 0.1 & 0.1 \\ 

\end{tabular*}

\end{center}
\end{minipage}
\end{table*}

%%%%%%%%%%%%%%%%%%%%%%%%%%%%%%%%%%%%%%%%%%%%%%%%
\thispagestyle{empty}
\appendix
\setcounter{section}{1}
\setcounter{table}{1}
\begin{table*}

\begin{minipage}{\textwidth}
\caption{Page 3.}  
\begin{center}

\begin{tabular*}{\textwidth}{@{\extracolsep{\fill} }lllllrrrrrr}
\hline
Name & Catalogue & \multicolumn{3}{l}{Catalogue Magnitudes}           & \multicolumn{6}{l}{Measured Magnitudes}           \\
 &  & J-Band    & H-Band  & K-Band  & H-Band & H L. Err. & H U. Err. & K-Band & K L. Err. & K U. Err. \\ \hline
G049.4883-00.3545A & UKIDSS GPS & 17.01 $\pm$ 0.02 & 14.709 $\pm$ 0.004 & 12.731 $\pm$ 0.002 & 12.1 & 0.3 & 0.4 & 10.30 & 0.06 & 0.06 \\
G049.4883-00.3545B & UKIDSS GPS & --   & 16.91 $\pm$ 0.02 & 14.053 $\pm$ 0.005 & 15 & 1 & -- & 12.0 & 0.2 & 0.2 \\
G049.4970-00.4718 & UKIDSS GPS & 14.620 $\pm$ 0.002 & 13.367 $\pm$ 0.001 & 12.481 $\pm$ 0.002 & 12.24 & 0.04 & 0.05 & 11.52 & 0.04 & 0.04 \\
G049.5373-00.3929 & 2MASS PSC & 9.54 $\pm$ 0.03 & 8.13 $\pm$ 0.04 & 7.58 $\pm$ 0.03 & 9.45 & 0.07 & 0.08 & 9.01 & 0.05 & 0.05 \\
G050.0721+00.5591 & UKIDSS GPS & 17.19 $\pm$ 0.02 & 14.873 $\pm$ 0.005 & 12.702 $\pm$ 0.002 & 14.9 & 0.2 & 0.3 & 12.80 & 0.05 & 0.05 \\
G050.2213-00.6063 & 2MASS PSC &  $>$ 13.7 &  $>$ 11.4 & 9.84 $\pm$ 0.07 & 14.18 & 0.05 & 0.05 & 10.92 & 0.05 & 0.05 \\
G050.2844-00.3925A & UKIDSS GPS & 18.39 $\pm$ 0.05 & 14.502 $\pm$ 0.003 & 11.411 $\pm$ 0.001 & 12.6 & 0.1 & 0.1 & 9.74 & 0.04 & 0.04 \\
G050.4802+00.7058 & \multicolumn{1}{c}{--} & --   & --   & --   & 13.22 & 0.09 & 0.10 & 11.35 & 0.04 & 0.04 \\
G052.2025+00.7217A & UKIDSS GPS & 16.46 $\pm$ 0.01 & 13.377 $\pm$ 0.001 & 10.106 $\pm$ 0.000 & 12.90 & 0.06 & 0.06 & 9.85 & 0.06 & 0.06 \\
G052.2078+00.6890 & 2MASS PSC &  $>$ 15.3 &  $>$ 13.7 & 12.21 $\pm$ 0.04 & 13.2 & 0.3 & 0.5 & 10.87 & 0.04 & 0.04 \\
G052.5405-00.9272 & 2MASS PSC &  $>$ 16.6 &  $>$ 14.5 & 11.85 $\pm$ 0.05 & 14 & 1 & 1 & 11.4 & 0.1 & 0.1 \\
G052.9217+00.4142 & UKIDSS GPS & 19.06 $\pm$ 0.09 & 15.715 $\pm$ 0.009 & 12.225 $\pm$ 0.001 & 14.4 & 0.3 & 0.4 & 11.38 & 0.05 & 0.06 \\
G052.9221-00.4892 & 2MASS PSC &  $>$ 15.9 & 14.1 $\pm$ 0.2 &  $>$ 11.2 & 13.6 & 0.4 & 0.6 & 11.26 & 0.03 & 0.03 \\
G053.1417+00.0705 & UKIDSS GPS & --   & --   & 12.720 $\pm$ 0.002 & 18 & 1 & -- & 13.08 & 0.09 & 0.10 \\
G053.5343-00.7943 & 2MASS PSC & 13.80 $\pm$ 0.03 & 10.70 $\pm$ 0.02 & 8.31 $\pm$ 0.02 & 12.13 & 0.04 & 0.04 & 9.64 & 0.02 & 0.02 \\
G055.1581-00.2991A & UKIDSS GPS & 18.59 $\pm$ 0.05 & 15.063 $\pm$ 0.005 & 12.174 $\pm$ 0.001 & 13.8 & 0.2 & 0.2 & 11.23 & 0.08 & 0.09 \\
G055.3704+00.1858 & 2MASS PSC & 11.33 $\pm$ 0.03 & 7.72 $\pm$ 0.02 &  $>$ 6.6 & 6.7 & 0.1 & 0.1 & 4.60 & 0.03 & 0.03 \\
G056.3694-00.6333 & UKIDSS GPS & --   & 15.322 $\pm$ 0.007 & 12.496 $\pm$ 0.002 & 15.3 & 0.5 & 0.8 & 12.6 & 0.1 & 0.1 \\
G056.4120-00.0277 & 2MASS PSC & 12.45 $\pm$ 0.02 & 9.93 $\pm$ 0.02 & 8.05 $\pm$ 0.02 & 8.64 & 0.04 & 0.04 & 6.81 & 0.04 & 0.04 \\
G057.5474-00.2717A & UKIDSS GPS & 16.52 $\pm$ 0.01 & 13.526 $\pm$ 0.002 & 11.131 $\pm$ 0.001 & 14.3 & 0.1 & 0.1 & 12.23 & 0.04 & 0.04 \\
G059.3614-00.2068 & UKIDSS GPS & 14.074 $\pm$ 0.002 & 11.847 $\pm$ 0.001 & 9.879 $\pm$ 0.000 & 11.28 & 0.04 & 0.04 & 9.44 & 0.02 & 0.02 \\
G059.6403-00.1812A & UKIDSS GPS & 15.323 $\pm$ 0.004 & 13.488 $\pm$ 0.001 & 11.422 $\pm$ 0.001 & 14.0 & 0.1 & 0.1 & 12.12 & 0.03 & 0.03 \\
G059.7831+00.0648 & UKIDSS GPS & --   & 16.32 $\pm$ 0.01 & 11.662 $\pm$ 0.001 & 16 & 2 & -- & 12.02 & 0.05 & 0.05 \\
G059.8329+00.6729 & UKIDSS GPS & 19.1 $\pm$ 0.1 & 16.29 $\pm$ 0.02 & 14.138 $\pm$ 0.007 & 16 & 1 & 1 & 14.2 & 0.3 & 0.3 \\
G060.5750-00.1861 & 2MASS PSC & 15.51 $\pm$ 0.10 & 13.9 $\pm$ 0.1 & 11.59 $\pm$ 0.06 & 14.0 & 0.2 & 0.3 & 11.81 & 0.09 & 0.09 \\
G063.1140+00.3416 & 2MASS PSC &  $>$ 16.9 & 13.17 $\pm$ 0.05 & 9.67 $\pm$ 0.02 & 10.84 & 0.05 & 0.06 & 7.13 & 0.06 & 0.06 \\
G065.7798-02.6121 & 2MASS PSC &  $>$ 14.8 & 12.52 $\pm$ 0.06 & 10.49 $\pm$ 0.04 & 9.91 & 0.01 & 0.01 & 9.63 & 0.04 & 0.04 \\
G068.2040+00.2387 & UKIDSS GPS & 17.60 $\pm$ 0.02 & 14.113 $\pm$ 0.002 & 11.448 $\pm$ 0.001 & 13.6 & 0.1 & 0.1 & 11.20 & 0.02 & 0.03 \\
G069.5395-00.9754 & \multicolumn{1}{c}{--} & --   & --   & --   & 18 & 1 & -- & 16 & 1 & 4 \\
G070.3329+01.5864 -- 1 & 2MASS PSC &  $>$ 15.0 & 13.28 $\pm$ 0.06 & 11.41 $\pm$ 0.05 & 15.3 & 0.3 & 0.4 & 14.1 & 0.2 & 0.2 \\
G070.3329+01.5864 -- 2 & 2MASS PSC &  $>$ 15.0 & 13.28 $\pm$ 0.06 & 11.41 $\pm$ 0.05 & 13.0 & 0.2 & 0.2 & 11.41 & 0.07 & 0.08 \\
G070.6740+01.1919 & 2MASS PSC & 10.33 $\pm$ 0.03 & 8.24 $\pm$ 0.02 & 6.56 $\pm$ 0.03 & 8.54 & 0.03 & 0.03 & 6.91 & 0.03 & 0.03 \\
G071.8944+01.3107 & 2MASS PSC &  $>$ 18.2 & 14.80 $\pm$ 0.08 & 11.74 $\pm$ 0.04 & 14.4 & 0.2 & 0.3 & 11.51 & 0.08 & 0.08 \\
G072.2479+00.2617B & UKIDSS GPS & 17.86 $\pm$ 0.03 & 15.429 $\pm$ 0.009 & 13.343 $\pm$ 0.003 & 15.2 & 0.2 & 0.3 & 13.25 & 0.03 & 0.03 \\
G073.0633+01.7958 & 2MASS PSC &  $>$ 15.4 &  $>$ 13.3 & 12.12 $\pm$ 0.04 & 13.6 & 0.2 & 0.2 & 11.57 & 0.06 & 0.06 \\
G073.6525+00.1944 & 2MASS PSC & 13.94 $\pm$ 0.03 & 11.47 $\pm$ 0.04 & 9.57 $\pm$ 0.02 & 11.31 & 0.03 & 0.03 & 9.40 & 0.04 & 0.04 \\
G073.6952-00.9996 & 2MASS PSC & 12.89 $\pm$ 0.04 & 10.63 $\pm$ 0.04 & 8.05 $\pm$ 0.02 & 11.16 & 0.02 & 0.02 & 8.41 & 0.03 & 0.03 \\
G075.6014+01.6394 & 2MASS PSC & 16.1 $\pm$ 0.1 & 14.01 $\pm$ 0.05 & 12.77 $\pm$ 0.04 & 14.54 & 0.09 & 0.09 & 13.48 & 0.03 & 0.03 \\
G075.7666+00.3424 -- 1 & UKIDSS GPS & 14.960 $\pm$ 0.003 & 12.412 $\pm$ 0.001 & 10.501 $\pm$ 0.001 & 10.6 & 0.1 & 0.1 & 8.52 & 0.03 & 0.03 \\
G075.7666+00.3424 -- 2 & UKIDSS GPS & --   & 14.855 $\pm$ 0.005 & 11.448 $\pm$ 0.001 & 14.2 & 0.3 & 0.3 & 11.06 & 0.05 & 0.05 \\
G076.0902+00.1412 & UKIDSS GPS & --   & 16.40 $\pm$ 0.02 & 12.310 $\pm$ 0.001 & 16.2 & 0.4 & 0.6 & 12.45 & 0.07 & 0.08 \\
G076.1877+00.0974 & UKIDSS GPS & --   & 15.67 $\pm$ 0.01 & 12.832 $\pm$ 0.002 & 15 & 2 & -- & 11.5 & 0.1 & 0.1 \\
G076.3829-00.6210 & 2MASS PSC & 10.36 $\pm$ 0.03 & 7.73 $\pm$ 0.02 & 5.87 $\pm$ 0.02 & 8.25 & 0.02 & 0.02 & 6.47 & 0.02 & 0.02 \\
G077.3455+04.4309 & UKIDSS GPS & 12.438 $\pm$ 0.001 & 11.180 $\pm$ 0.000 & 10.145 $\pm$ 0.000 & 9.4 & 0.1 & 0.1 & 8.23 & 0.06 & 0.07 \\
G077.4052-01.2136 & 2MASS PSC &  $>$ 17.1 & 15.5 $\pm$ 0.1 & 12.55 $\pm$ 0.04 & 15.8 & 0.2 & 0.3 & 12.82 & 0.09 & 0.10 \\
G077.4622+01.7600 -- 1 & UKIDSS GPS & --   & --   & 10.914 $\pm$ 0.001 & 13.09 & 0.08 & 0.09 & 11.46 & 0.03 & 0.03 \\
G077.4622+01.7600 -- 2 & 2MASS PSC &  $>$ 12.9 & 11.02 $\pm$ 0.09 &  $>$ 9.1 & 14.24 & 0.07 & 0.08 & 12.88 & 0.06 & 0.06 \\
G077.8999+01.7678 & UKIDSS GPS & 16.258 $\pm$ 0.008 & 13.213 $\pm$ 0.001 & 10.345 $\pm$ 0.000 & 13.07 & 0.03 & 0.03 & 10.10 & 0.04 & 0.04 \\
G077.9550+00.0058 & UKIDSS GPS & 14.731 $\pm$ 0.003 & 13.007 $\pm$ 0.001 & 11.882 $\pm$ 0.001 & 13.15 & 0.06 & 0.06 & 12.09 & 0.03 & 0.04 \\
G078.1224+03.6320 & UKIDSS GPS & 14.614 $\pm$ 0.002 & 13.130 $\pm$ 0.001 & 12.047 $\pm$ 0.001 & 15.8 & 0.2 & 0.3 & 13.0 & 0.2 & 0.2 \\
G078.4373+02.6584B & UKIDSS GPS & --   & 14.936 $\pm$ 0.004 & 11.281 $\pm$ 0.001 & 14.7 & 0.2 & 0.2 & 11.33 & 0.05 & 0.05 \\ 
G078.4705-00.1830 & UKIDSS GPS & 19.8 $\pm$ 0.1 & 16.27 $\pm$ 0.01 & 13.426 $\pm$ 0.003 & 17 & 1 & 1 & 14.04 & 0.05 & 0.05 \\
G078.4754+01.0421 & 2MASS PSC &  $>$ 17.2 & 12.95 $\pm$ 0.04 & 8.72 $\pm$ 0.02 & 13.7 & 0.1 & 0.1 & 9.10 & 0.06 & 0.06 \\

\end{tabular*}

\end{center}
\end{minipage}
\end{table*}